\documentclass[%
 reprint,
superscriptaddress,
 amsmath,amssymb,
prb,
]{revtex4-2}

\usepackage{graphicx, subfloat}
\usepackage{dcolumn}
\usepackage{bm}
\usepackage{amsfonts,braket}
\usepackage{hyperref}
\usepackage{color}
\usepackage[utf8]{inputenc}
\usepackage{csquotes}
\usepackage[T1]{fontenc}
\usepackage{multirow}
\usepackage{tabularx}
\usepackage{siunitx, booktabs}
\usepackage{placeins}
\graphicspath{{figures/}}

\newcommand{\subkn}[2]{_{\mathbf{#1}{#2}}}
\newcommand{\kn}{_{\mathbf{k}{n}}}
\newcommand{\qv}{_{\mathbf{q}{\nu}}}
\newcommand{\wqv}{\omega_{\mathbf{q}{\nu}}}

\newcommand{\qpoint}{$\mathbf{q}$-point }
\newcommand{\kpoint}{$\mathbf{k}$-point }
\newcommand{\qpoints}{$\mathbf{q}$-points }

\DeclareMathOperator{\e}{e}
\newcommand{\fro}{Fr\"ohlich }

\newcolumntype{d}{D{.}{.}{-1}}
\newcolumntype{Z}{D..{4.2}}
\newcolumntype{Y}{>{\centering\arraybackslash}X}
\newcolumntype{R}[1]{>{\raggedleft\arraybackslash}p{#1}}
\newcolumntype{C}[1]{>{\centering\arraybackslash}p{#1}}

\begin{document}


\title{Effect of spin-orbit coupling on the zero-point renormalization of the electronic band gap in cubic materials: First-principles calculations and generalized Fr\"ohlich model}

\author{V\'eronique Brousseau-Couture}
\email{veronique.brousseau.couture@umontreal.ca}
\affiliation{%
D\'epartement de physique, Universit\'e de Montr\'eal, C.P. 6128, Succursale Centre-Ville, Montr\'eal, Qu\'ebec, Canada H3C 3J7
}%

\author{Xavier Gonze}%
\affiliation{%
Institute of Condensed Matter and Nanosciences, UCLouvain, B-1348 Louvain-la-Neuve, Belgium
}%

\author{Michel C\^ot\'e}
\affiliation{%
D\'epartement de physique, Universit\'e de Montr\'eal, C.P. 6128, Succursale Centre-Ville, Montr\'eal, Qu\'ebec, Canada H3C 3J7
}%

\date{\today}

\begin{abstract}
The electronic structure of semiconductors and insulators is affected by ionic motion through electron-phonon interaction, yielding temperature-dependent band gap energies and zero-point renormalization (ZPR) at absolute zero temperature. For polar materials, the most significant contribution to the band gap ZPR can be understood in terms of the \fro model, which focuses on the non-adiabatic interaction between an electron and the macroscopic electrical polarization created by a long-wavelength optical longitudinal phonon mode. On the other hand, spin-orbit interaction (SOC) modifies the bare electronic structure, which will, in turn, affect the electron-phonon interaction and the ZPR. We present a comparative investigation of the effect of SOC on the band gap ZPR of twenty semiconductors and insulators with cubic symmetry using first-principles calculations. We observe a SOC-induced decrease of the ZPR, up to 30\%, driven by the valence band edge, which almost entirely originates from the modification of the bare electronic eigenenergies and the decrease of the hole effective masses near the $\Gamma$ point. We also incorporate SOC into a generalized \fro model, addressing the Dresselhaus splitting which occurs in non-centrosymmetric materials, and confirm that the predominance of non-adiabatic effects on the band gap ZPR of polar materials is unchanged when including SOC. Our generalized \fro model with SOC provides a reliable estimate of the SOC-induced decrease of the polaron formation energy obtained from first principles and brings to light some fundamental subtleties in the numerical evaluation of the effective masses with SOC for non-centrosymmetric materials. We finally warn about a possible breakdown of the parabolic approximation, one of the most fundamental assumptions of the \fro model, within the physically relevant energy range of the \fro interaction for materials with high phonon frequencies treated with SOC.\\
\newline
\textit{This is a post-peer-review version of the article published in Physical Review B, which includes the Supplemental Material in the main file.}
\end{abstract}

\maketitle

\section{\label{sec:intro}Introduction}
Electron-phonon interaction (EPI) has been widely investigated from a theoretical point of view since the late 1940s, through pioneering works of Pekar~\cite{pekar_local_1946}, Landau and Pekar~\cite{landau_effective_1948}, Fr\"ohlich~\cite{frohlich_properties_1950}, and Feynman~\cite{feynman_slow_1955}, and numerous subsequent works~\cite{devreese_frohlich_2009, giustino_electron-phonon_2017}. Using model Hamiltonians, those first theories essentially addressed the interaction of an electron in an isotropic, continuous medium with the macroscopic polarization induced by longitudinal optical (LO) long-range lattice vibrations, which yields in a correlated state called a polaron. The \fro model since became the foundation stone of modern large-polaron studies. From a first-principles point of view, the works of Allen, Heine and Cardona~\cite{allen_theory_1976, allen_theory_1981, allen_temperature_1983} (AHC) in the 1980s clarified earlier theories by Fan~\cite{fan_temperature_1951} and Anton\v{c}\'{i}k~\cite{antoncik_theory_1955}. They provided a unified formalism for the EPI self-energy, rooted in many-body perturbation theory, which addresses all types of lattice vibrations. 

Despite their fundamentally different perspectives, the model Hamiltonian and first-principles approaches address the same problem, namely, the consequences of EPI on the electronic structure. Amongst numerous effects on transport and optical properties of materials~\cite{giustino_electron-phonon_2017}, EPI modifies the quasiparticle energy and introduces finite quasiparticle lifetimes, which depend on the phonon population at a given temperature. As a consequence, the electronic structure does not only acquire a temperature dependence: it is affected even at absolute zero temperature, through the zero-point motion of the ions. This $T=0$~K correction is known as the zero-point renormalization (ZPR). From the \fro model perspective, the band edge ZPR corresponds to the polaron formation energy.

In recent years, considerable efforts have been directed towards tackling the \fro interaction within the full complexity of real materials as captured by first-principles methods (see Ref.~\cite{miglio_predominance_2020, guster_frohlich_2021} and references therein). Among others, Sio \textit{et al.}~\cite{sio_polarons_2019,sio_ab_2019} developed a first-principles theory of polarons, later reformulated using a variational principle~\cite{vasilchenko_variational_2022}. More recently, Lafuente-Bartolome \textit{et al.} proposed a self-consistent many-body Green's function theory which simultaneously addresses phonon-induced band structure renormalization and small polaron formation~\cite{lafuente-bartolome_unified_2022,lafuente-bartolome_abinitio_2022}. From another perspective, Houtput and Tempere~\cite{houtput_beyond-frohlich_2020} derived anharmonic corrections to the \fro Hamiltonian, and Kandolf \textit{et al.}~\cite{kandolf_manybody_2022} and Macheda \textit{et al.}~\cite{macheda_frohlich_2022} investigated the \fro interaction in doped solids. Other works proposed models retaining 
certain fundamental assumptions of the original \fro model while lifting some of its hypotheses. Schlipf \textit{et al.}~\cite{schlipf_carrier_2018} addressed the case of multiple phonon branches, 
relying on the first-principles \fro vertex proposed by Verdi and Giustino~\cite{verdi_frohlich_2015}. Miglio \textit{et al.}~\cite{miglio_predominance_2020} introduced a generalized \fro model (gFr), based on a simplified electron-phonon vertex, that allows for multiple phonon branches, degenerate band extrema and anisotropic band warping. The authors used this model to reveal the predominance of non-adiabatic effects in the ZPR of semiconductors and insulators and explain why including such effects in calculations is essential to obtain an agreement between the first-principles band gap ZPR (ZPR$_{\rm{g}}$) and experimental data. Their gFr model was recently used to obtain polaron effective masses and localization lengths in cubic materials~\cite{guster_frohlich_2021}, as well as to investigate the domain of applicability of the \fro model using a high-throughput computational framework~\cite{demelo_highthroughput_2022}.

One question which remains unaddressed in Ref.~\cite{miglio_predominance_2020} is the effect spin-orbit coupling (SOC). It is well known that SOC lifts the spin degeneracy of the Bloch states throughout the Brillouin zone, except at time-reversal invariant $\mathbf{k}$~points. For the valence band maximum (VBM) of cubic materials, which is triply-degenerate when neglecting SOC, this leads to a \mbox{$4+2$} degeneracy: the two split-off bands are moved to lower energies compared to the heavy hole and light hole bands, which remain degenerate at the $\Gamma$ point. This loss of degeneracy could affect the ZPR$_{\rm{g}}$ predicted by the gFr model. In addition to the electronic eigenvalues, the inclusion of a SOC term in the external potential of the first-principles Hamiltonian will also have repercussions on the first-order Hamiltonian perturbed by atomic displacements, which is a key quantity for computing the ZPR. 

SOC has often been neglected throughout literature when investigating the \fro interaction since strong polaronic effects are most likely to occur in materials where the LO phonon frequency is large. Such systems typically contain light atoms (e.g. oxides), for which SOC can reasonably be expected to be weak. 
Some theoretical studies have addressed the consequences of SOC on EPI in 2D materials~\cite{kandemir_electron-phonon_2018, li_spin-orbit_2007, chen_phonon-limited_2007}, mainly through Rashba-Holstein~\cite{cappelluti_electron-phonon_2007, grimaldi_large_2010} and Rashba-\fro~\cite{grimaldi_energy_2008,vardanyan_two-dimensional_2012} model Hamiltonians. To the best of our knowledge, only Trebin and R\"ossler~\cite{trebin_polarons_1975} explicitly investigated the effect of SOC on the \fro polaron for triply-degenerate band extrema in 3D materials. However, they relied on an isotropic model Hamiltonian, thus neglecting the effect of band warping.

From the first-principles perspective, density-functional perturbation theory calculations including SOC have been available for about 15 years~\cite{verstraete_density_2008, dal_corso_density_2007}. Other formalisms relying on finite differences and distorted supercells~\cite{monserrat_electronphonon_2018}, as well as the recent special displacement method~\cite{zacharias_theory_2020}, have also been used to investigate this question. Nevertheless, SOC remains commonly neglected in ZPR calculations to this day. Full first-principles EPI calculations with SOC are typically done on a case-by-case basis~\cite{schlipf_carrier_2018, zheng_rashba_2015, molina-sanchez_temperature-induced_2016}. 
Rashba materials~\cite{monserrat_temperature_2017}, for which SOC is known to have a profound impact on either the electronic structure or the phonon frequencies, and topological materials~\cite{monserrat_temperature_2016, brousseau-couture_temperature_2020}, in which SOC is necessary to induce the band inversion, have naturally been investigated by including SOC in first-principles EPI calculations. Some compound-specific comparative studies have been made, for example, in PbTe~\cite{cao_thermally_2019,querales-flores_temperature_2019}, CH$_3$NH$_3$PbI$_3$~\cite{Saidi_temperature_2016} and BAs~\cite{bravic_finite_2019}, as well as when investigating the superconducting coupling constant~\cite{heid_effect_2010, golab_electron-phonon_2019,tutuncu_effects_2017, hu_effects_2016}. Yet, even in the most simple case of cubic materials, the effect of SOC on EPI and the ZPR has not received the thorough investigation it deserves.

In this article, we investigate the effect of SOC on the ZPR of twenty semiconductors using the non-adiabatic AHC framework. We focus on representative cubic materials, as their well-characterized electronic structure provides a simple framework to investigate the mechanisms at play. Their triply-degenerate VBM also proves ideal to investigate the effect of SOC on the polaron formation energy of degenerate extrema within the gFr model. We evaluate the first-principles ZPR with the AHC methodology and extend the generalized \fro model of Miglio \textit{et al.}~\cite{miglio_predominance_2020} to include SOC. First-principles calculations show that
spin-orbit coupling reduces the zero-point renormalization of the valence band edge by 15\%–30\% for the heavier materials, e.g. the tellurides. We address the SOC-induced Dresselhaus splitting~\cite{dresselhaus_spin-orbit_1955} occuring in non-centrosymmetric materials, which shifts the band extrema slightly away from its location without SOC in reciprocal space. The leading mechanism driving the observed SOC-induced decrease of the ZPR$_{\rm{g}}$ is found to be the variation of the electronic eigenenergies of the occupied bands and the decrease of the hole effective masses near the $\Gamma$ point. We also confirm the claims of Miglio \textit{et al.}~\cite{miglio_predominance_2020} regarding the predominance of non-adiabatic effects in the ZPR$_{\rm{g}}$ of polar materials. We relate the results from the two approaches and bring to light some limitations of the approximations inherent to the gFr model when SOC is considered.

Section~\ref{sec:theory} presents an overview of the theoretical concepts used throughout this work. We first review the AHC formalism for EPI (Sec.~\ref{sec:theo-ahc}), then briefly discuss some key consequences of SOC in the first-principles perspective (Sec.~\ref{sec:theo-soc}) before demonstrating how to incorporate SOC into the gFr model of Ref.~\cite{miglio_predominance_2020} (Sec.~\ref{sec:theo-gfr}) and investigating the consequences of Dresselhaus splitting on our results (Sec.~\ref{sec:gfr-with-dresselhaus}). Section~\ref{sec:computation} provides the relevant technical details regarding our calculations. We respectively analyze our first-principles and gFr model results in Secs.~\ref{sec:res-fp} and~\ref{sec:res-gfr}, then summarize our findings in Sec.~\ref{sec:concl}.

\section{\label{sec:theory}Methodology}

\subsection{AHC formalism}\label{sec:theo-ahc}
\begin{figure}
    \centering
    \includegraphics[width=\columnwidth]{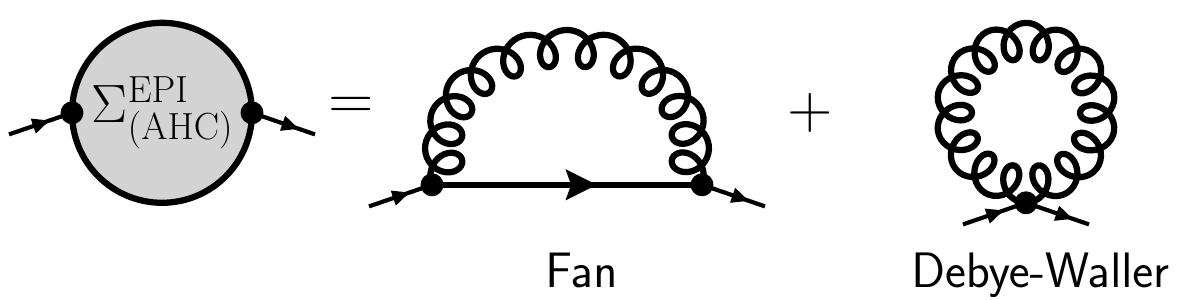}
    \caption{\textbf{Fan and Debye-Waller diagrams contributing to the AHC self-energy.}}
    \label{fig:ahcfeynman}
\end{figure}
In the following, we briefly summarize the key concepts of the nonadiabatic Allen-Heine-Cardona (AHC) framework~\cite{allen_theory_1976, allen_theory_1981, allen_temperature_1983}. We work with the Hartree atomic unit system, such that $\hbar=m_e=c=|e|=1$.

Within the many-body perturbation theory formalism, the electron-phonon interaction at temperature $T$ affects the electronic Green's function through a frequency-dependent electron-phonon self-energy, $\Sigma\kn(\omega, T)$, where $\mathbf{k}$ and $n$ are respectively the electron wavevector and band index. At the lowest order of perturbation, known as AHC theory~\cite{allen_temperature_1983}, the self-energy contains two terms, called the Fan and Debye-Waller (DW) contributions:
\begin{equation}\label{eq:ahcsigma}
    \Sigma\kn^{\rm{AHC}}(\omega, T) = \Sigma\kn^{\rm{Fan}}(\omega, T) + \Sigma\kn^{\rm{DW}}(T).
\end{equation}
The dynamical Fan self-energy contains two first-order vertices treated at second order in perturbation theory, while the static Debye-Waller self-energy has one second-order vertex treated at first order in perturbation theory. The Feynman diagrams corresponding to these contributions are shown in Fig.~\ref{fig:ahcfeynman}. 

Note that we implicitly suppose that the full self-energy matrix can be approximated by its diagonal counterpart, i.e. $\Sigma_{\mathbf{k}n n'}\propto \delta_{n n'}$. The non-diagonal contributions hybridize the unperturbed electronic eigenstates within the interacting Green's function~\cite{lihm_phonon-induced_2020} and become important when the band gap nearly vanishes. These can be safely neglected here as we work with semiconductors and insulators. 

Within this framework, the temperature dependence of an electronic eigenstate with eigenvalue $\varepsilon\kn$ then reads
\begin{equation}\label{eq:tdep-eigen}
    \varepsilon\kn(T) = \mathfrak{Re}\left[\Sigma\kn^{\rm{AHC}}(\omega=\varepsilon\kn(T), T)\right] + \varepsilon\kn^0.
\end{equation}

From this point, we work exclusively at $T=0$~K. The ZPR of an electronic eigenstate $\ket{\mathbf{k}n}$ is obtained from Eq.~(\ref{eq:tdep-eigen}),
\begin{equation}
    \textrm{ZPR}\kn = \varepsilon\kn(T=0) - \varepsilon\kn^0,
\end{equation}
while the band gap ZPR is the difference between the ZPR of the conduction and valence band edges (respectively, ZPR$_{\rm{c}}$ and ZPR$_{\rm{v}}$), \begin{equation}
   \textrm{ZPR}_{\rm{g}} =  \textrm{ZPR}_{\rm{c}} - \textrm{ZPR}_{\rm{v}}.
\end{equation}

We apply the on-the-mass-shell approximation to Eq.~(\ref{eq:ahcsigma}), thus evaluating the Fan self-energy at the poles of the Green's function, namely, at the bare electronic eigenvalue, $\varepsilon\kn^0$,
\begin{equation}
    \Sigma\kn^{\rm{Fan}}(\varepsilon\kn(T=0), T=0)\approx\Sigma\kn^{\rm{Fan}}(\omega=\varepsilon\kn^0, T=0).
\end{equation}
Furthermore approximating the interacting electronic Green's function by the noninteracting Kohn-Sham wavefunction obtained from density-functional theory (DFT), one obtains the standard expression for the non-adiabatic Fan self-energy~\cite{giustino_electron-phonon_2017},
\begin{equation}\label{eq:sigmafan}
    \begin{split}
   & \Sigma_{\mathbf{k}n}^{\text{Fan}}(\varepsilon\kn^0, T=0) = 
   \sum\limits\qv^{\text{BZ}}\sum\limits_{n'}
   |\bra{\mathbf{k+q}n'}\nabla\qv V^{\rm{KS}}
   \ket{\mathbf{k}n}|^2 \\
   & \times \left[ \frac{1 -
   f_{\mathbf{k+q}n'}}{\varepsilon\kn^0-\varepsilon^0\subkn{k+q}{n'}-
   \wqv + i\eta_\mathbf{k}} + \frac{
   f_{\mathbf{k+q}n'}}{\varepsilon^0\kn-\varepsilon^0_{\mathbf{k+q}n'}+
   \wqv + i\eta_\mathbf{k}} \right].
    \end{split}
\end{equation}
The contributions of all phonon modes with frequency $\omega\qv$ are summed for all wavevector $\mathbf{q}$ and branch index $\nu$ in the Brillouin zone (BZ). In Eq.~(\ref{eq:sigmafan}) and throughout this work, all phonon modes summations are implicitly normalized by the number of phonon wavevectors used to sample the Brillouin zone. Since we work at $T=0$~K, the Fermi-Dirac occupation functions, $f\kn$, are either 1 for the occupied states or 0 for the conduction bands. The small imaginary parameter $\eta_{\mathbf{k}} = \eta\,\textrm{sgn}(\varepsilon\kn^0-\mu)$, with $\mu$ the chemical potential and $\eta$ real and positive, shifts the poles of the Green's function in the complex plane to maintain causality. Without SOC, the electronic bands are implicitly spin degenerate.

The electron-phonon matrix elements squared,
\begin{equation}\label{eq:gkkfan}
    \left|g_{\mathbf{k}n n'}^{\rm{Fan}}(\mathbf{q}\nu)\right|^2 \triangleq |\bra{\mathbf{k+q}n'}\nabla\qv V^{\rm{KS}}\ket{\mathbf{k}n}|^2,
\end{equation}
capture the probability that an electron in eigenstate $\varepsilon\kn^0$ interacts with a $\mathbf{q}\nu$-phonon, given the self-consistent first-order variation of the Kohn-Sham potential (labeled with superscript \enquote{KS}) induced by the collective atomic motion along this phonon mode~\cite{giustino_electron-phonon_2017, ponce_temperature_2014}. The operator $\nabla\qv$ expressed in the position basis can be written as
\begin{equation}\label{eq:grad_operator}
    \begin{split}
    \nabla\qv &=
    \frac{1}{\sqrt{2\wqv}}\sum\limits_{\kappa\alpha}U_{\nu, \kappa\alpha}(\mathbf{q}) \sum\limits_l \e^{i\mathbf{q}\cdot\bm{R}_l}\frac{\partial }{\partial \bm{R}_{l\kappa\alpha}}\\ &=
    \frac{1}{\sqrt{2\wqv}}\sum\limits_{\kappa\alpha}U_{\nu, \kappa\alpha}(\mathbf{q}) \partial_{\kappa\alpha}(\mathbf{q}),
\end{split}
\end{equation}
where $\bm{R}_{l\kappa\alpha}$ denotes the displacement of atom $\kappa$, located in unit cell $l$, in cartesian direction $\alpha$. The phonon eigendisplacement vector, $U_{\nu, \kappa\alpha}(\mathbf{q})$, verifies the generalized eigenvalue equation
\begin{equation}
    M_\kappa \omega\qv^2U_{\nu,\kappa\alpha}(\mathbf{q}) = \sum\limits_{\kappa' \alpha'} \Phi_{\kappa \kappa'}^{\alpha \alpha'}(\mathbf{q}) U_{\nu, \kappa' \alpha'}(\mathbf{q})
\end{equation}
and the normalization condition
\begin{equation}
    \sum\limits_{\kappa\alpha}M_\kappa U_{\nu,\kappa\alpha}^\ast(\mathbf{q})U_{\nu',\kappa\alpha}(\mathbf{q}) = \delta_{\nu\nu'},
\end{equation}
where $M_\kappa$ is the atomic mass of atom $\kappa$. The dynamical matrix, $\Phi_{\kappa \kappa'}^{\alpha \alpha'} (\mathbf{q})$, is the Fourier transform of the second derivative of the total energy with respect to two atomic displacements,
\begin{equation}
    \Phi_{\kappa \kappa'}^{\alpha \alpha'}(\mathbf{q}) = \sum\limits_l \e^{i\mathbf{q}\cdot\bm{R}_l}\frac{\partial^2 E}{\partial\bm{R}_{l\kappa\alpha}\partial\bm{R}_{0\kappa'\alpha'}}.
\end{equation}

For its part, the Debye-Waller self-energy is formally defined as~\cite{giustino_electron-phonon_2017} 
\begin{equation}\label{eq:sigmadw_def}
    \Sigma^{\text{DW}}\kn =
\sum\limits\qv\frac{1}{2}\bra{\mathbf{k}n}
\nabla\qv\nabla_{-\mathbf{q}\nu} V^{\rm{KS}} \ket{\mathbf{k}n}.
\end{equation}
The direct evaluation of the second-order derivative of the Kohn-Sham potential with respect to atomic displacements entering Eq.~(\ref{eq:sigmadw_def}) is a computational bottleneck in the density-functional perturbation theory approach. By applying the rigid-ion approximation, i.e. assuming that the potentials created by each nucleus are independent of each other, one can replace the second-order derivatives by the same first-order derivatives entering $\Sigma^{\rm{Fan}}\kn$~\cite{ponce_temperature_2014}, yielding
\begin{equation}\label{eq:sigmadw}
    \Sigma^{\rm{DW, RIA}}\kn(T=0) = \sum\limits\qv^{\rm{BZ}}\sum\limits_{n'\neq n}-\frac{1}{4\omega\qv}\frac{\left|g^{\rm{DW}}\subkn{k}{n n'} (\mathbf{q}\nu)\right|^2}{\varepsilon\kn^0-\varepsilon\subkn{k}{n'}^0+i\eta},
\end{equation}
where RIA stands for rigid ion approximation and where
\begin{equation}\label{eq:gkkdw}
\begin{split}
    &\left|g^{\rm{DW}}_{\mathbf{k}n n'} (\mathbf{q}\nu)\right|^2 = \\
    &\sum\limits_{\kappa\kappa'} \sum\limits_{\alpha\alpha'}\left[U_{\nu,\kappa\alpha}\,(\mathbf{q})U_{\nu,\kappa\alpha'}\,(\mathbf{q})^\ast+U_{\nu,\kappa'\alpha}\,(\mathbf{q})U_{\nu,\kappa'\alpha'}\,(\mathbf{q})^\ast\right]\\
    &\times\bra{\mathbf{k}n}V^{(1)}_{\kappa\alpha}(0)^\ast\ket{\mathbf{k+q}n'}\bra{\mathbf{k+q}n'}V^{(1)}_{\kappa'\alpha'}(0)\ket{\mathbf{k}n},
    \end{split}
\end{equation}
with
\begin{equation}
   V^{(1)}_{\kappa\alpha}(0) = \partial_{\kappa\alpha}(\mathbf{q}=\bm{0})V^{\rm{KS}}, 
\end{equation}
following the definition of the operator $\partial_{\kappa\alpha}(\mathbf{q})$ in the second line Eq.~(\ref{eq:grad_operator}). The consequences of the rigid-ion approximation on the ZPR have been discussed in Ref.~\cite{ponce_temperature_2014} for crystals and in Ref.~\cite{gonze_theoretical_2011} for molecules.

\subsection{Spin-orbit interaction}\label{sec:theo-soc}
We now examine how SOC can affect Eq.~(\ref{eq:sigmafan}) and~(\ref{eq:sigmadw}).
Upon inclusion of SOC, the electronic wavefunction becomes a spinor,
\begin{equation}
    \ket{\mathbf{k}n} = \begin{pmatrix} \ket{\mathbf{k}n\uparrow} \\ \ket{\mathbf{k}n\downarrow}\end{pmatrix},
\end{equation}
and the Hamiltonian, a \mbox{2$\times$2} matrix,
\begin{equation}
    \hat{H}_{\mathbf{k}} = 
        \begin{pmatrix}
            H_{\mathbf{k}\uparrow\uparrow} & H_{\mathbf{k}\uparrow\downarrow}\\
            H_{\mathbf{k}\downarrow\uparrow} & H_{\mathbf{k}\downarrow\downarrow}
        \end{pmatrix}.
\end{equation}      
In real space, the general form of the SOC contribution to the electronic Hamiltonian writes~\cite{dresselhaus_spin-orbit_1955}
\begin{equation}\label{eq:generalsoc}
    \hat{H}^{\rm{SOC}}(\bm{r}) = \frac{1}{4}\left(\nabla V(\bm{r})\times \bm{\hat{P}}\right)\cdot \bm{\sigma},
\end{equation}
where $\bm{\hat{P}}$ is the momentum operator and $\bm{\sigma}$ are the Pauli matrices.

For a plane wave basis set and norm-conserving pseudopotentials, SOC only enters the Hamiltonian through the electron-ion term. Assuming that the pseudopotentials are fully separable and substituting the Coulomb potential in Eq.~(\ref{eq:generalsoc}), one recovers the typical $\bm{L}\cdot\bm{S}$ term from introductory quantum mechanics. For a single atom, one gets~\cite{verstraete_density_2008} 
\begin{equation}
\begin{split}
    V^{\rm{e-ion}}(\bm{r, r'}) =& \sum\limits_l V_l^{\text{SR}}(r, r') \ket{ls}\bra{ls} \\&+ \sum\limits_l V_l^{\text{SOC}}(r, r') \bm{L}\cdot\bm{S}\ket{ls}\bra{ls},
    \end{split}
\end{equation}
where  $V_l^{\text{SR}}(r, r')$  and $V_l^{\text{SOC}}(r, r')$ follow the Kleinman-Bylander construction~\cite{kleinman_efficacious_1982},
\begin{equation}
    V_l^{x} = f_l^x(r)E_l^{\rm{KB},x}f_l^x(r'),\;\;x\in\{\rm{SR},\,\rm{SOC}\},
\end{equation}
where $E_l^{\rm{KB},x}$ is the Kleinman-Bylander energy~\cite{gonze_first-principles_2002}.
SR stands for the scalar-relativistic contribution to the electron-ion potential (hence, without SOC), and $\ket{ls}\bra{ls}$ is the projector on the tensor product subspace of angular momentum $L$ and spin $S$, which has dimension $2(2l+1)$. Detailed expressions for $V_l^{\text{SR}}$ and $V_l^{\text{SOC}}$ can be found in Ref.~\cite{hartwigsen_relativistic_1998}. No magnetism is considered, such that the electronic density is given by a single scalar function, $\rho(\bm{r})$. 

The consequences of SOC on the explicit density-functional perturbation theory equations have been derived in Refs.~\cite{verstraete_density_2008} and~\cite{dal_corso_density_2007} for norm-conserving pseudopotentials. In our case, the general form of the equations presented in Sec.~\ref{sec:theo-ahc} remain unchanged, but all the relevant physical quantities, i.e. $\omega\qv$, $\varepsilon\kn^0$ and the electron-phonon matrix elements squared, Eqs.~(\ref{eq:gkkfan}) and~(\ref{eq:gkkdw}), now capture the effect of SOC. There is no implicit sum on the spin degree of freedom, as the spinorial electronic wavefunctions mix the spin-up and spin-down components.

\subsection{Generalized \fro model}\label{sec:theo-gfr}
In the following, we discuss how to incorporate SOC into the generalized \fro model developed in Ref.~\cite{miglio_predominance_2020}. For completeness, we start by reviewing the key elements of this model. First neglecting SOC, the Hamiltonian at the first order of interaction writes:
\begin{equation}\label{eq:gfrHamiltonian}
\begin{split}
H =& \sum\limits\subkn{k}{n\sigma} \frac{\theta k^2}{2m^\ast_n(\hat{\mathbf{k}})} c^\dagger\subkn{k}{n\sigma} c\subkn{k}{n\sigma} +
\sum\limits\subkn{q}{j}\omega\subkn{0}{j}(\hat{\mathbf{q}})\left(a^\dagger\subkn{q}{j} a\subkn{q}{j}
+ \frac{1}{2}\right)\\&
+ \sum\limits\subkn{k}{n n'\sigma}\sum\subkn{q}{j}
g^{\rm{gFr}}\subkn{k}{n n'}(\mathbf{q}j) c^\dagger\subkn{k+q}{n'\sigma} c\subkn{k}{n\sigma} \left(a\subkn{q}{j} +
            a^\dagger\subkn{-q}{j}\right).
            \end{split}
\end{equation}
The first term corresponds to parabolic bare electronic eigenenergies $\varepsilon\kn$ with direction-dependent effective mass $m^\ast_n(\hat{\mathbf{k}})$, while the second term allows for multiple phonon branches $j$ with direction-dependent Einstein frequency $\omega\subkn{0}{j}(\hat{\mathbf{q}})$, evaluated at the zone center $\Gamma$. The $c^\dagger\subkn{k}{n\sigma}$, $c\subkn{k}{n\sigma}$, $a^\dagger\subkn{q}{j}$, $a\subkn{q}{j}$ are respectively the creation and annihilation operators for electrons and phonons, while $\hat{\mathbf{k}}$ and $\hat{\mathbf{q}}$ are unit vectors. The parameter $\theta$ gives the sign of the effective mass: $\theta=-1$ for the holelike bands and $\theta=1$ for the electronlike bands. The sum on spin index $\sigma$ implies that all electronic states are doubly degenerate. 

The last term couples the electron and phonon subsystems through the \fro interaction, with matrix element
\begin{multline}\label{eq:gfrcoupling}
    g^{\rm{gFr}}\subkn{k}{n n'}(\mathbf{q}j) = \frac{1}{q}\frac{4\pi}{\Omega_0}\left(\frac{1}{2\,\omega\subkn{0}{j}(\hat{\mathbf{q}})V_{\rm{BvK}}}\right)^{1/2}\frac{\hat{\mathbf{q}}\cdot\bm{p}_j(\mathbf{\hat{q}})}{\epsilon^\infty(\mathbf{\hat{q}})}\\
    \times \sum_m s_{n' m}(\mathbf{\hat{k}}')s_{n m}^\ast(\mathbf{\hat{k}}),
\end{multline}
where $\mathbf{k}' = \mathbf{k+q}$, $\Omega_0$ is the primitive unit cell volume, $V_{\rm{BvK}}$ is the Born-von Karman normalization volume associated with the $\mathbf{k}$ and $\mathbf{q}$ samplings, and $\epsilon^\infty$ is the macroscopic optic dielectric constant, obtained from the dielectric tensor,
\begin{equation}
    \epsilon^\infty(\mathbf{\hat{q}}) = \sum\limits_{\alpha\beta}\hat{q}_\alpha \epsilon^\infty_{\alpha\beta}\hat{q}_\beta.
\end{equation}
Here, $\bm{p}_j(\mathbf{\hat{q}})$ is the mode-polarity vector of the \mbox{$j$-phonon} mode~\cite{miglio_predominance_2020}, constructed from the Born effective charges, $Z^\ast_{\kappa\alpha,\alpha'}$, and the phonon eigendisplacement vectors, $U_{j,\kappa \alpha}(\mathbf{q})$, summing over all Cartesian directions $\alpha$ and all atoms $\kappa$ in the unit cell,
\begin{equation}
    p_{j,\alpha'}(\mathbf{\hat{q}}) = \lim_{q\rightarrow 0}\sum\limits_{\kappa\alpha} Z^\ast_{\kappa\alpha,\alpha'}U_{j,\kappa \alpha}(q\hat{\mathbf{q}}).
\end{equation}
Note also that our formulation of the \fro matrix element, Eq.~(\ref{eq:gfrcoupling}), relies on the Born and Huang convention for the phonon eigenvectors~\cite{born_dynamical_1954}, which implies the following relation:
\begin{equation}\label{eq:signconvention}
    U_{j,\kappa\alpha}(-\mathbf{q}) = U^\ast_{j,\kappa\alpha}(\mathbf{q}),
\end{equation}
such that Eq.~(\ref{eq:gfrHamiltonian}) is hermitian. See Ref.~\cite{guster_erratum_2022} for a thorough discussion of the different phase conventions in the literature.

The unitary matrix $s_{nm}(\hat{\mathbf{k}})$ 
describes the direction-dependent overlap between the electronic states at the band extrema located at $\Gamma$ and states along the $\mathbf{k}$ direction, in the $k\rightarrow 0$ limit, computed from the periodic part  of the wavefunction (indicated by the subscript $P$):
\begin{equation}\label{eq:smatrix}
    s_{nm}(\hat{\mathbf{k}}) = \lim_{k\rightarrow0}\braket{k\hat{\mathbf{k}}n|\Gamma m}_P.
\end{equation}
While we set the band extrema at $\Gamma$ for convenience, the previous definition allows for a band extrema located at any wavevector in the Brillouin zone.

In all previous expressions, the sums on electronic bands index $n$, $n'$ and $m$ run on the degenerate subset of bands connected to extrema, thus allowing interband couplings within this subset. The sum on phonon branches $j$ is restricted to LO modes, as $\bm{p}_j(\mathbf{\hat{q}})$ is zero otherwise.

When SOC is considered, the Hamiltonian is no longer diagonal in spin space. We can, however, define new electronic creation and annihilation operators, $\tilde{c}^\dagger\kn$ and $\tilde{c}\kn$,
 such that the electronic part of the Hamiltonian can be written as
\begin{equation}
    H^{\rm{el}} = \sum\limits\kn \varepsilon^{\rm{SOC}}\kn\tilde{c}^\dagger\kn\tilde{c}\kn.
\end{equation}
In order to formulate a \fro Hamiltonian for the SOC case, we will suppose that the relevant part of the electronic structure is at a band extremum, with quadratic departure from the extremal eigenenergy
as a function of the wavevector. 
This is the same hypothesis as for the generalized
\fro model without SOC.
Generally speaking, this hypothesis is correct 
when the band extrema are non-degenerate (except for the spin degeneracy) when the SOC is not present.
It will hold also when the starting band extremum is degenerate, provided the typically spin-orbit coupling energy is much bigger than the phonon energy, so that,
after applying the SOC, one is left with a new band extremum with quadratic departure of the eigenenergy 
in a sufficiently large zone, where the phonon energy is relevant.

Supposing this hypothesis to be valid, we take the extremum eigenvalue as zero of energy, and expand the
eigenvalue as 
\begin{equation}\label{eq:eigensoc-gfr}
\varepsilon^{\rm{SOC}}\kn = \frac{\theta k^2}{2 \widetilde{m}_n^\ast(\hat{\mathbf{k}})},
\end{equation}
which captures the modification of the electronic effective masses near the band extrema induced by SOC, $\widetilde{m}_n^\ast(\hat{\mathbf{k}})$ replacing $m_n^\ast(\hat{\mathbf{k}})$. We also neglect spin-phonon interaction, thus assuming that SOC affects the vibrational properties through electronic properties only. Within these assumptions, we recover an expression identical to Eq.~(\ref{eq:gfrHamiltonian}), in which the sum on $\sigma$ has been absorbed inside the new electronic operators. The starting point of Ref.~\cite{miglio_predominance_2020} can therefore be taken as implicitly incorporating the effects of SOC on the electronic and vibrational properties. From now on, we simplify the notation by dropping all tilde on electronic quantities which include SOC, i.e. $\widetilde{m}_n^\ast(\hat{\mathbf{k}})\rightarrow m_n^\ast(\hat{\mathbf{k}})$.    

We now follow the same procedure as described in Section~5 of the Supplementary Notes of Ref.~\cite{miglio_predominance_2020}: we substitute $g^{\rm{gFr}}\subkn{k}{n n'}(\mathbf{q}j)$ for the matrix elements in the general expression for $\Sigma^{\rm{Fan}}\kn$ (Eq.~(\ref{eq:sigmafan})) and, as per the original \fro model, take the continuum macroscopic limit, replacing the discrete sum over $\mathbf{q}$ by an integral over the $\mathbf{q}$ coordinate \mbox{($\sum_{\mathbf{q}}\,f(\mathbf{q})/V_{\rm{BvK}}\rightarrow \Omega_0/(2\pi)^3\int d^3\mathbf{q}\,f(\mathbf{q})$)}, thus extending the Brillouin zone boundaries to infinity. Since we only consider interband contributions within the degenerate subset of bands connected to the extrema, only the second (first) term inside the brakets of Eq.~(\ref{eq:sigmafan}) contribute for the holelike (electronlike) bands. Taking the $q\rightarrow0$ limit of the denominator for purely parabolic electronic bands,
 we are left with
 \begin{equation}\label{eq:gfr_beforeradial}
    \begin{split}
    \textrm{ZPR}^{\rm{gFr}}_{n_\theta} = -\frac{\theta}{\pi\Omega_0}\int\,d^3\mathbf{q}&\sum\limits_{j n}\frac{1}{q^2}\frac{|s_{n,n_\theta}(\hat{\mathbf{q}})|^2}{\omega_{\bm{0}j}(\hat{\mathbf{q}})}\left(\frac{\hat{\mathbf{q}}\cdot\bm{p}_j(\mathbf{\hat{q}})}{\epsilon^\infty(\mathbf{\hat{q}})}\right)^2\\
    &\times \frac{1}{\frac{q^2}{2m^\ast_{n}(\hat{\mathbf{q}})} + \omega_{\bm{0}j}(\hat{\mathbf{q}})} \,,
    \end{split}
 \end{equation}
 where $n_{\theta=-1}$ is the band index of the VBM and $n_{\theta=1}$, that of the conduction band minimum (CBM).
Using spherical coordinates, the radial part of this three-dimensional integral has an analytic solution of the form
 \begin{equation}\label{eq:radialintegral}
     \int\limits_0^\infty \,dq\, \frac{1}{C_1 q^2+C_2} = \frac{1}{\sqrt{C_1 C_2}}\frac{\pi}{2},
 \end{equation} 
 where the parameters $C_1=(2m^\ast_{n}(\hat{\mathbf{q}}))^{-1}$ and $C_2=\omega_{\bm{0}j}(\mathbf{\hat{q}})$ are positive. Recall that, for the VBM, the negative curvature of the electronic bands is parametrized by $\theta$. This yields
\begin{multline}\label{eq:gfr1}
    \textrm{ZPR}^{\rm{gFr}}_{n_\theta} = -\frac{\theta}{\sqrt{2}\Omega_0}\oint\limits_{4\pi}\,d\hat{\mathbf{q}}\sum\limits_{j n}|s_{n,{n_\theta}}(\hat{\mathbf{q}})|^2\frac{(m^\ast_{n}(\hat{\mathbf{q}}))^{1/2}}{\omega_{\bm{0}j}(\hat{\mathbf{q}})^{3/2}}\\
    \times \left(\frac{\hat{\mathbf{q}}\cdot\bm{p}_j(\mathbf{\hat{q}})}{\epsilon^\infty(\mathbf{\hat{q}})}\right)^2 .
\end{multline}
With the previous definition of $\theta$, we thus obtain a positive (negative) ZPR for the VBM (CBM). 

Up to this point, no special treatment was made to consider SOC in the treatment of the \fro interaction outside incorporating it implicitly in the static electronic and vibrational properties, i.e. $m_n^\ast(\hat{\mathbf{q}})$, $\omega_{\bm{0}j}(\mathbf{\hat{q}})$, $\bm{p}_j(\mathbf{\hat{q}})$ and $\epsilon^\infty(\mathbf{\hat{q}})$ are computed with SOC. 

We finally argue that the treatment of $|s_{n,n_\theta}(\hat{\mathbf{q}})|^2$ based on the point group symmetry argument detailed in the Supplementary Information of Ref.~\cite{miglio_predominance_2020} remains valid in the presence of SOC. For this paper, we will treat the $3\times2\rightarrow 4+2$ degeneracy arising from a cubic space group, taking the VBM of cubic materials as a typical example. The argument could be generalized to any space group symmetry using group theory.

As the degeneracy arises from symmetry, i.e. it is not accidental, the degenerate electronic wavefunctions at the extrema can be decomposed in a basis of orthonormal eigenfunctions that form an irreducible representation of the symmetry group, $\mathcal{G}=T_d$ for the zincblende structure and $\mathcal{G}=O_h$ for the diamond structure. Without SOC, this basis contains three eigenfunctions, denoted $\{\ket{X}, \ket{Y}, \ket{Z}\}$, which each are doubly degenerate in the spin space. When considering SOC, the basis functions $\{\ket{X\uparrow}, \ket{Y\uparrow}, \ket{Z\uparrow}, \ket{X\downarrow}, \ket{Y\downarrow}, \ket{Z\downarrow}\}$ no longer form a good basis choice as they do not form an irreducible representation of the double group, $\mathcal{G}\otimes D_{1/2}$. One rather has to use linear combinations of those states, namely, the fourfold $\{\ket{j=3/2}\}$ states for the degenerate heavy hole and light hole bands, which form the VBM, and the twofold $\{\ket{j=1/2}\}$ states should one wish to evaluate the ZPR for the split-off bands.

We now express the eigenstates entering the $s_{nm}(\hat{\mathbf{q}})$ overlap integrals (Eq.~(\ref{eq:smatrix})) in this basis. The $\ket{\Gamma v}$
state, where $v=n_{\theta=-1}$ is the band index of the VBM, can be written as
\begin{equation}
    \ket{\Gamma v} = \sum\limits_{m\in\{\pm3/2, \pm 1/2\}} u_{vm}\ket{\frac{3}{2}m},
\end{equation}
where $u_{vm}$ is the coefficient of the basis function $\ket{3/2\;m}$, and the $\ket{q\hat{\mathbf{q}}n}$ state becomes
\begin{equation}
\begin{split}
    \ket{q\hat{\mathbf{q}}n} &= \sum\limits_{m\in\{\pm3/2, \pm 1/2\}}\ket{\frac{3}{2}m}\braket{\frac{3}{2}m|q\hat{\mathbf{q}}n},\\
    &= \sum\limits_{m\in\{\pm3/2, \pm 1/2\}}
    s_{nm}(\hat{\mathbf{q}})\ket{\frac{3}{2}m},
    \end{split}
\end{equation}
where $s_{nm}(\hat{\mathbf{q}})$ is the overlap integral with the basis function $\ket{\frac{3}{2}m}$. The $q\rightarrow 0$ limit is implied.

Substituting the last two equations in Eq.~(\ref{eq:gfr1}), we obtain an expression which is identical to the Supplementary Eqs.~(24) and~(25) of Ref.~\cite{miglio_predominance_2020}, with the double sum on $m,m'\in\{X,Y,Z\}$ replaced by a double sum on $m,m'\in\{\pm3/2, \pm1/2\}$. The remaining of the argument thus holds, yielding a final expression for the $\textrm{ZPR}^{\rm{gFr}}$ which has the same form as their Eq.~(6):
\begin{equation}\label{eq:zpr-gfr}
    \textrm{ZPR}^{\rm{gFr}}_{n_\theta} = -\sum\limits_{j n}\frac{\theta}{\sqrt{2}\Omega_0 n_{\rm{deg}}}\oint\limits_{4\pi}d\hat{\mathbf{q}}\frac{(m^\ast_{n}(\hat{\mathbf{q}}))^{1/2}}{\omega_{\bm{0}j}(\hat{\mathbf{q}})^{3/2}}
    \left(\frac{\hat{\mathbf{q}}\cdot\bm{p}_j(\mathbf{\hat{q}})}{\epsilon^\infty(\mathbf{\hat{q}})}\right)^2,
\end{equation}
in which $n_{\rm{deg}}$ is now the degree of degeneracy of the band extrema in presence of SOC. As the $n$ summation is made over degenerate states, the division by $n_{\rm{deg}}$ yields 
an average over degenerate states. Note that this last expression can be further simplified when applied to cubic systems, as in that case the phonon frequencies and mode-polarity vector do not not depend on the wavevector orientation, and the dielectric tensor is isotropic (see Eq.~(45) of Ref.~\cite{guster_frohlich_2021}). 

For materials whose vibrational properties are not significantly affected by SOC, such as cubic semiconductors and insulators, the modification of the electronic effective masses induced by SOC will have a dominant effect on the $\textrm{ZPR}^{\rm{gFr}}$. 
The reduced dimensionality of the irreducible representation at the extrema plays also a role. However, 
it is not directly due to the
smaller number of degenerate states contributing to the $\textrm{ZPR}^{\rm{gFr}}$. Indeed, Eq.~(\ref{eq:zpr-gfr}) makes it clear that an average over degenerate bands is to be computed, not a simple sum of contributions. 
The modification of the $\textrm{ZPR}^{\rm{gFr}}$ due to spin-orbit coupling
 is analytically obtained in the isotropic degenerate model of Trebin and R\"ossler [32], see their Eq.(13), combined with the effective masses from their Eqs.~(6a) and (6b).

\subsection{Generalized \fro model in the presence of Dresselhaus splitting}\label{sec:gfr-with-dresselhaus}

In Sec.~\ref{sec:theo-gfr}, we implicitly assumed that the location of the band extrema in reciprocal space is unchanged by the inclusion of SOC, i.e. it remains at the high-symmetry, degenerate $\mathbf{k}$-point. However, for non-centrosymmetric materials such as those of zincblende structure, SOC acts as an effective magnetic field which splits the previously spin-degenerate states. As a consequence, the band extrema are slightly displaced from their location without SOC, both in energy and momentum (see Fig.~\ref{fig:cds-dresselhaus} of Appendix~\hyperref[sec:dressmodel]{B}), thus breaking, in principle, one of our underlying hypotheses. This effect was originally discussed by Dresselhaus~\cite{dresselhaus_spin-orbit_1955} in~1955. In the following, we analyze the consequences of Dresselhaus splitting on Eq.~(\ref{eq:zpr-gfr}) for the degenerate $\mathbf{k}$-point (i.e. the $\Gamma$ point in the current work). See Appendix~\hyperref[sec:dressmodel]{B} for more details about the Dresselhaus Hamiltonian and its consequences on the band structure of zincblende materials.

In the presence of Dresselhaus splitting, the electronic dispersion of band $n$ at \mbox{$\mathbf{k}=\Gamma+\mathbf{q}$} is no longer \mbox{$\theta q^2/(2m_n^\ast(\hat{\mathbf{q}}))$}, following Eq.~(\ref{eq:eigensoc-gfr}), but rather 
 \begin{equation}\label{eq:sm-dressdispersion}
     \varepsilon\subkn{q}{n}^{\rm{SOC}} = \frac{\theta (q-k^0_n(\hat{\mathbf{q}}))^2}{2m^\ast_{n}(\hat{\mathbf{q}})} - \theta\Delta E_n(\hat{\mathbf{q}}),
 \end{equation}
where $k^0_n(\hat{\mathbf{q}})$ and $\Delta E_n(\hat{\mathbf{q}})$ are respectively the momentum and energy offsets characterizing the Dresselhaus splitting of band $n$ along direction $\hat{\mathbf{q}}$. As for the effective masses, we define $\Delta E_n(\hat{\mathbf{q}})$ as positive and let $\theta$ parametrize the sign of the energy offset. Equation~(\ref{eq:gfr_beforeradial}) therefore becomes
\begin{multline}\label{eq:gfr_beforeradial-dresselhaus}
    \textrm{ZPR}^{\rm{gFr}}_{n_\theta} = -\frac{\theta}{\pi\Omega_0}\int\,d^3\mathbf{q}\sum\limits_{j n}\frac{1}{q^2}\\ \times\frac{f(\hat{\mathbf{q}})}{\frac{(q-k^0_n(\hat{\mathbf{q}}))^2}{2m^\ast_{n}(\hat{\mathbf{q}})} + \omega_{\bm{0}j}(\hat{\mathbf{q}})-\Delta E_n(\hat{\mathbf{q}})} \,,
\end{multline}
where $f(\hat{\mathbf{q}})$ is a purely angular function which includes the two rightmost fractions of the first line of Eq.~(\ref{eq:gfr_beforeradial}).

As in Sec.~\ref{sec:theo-gfr}, we express the integral in spherical coordinates. However, instead of the usual integral boundaries, we rather integrate on half the sphere (note the lower bound of the $\cos\theta$ integral), while simultaneously extending the lower bound of the radial integral to $-\infty$:
\begin{equation}
\begin{split}
I &= \int\,d^3\mathbf{q}\sum\limits_{j n}\frac{1}{q^2}\frac{f(\hat{\mathbf{q}})}{\frac{(q-k^0_n(\hat{\mathbf{q}}))^2}{2m^\ast_{n}(\hat{\mathbf{q}})} + \omega_{\bm{0}j}(\hat{\mathbf{q}})-\Delta E_n(\hat{\mathbf{q}})} \\
&= \int\limits_{0}^1\!\! d(\cos\theta)\!\!\int\limits_{0}^{2\pi}\!\!d\phi\!\!\int\limits_{-\infty}^\infty\!\! dq\sum\limits_{j n}\frac{f(\hat{\mathbf{q}})}{\frac{(q-k^0_n(\hat{\mathbf{q}}))^2}{2m^\ast_{n}(\hat{\mathbf{q}})} + \omega_{\bm{0}j}(\hat{\mathbf{q}})-\Delta E_n(\hat{\mathbf{q}})}.
\end{split}
\end{equation}

With the change of variable $q'(\hat{\mathbf{q}})=q-k_0(\hat{\mathbf{q}})$, this integral becomes
\begin{equation}
I = \int\limits_{0}^1\!\! d(\cos\theta)\!\!\int\limits_{0}^{2\pi}\!\!d\phi\!\!\int\limits_{-\infty}^\infty\!\! dq'\sum\limits_{j n}\frac{f(\hat{\mathbf{q}})}{\frac{(q'(\hat{\mathbf{q}}))^2}{2m^\ast_{n}(\hat{\mathbf{q}})} + \omega_{\bm{0}j}(\hat{\mathbf{q}})-\Delta E_n(\hat{\mathbf{q}})}.
\end{equation}
Note that, by construction, the unit vectors $\hat{\mathbf{q}}$ and $\hat{\mathbf{k}}^0_n$ point in the same direction, hence $\hat{\mathbf{q}}'=\hat{\mathbf{q}}$. Transforming back with $q'=q$, we note that the contribution of the momentum offset has been exactly eliminated. One can then return the radial and angular integral boundaries to their usual values and perform the radial integral, which takes a form similar to Eq.~(\ref{eq:radialintegral}):
\begin{equation}\label{eq:sm-radialintegral-dress}
      \int\limits_{0}^\infty dq\, \frac{1}{C_1q^2+C_2'},   
\end{equation}
where the parameter $C_2$ has been replaced by \mbox{$C_2'=\omega_{\bm{0}j}(\mathbf{\hat{q}})-\Delta E_n(\hat{\mathbf{q}})$}. The parameter $C_1$ and the angular function $f(\hat{\mathbf{q}})$, respectively defined below Eq.~(\ref{eq:radialintegral}) and in Eq.~(\ref{eq:gfr_beforeradial-dresselhaus}), therefore remain unchanged by Dresselhaus splitting. As a consequence,  the treatment of the overlap matrices 
which enter $f(\hat{\mathbf{q}})$, discussed in Sec.~\ref{sec:theo-gfr}, is also unaltered.


In light of this analysis, one finds that, in presence of Dresselhaus splitting, Eq.~(\ref{eq:zpr-gfr}) generalizes exactly to
\begin{equation}\label{eq:zpr-gfr-dresselhaus-exact}
\begin{split}
    \textrm{ZPR}^{\rm{gFr}}_{n_\theta} = -\sum\limits_{j n}\frac{\theta}{\sqrt{2}\Omega_0 n_{\rm{deg}}}&\oint\limits_{4\pi}d\hat{\mathbf{q}}\left(\frac{\hat{\mathbf{q}}\cdot\bm{p}_j(\mathbf{\hat{q}})}{\epsilon^\infty(\mathbf{\hat{q}})}\right)^2\\ 
    \times&\frac{(m^\ast_{n}(\hat{\mathbf{q}}))^{1/2}}{\omega_{\bm{0}j}(\hat{\mathbf{q}})\sqrt{\omega_{\bm{0}j}(\hat{\mathbf{q}})-\Delta E_n(\hat{\mathbf{q}})}}.
\end{split}
\end{equation}
Supposing that $\Delta E_n(\hat{\mathbf{q}})$ is small compared to $\omega_{\bm{0}j}(\hat{\mathbf{q}})$, one can further simplify the last expression by Taylor expanding the inverse square root, yielding
   \begin{equation}\label{eq:zpr-gfr-dresselhaus}
   \begin{split}
    \textrm{ZPR}^{\rm{gFr}}_{n_\theta} = -\sum\limits_{j n}\frac{\theta}{\sqrt{2}\Omega_0 n_{\rm{deg}}}\oint\limits_{4\pi}d\hat{\mathbf{q}}&\left(\frac{\hat{\mathbf{q}}\cdot\bm{p}_j(\mathbf{\hat{q}})}{\epsilon^\infty(\mathbf{\hat{q}})}\right)^2\frac{(m^\ast_{n}(\hat{\mathbf{q}}))^{1/2}}{\omega_{\bm{0}j}(\hat{\mathbf{q}})^{3/2}}\\
   \times& \left(1 + \frac{\Delta E_n(\mathbf{\hat{q}})}{2\omega_{\bm{0}j}(\hat{\mathbf{q}})}\right).
\end{split}
\end{equation}

Comparing this expression with Eq.~(\ref{eq:zpr-gfr}), one finds that the energy offset stemming from the Dresselhaus splitting slightly enhances the ZPR at the $\Gamma$ point (recall that $\Delta E_n(\hat{\mathbf{q}})$ is defined as positive). The angular-dependent energy offset can therefore be interpreted as direction-dependent modulation of the integrand. Should the contribution of the energy offsets be neglected, one recovers Eq.~(\ref{eq:zpr-gfr}), which can now be seen as a lower bound to the true value of ZPR$^{\rm{gFr}}$. An upper bound could also be obtained by estimating the largest value of $\Delta E_n(\hat{\mathbf{q}})$ for a given physical system. Equation~(\ref{eq:zpr-gfr-dresselhaus-exact}) could also, in principle, be used to investigate the Fr\"ohlich-induced ZPR of the degenerate band crossing points in Rashba systems~\cite{manchon_new_2015}, provided that the LO frequency is larger than the largest energy offset. Greater care would be required otherwise, as Eq.~(\ref{eq:zpr-gfr-dresselhaus-exact}) would have poles.

Lastly, if we consider the Dresselhaus-splitted bands to be independent of each other, the known polaron effective mass enhancement induced by the isotropic \fro interaction~\cite{mahan_many-particle_1990} should remain valid even in presence of anisotropic bands (see the numerical results of Ref.~\cite{guster_frohlich_2021}). This suggests that EPI would attenuate the (already small) magnitude of the energy and momentum offsets stemming from SOC. However, the effect of the possible couplings between the almost-degenerate bands, as well as continuity conditions between the single band picture at the true extrema and the degeneracies at the $\Gamma$ point remain to be investigated. Besides, as the momentum offset plays no role in Eqs.~(\ref{eq:zpr-gfr-dresselhaus-exact}) and~(\ref{eq:zpr-gfr-dresselhaus}), one can conclude than noncentrosymetric cubic materials retain the \fro physical picture of $\Gamma$-centered parabolic bands, which has been well corroborated by the experimental literature. See, for example, Fig.~2(b) of Ref.~\cite{cho_observation_2021}.

\section{\label{sec:computation}Computational details}

\subsection{First-principles calculations}

All first-principles calculations were performed with the \textsc{Abinit} software package~\cite{gonze_abinitproject:_2020}. The bulk ground state properties were obtained from density functional theory, while vibrational properties and electron-phonon coupling were computed within density-functional perturbation theory~\cite{gonze_first-principles_1997,gonze_dynamical_1997}. When SOC is taken into account, it is included both in the ground state and the density-functional perturbation theory calculations. We use norm-conserving pseudopotentials from the Pseudo-Dojo project~\cite{van_setten_pseudodojo:_2018} and rely on the generalized gradient approximation of the Perdew-Burke-Ernzerhof functional (PBE-GGA)~\cite{perdew_generalized_1996}. The lattice parameters were optimized in the absence of SOC until all forces on the atoms were below \mbox{$10^{-7}$ hartree/bohr$^3$}, except for Ge, where we used the experimental lattice parameter, as otherwise the obtained optimized lattice parameter for the PBE-GGA functional predicts a metallic ground state. 

In order to isolate the effect of SOC on the EPI, we kept the lattice parameter fixed to the theoretical value without SOC. The electron-phonon self-energy was computed with the \texttt{ElectronPhononCoupling} Python module~\cite{antonius_electronphononcoupling:_2018}. All relevant calculation parameters, including the relaxed lattice parameters, the maximal plane wave energy, the Monkhorst-Pack sampling of the Brillouin zone for \mbox{$\mathbf{k}$-points} and \mbox{$\mathbf{q}$-points}, and the broadening parameter $\eta$ for the self-energy can be found in Table~S2 of the Supplemental Material~\cite{supplemental_zprsoc}.

We evaluate the sum on band index $n'$ in the self-energy (Eq.~(\ref{eq:sigmafan}) and~(\ref{eq:sigmadw})) using a semi-static approach~\cite{ponce_temperature_2014, antonius_dynamical_2015}: we replace the explicit evaluation of the non-adiabatic contribution of the high energy bands, namely, bands where the phonon frequencies are negligible compared to the difference between the electronic eigenenergies, by the solution of a Sternheimer equation~\cite{gonze_theoretical_2011} for the subspace orthonormal to the active subspace. We chose the explicit number of bands in the active subspace such that the energy difference between the CBM and the highest band was at least 20 eV. We finally obtain the converged ZPR values in the $N_q\rightarrow\infty$ limit using the linear extrapolation method described in Ref.~\cite{ponce_temperature_2015}. 

\subsection{Generalized \fro model}

The generalized \fro model (Eq.~(\ref{eq:zpr-gfr})) relies on the evaluation of angular-averaged square-root effective masses. From the first-principles perspective, electronic effective masses are typically computed either from finite differences or from density-functional perturbation theory. In the absence of SOC, we use the latter to evaluate the effective mass tensor, or, in the case of degenerate states, the transport-equivalent effective mass tensor defined in Ref.~\cite{laflamme_janssen_precise_2016}. 

When SOC is taken into account, the calculation of the effective mass tensor from density-functional perturbation theory is not currently implemented in the \textsc{Abinit} code for norm-conserving pseudopotentials. 
Hence, we evaluate the effective masses with SOC using order-4 central finite differences from the first-principles electronic eigenvalues. The VBM of zincblende materials constitutes a special case, as the electronic dispersion displays Dresselhaus splitting due to the lack of inversion symmetry~\cite{dresselhaus_spin-orbit_1955}. Therefore, we model the dispersion with the Dresselhaus Hamiltonian~\cite{dresselhaus_spin-orbit_1955} and obtain the effective masses from quadratic fits.  For comparison, we also compute the angular-averaged effective masses for the VBM using the electronic dispersion obtained from the Luttinger-Kohn~\cite{luttinger_motion_1955} Hamiltonian in the presence of SOC. See Appendixes~\hyperref[sec:lkmodel]{A} and~\hyperref[sec:dressmodel]{B} for more details about our treatment of the VBM. When Dresselhaus splitting is noticeable near the CBM of zincblende materials, we evaluate the effective masses from quadratic fits using the first-principles electronic dispersion. We finally note that, despite not being very accurate, the electronic effective masses computed with GGA-PBE are sufficient for the purpose of this work, which focuses on EPI. Lastly, as the effective masses computed from PBE for GaAs at the theoretically relaxed lattice parameter are particularly small, we also provide results computed at the experimental lattice parameter~\cite{kittel_condensed_2004}.

\section{Results and discussion}

\subsection{\label{sec:res-fp}First-principles}

\subsubsection{Effect of SOC on the VBM and CBM ZPR}\label{sec:res-vbmcbm}

\begin{figure*}
    \centering
    \includegraphics[width=\linewidth]{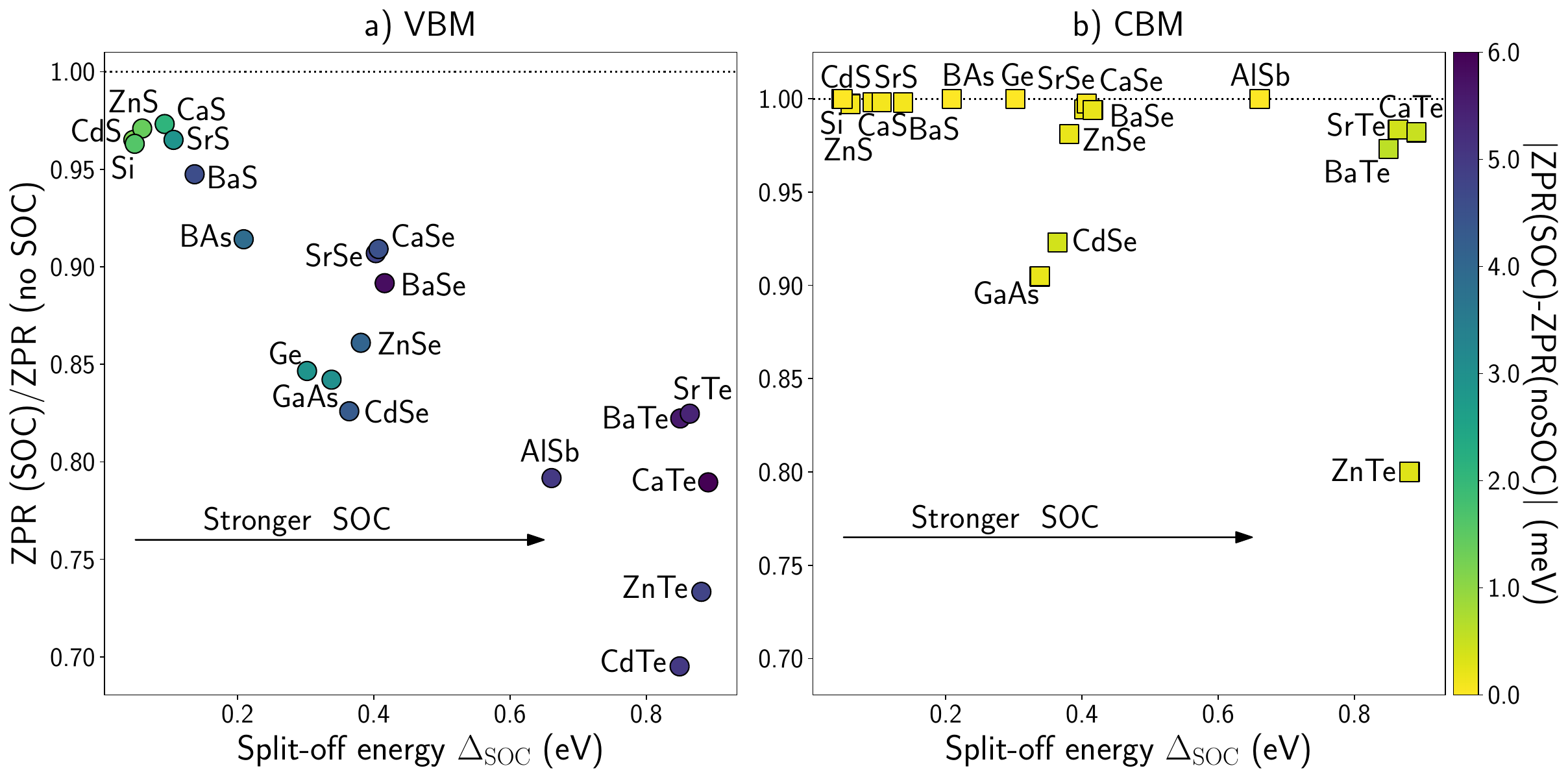}
    \caption{\textbf{Effect of SOC on the ZPR from first principles}, for a) the VBM (circle markers) and b) the CBM (square markers). The ZPR reduction ratio, ZPR(SOC)/ZPR(no SOC), is presented with respect to the split-off energy, $\Delta_{\rm{SOC}}$, which is an indicator of the SOC's strength. The color scale shows the absolute value of the SOC-induced correction to the ZPR, $|$ZPR(SOC)-ZPR(no SOC)$|$, in meV. For the VBM, the absolute difference remains under 6~meV but increases with $\Delta_{\rm{SOC}}$, yielding a relative decrease that reaches up to 30\% for the heavier materials (AlSb and the tellurides). For the CBM, most materials show a relative decrease below 5\%, with the exception of ZnTe, CdSe and GaAs. The absolute value of the difference remains negligible, below 1~meV. CdTe is absent from the CBM figure (see text). Numerical values are provided in Table~S3 of the Supplemental Material~\cite{supplemental_zprsoc}.}
    \label{fig:fp_results_lin}
\end{figure*}
The effect of SOC on the ZPR computed from first principles for our twenty materials is shown in Fig.~\ref{fig:fp_results_lin}, for both the VBM (left, circle markers) and the CBM (right, square markers). Both subfigures display the ZPR reduction ratio, ZPR(SOC)/ZPR(no SOC), with respect to the split-off energy, $\Delta_{\rm{SOC}}$, which is a direct indicator of the SOC's strength. The color scale indicates the absolute value of the SOC-induced correction to the ZPR for each band extrema, $|$ZPR(SOC)-ZPR(no SOC)$|$. Numerical values can be found in Table~S3 of the Supplemental Material~\cite{supplemental_zprsoc}.

On the one hand, one can observe that the relative decrease of the valence band edge ZPR, ZPR$_{\rm{v}}$, qualitatively increases with $\Delta_{\rm{SOC}}$. Recall that, for zincblende materials, the leading orbital character of the VBM is $p$ states from the anion. Hence, the less affected materials regroup the lighter anions, namely, all sulfides and Si, for which the relative decrease is below 5\%. The selenides, arsenides and Ge display a relative decrease ranging from \mbox{$\sim$10--15\%}, while the heavier materials in our set, AlSb and the tellurides, see their ZPR$_{\rm{v}}$ reduced by \mbox{$\sim$15--30\%}. However, the numerical value of $|$ZPR$_{\rm{v}}$(SOC)-ZPR$_{\rm{v}}$(no SOC)$|$ remains small, under 6~meV for all materials. Nevertheless, the color scale clearly indicates that the absolute value of the correction increases with $\Delta_{\rm{SOC}}$. Small absolute differences were to be expected, as stronger SOC is naturally present in heavier materials, which typically display smaller ZPR. 

However, the effect of SOC we observe on the ZPR$_{\rm{v}}$, hence on the real part of the self-energy, is not nearly as significant as the relative impact reported in the literature for the hole mobility~\cite{ma_first-principles_2018,ponce_towards_2018,ponce_first-principles_2021}, which can be over 10\% in weak-SOC materials like Si and reach more than 50\% in heavier materials. In this context, taking SOC into account reduces the number of scattering channels, hence increasing the mobility. In contrast, despite their respective contributions being reduced by SOC, all phonon wavevectors still contribute to the ZPR. Recall that the mobility depends on the electron-phonon self-energy through the relaxation time of the electronic states, which goes as the inverse of the imaginary part of $\Sigma\kn^{\rm{AHC}}$~\cite{giustino_electron-phonon_2017}. Thus, one could expect the inverse relaxation time of a given electronic state when including SOC to decrease by a similar ratio as the mobility when SOC is neglected. Nevertheless, it is not entirely clear how those two ratios should correlate, as the inverse relaxation time is defined for each electronic state, while the mobility is a global quantity integrated on the BZ, hence all the neighboring states around the $\Gamma$ point contribute to the hole mobility. While we have not attempted a full study of the imaginary part of the self-energy for all materials, we observe a relative decrease of the imaginary part of $\Sigma\kn$ near the VBM which is larger than the ones reported in Fig.~\ref{fig:fp_results_lin}a) for the ZPR of AlSb, ZnTe, CdTe and Si. This agrees with the trends reported in the literature for the mobility. See Sec.~S2 and Fig.~S3 of the Supplemental Material~\cite{supplemental_zprsoc} for more details. As the real and imaginary parts of $\Sigma\kn^{\rm{AHC}}$ are related to one another by the Kramers-Kronig relations, further investigation will be required to fully understand the effect of SOC on the full electron-phonon self-energy. 

The CBM, on the other hand, displays very little modification of the ZPR from SOC. The relative decrease remains under 5\% for all materials, including AlSb and some tellurides. These results are in line with the atomic picture, in which a $s$-like band such as the CBM of zincblende materials is not affected by SOC since $l=0$. This argument does not hold for the $d$-like CBM of the rocksalt alkaline earth chalcogenides; in that case, the conduction band edge ZPR (ZPR$_{\rm{c}}$) decrease remains negligible as the CBM is well-isolated in energy from the other bands. The only three exceptions to this trend are ZnTe (20\%), GaAs (9\%) and CdSe (8\%). 
CdTe is absent from the CBM figure as its ZPR$_{\rm{c}}$ changes sign when including SOC, going from --0.4 to 0.4 meV. 
Nevertheless, these seemingly large relative corrections are not numerically significant: the absolute correction remains under 1~meV for all materials (see the color scale), which is within the typical numerical accuracy of this type of calculations.

\begin{figure}
    \centering
    \includegraphics[width=\linewidth]{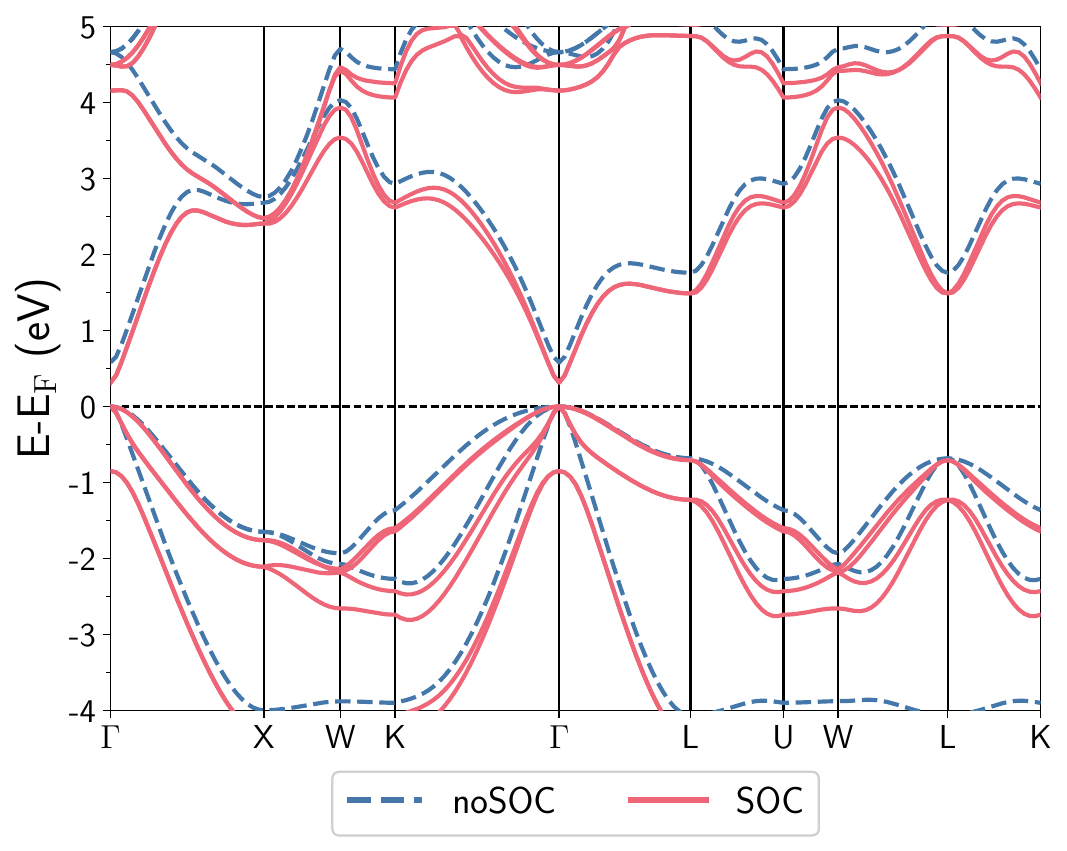}
    \caption{\textbf{Band structure of zincblende CdTe,} with SOC (solid red lines) and without SOC (dashed blue lines). As a guide to the eye, the SOC energy bands have been shifted such that the VBM coincides with its equivalent without SOC. The Fermi energy without SOC has been set to zero.}
    \label{fig:cdtebs}
\end{figure}
To understand these different tendencies from a qualitative point of view, one can picture a typical zincblende band structure. In Fig.~\ref{fig:cdtebs}, the energy bands of CdTe with SOC (solid red lines) have been shifted such that the VBM coincides with the VBM without SOC (dashed blue lines). The Fermi energy without SOC has been set to zero. When comparing both sets of bands, one can see that the general shape of the unoccupied subset of bands is scarcely affected by SOC, other than a relatively small energy shift between spin-split bands, located mostly close to the Brillouin zone boundaries. Recall that the Fan and Debye-Waller contributions to the self-energy have opposite signs (see Eq.~(\ref{eq:sigmafan}) and~(\ref{eq:sigmadw})) and almost equal magnitudes. Nevertheless, when considering the contribution from either the occupied or unoccupied subset of bands to the AHC self-energy, the Fan term typically governs the net sign of the renormalization in semiconductors. Hence, as
the CBM is mostly repelled by couplings with nearby conduction states of higher energy, one can deduce that the ZPR of the CBM will barely be affected by SOC, since the effective mass does not change significantly and the band extrema is well isolated in energy. The three apparent outliers, CdSe, ZnTe and GaAs, all feature a direct fundamental gap at the zone center, small CBM effective masses and a relatively small band gap energy (albeit always much larger than the highest phonon frequency, GaAs with SOC being the exception). In the light of the Kane model~\cite{kane_band_1957}, recall that the CBM and VBM at the $\Gamma$ point are linked through an avoided crossing. The CBM is thus indirectly affected by SOC, through its interplay with the heavy hole and 
light hole $p$-like bands. A small band gap reinforces this interaction, resulting in a greater relative decrease of the ZPR$_{\rm{c}}$ for these three materials. We do not observe such an effect in indirect band gap materials like AlSb. 

 When rather considering the VBM, one can observe the loss of degeneracy predicted by group theory, as the two split-off bands have been shifted by $\Delta_{\rm{SOC}}$ below the heavy hole and light hole bands. Figure~\ref{fig:cdtebs} also reveals two main consequences of SOC on the occupied bands: on the one hand, the effective masses in the vicinity of the zone center are reduced, and on the other hand, the energy shift between the spin-split bands occurs throughout the Brillouin zone, thus globally lowering the eigenenergies of the occupied states with respect to the VBM energy.

With this simple picture in mind, our results suggest that the effect of SOC can be safely neglected for band extrema which are well-isolated in energy, should the modification of the effective mass remain small. In contrast, degenerate extrema or densely entangled bands must be treated more carefully. 
We finally emphasize that one must not systematically overlook the apparent small magnitude of the SOC corrections to the ZPR$_{\rm{v}}$. 
What may at first be perceived as \enquote{only a few meV} nevertheless captures a significant relative decrease for the heavier materials, that reaches 15--30\% of the predicted ZPR$_{\rm{v}}$ without SOC. This effect cannot be neglected when aiming for predictive results, especially if one seeks to validate numerical predictions with experimental data.

\subsubsection{Experiment vs first principles}\label{sec:fpvsexp}

\begin{figure}
    \centering
    \includegraphics[width=\linewidth]{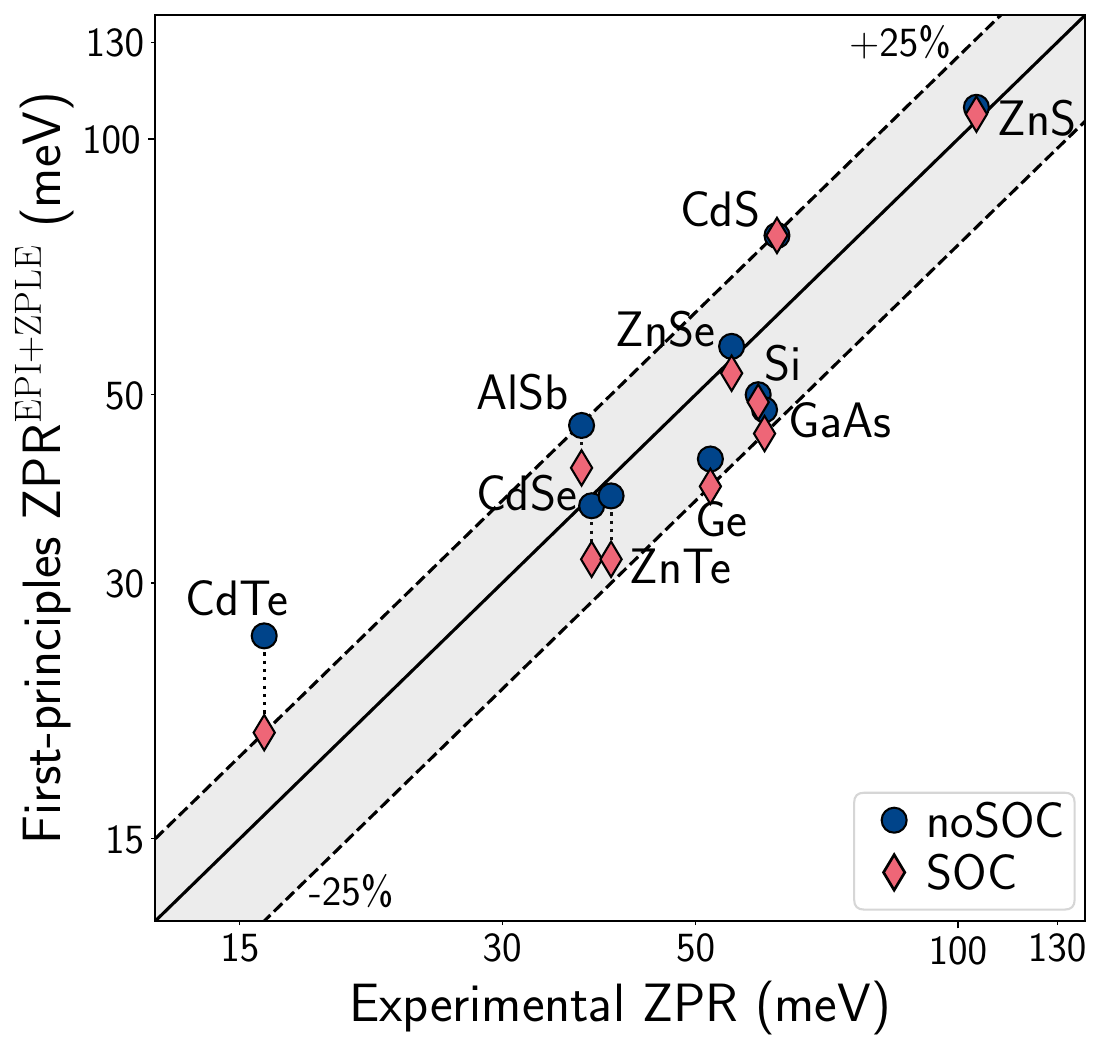}
    \caption{\textbf{Comparison between first-principles AHC band gap ZPR and experimental values,} without SOC (blue circles) and with SOC (red diamonds). The dashed lines, emphasized by the shaded gray area, indicate the theoretical ZPR$_{\rm{g}}$ agrees with the experimental one within 25\%. Note that the scales are logarithmic. When SOC is included, all materials lie within the shaded area. Numerical experimental values are provided in Table~S5 of the Supplemental Material~\cite{supplemental_zprsoc}.
    }
    \label{fig:experimentlin}
\end{figure}

We now examine how does the inclusion of SOC affects the global agreement between ZPR$_{\rm{g}}$ and available experimental data.  
To make a fair comparison with experimental data, the first-principles data shown in Fig.~\ref{fig:experimentlin} include the theoretical contribution of the zero-point expansion of the lattice~\cite{allen_theory_2020,brousseau-couture_zero-point_2022}. This term originates from the phonon contribution to the total free energy of the crystal, which increases the $T=0$~K lattice parameter compared to the static equilibrium value and, in turn, affects the band gap energy (see Appendix~\hyperref[sec:sm-zple]{D} for more details). Note that the scales are logarithmic. The shaded gray area highlights the region where both ZPR (first-principles and experimental) agree within 25\% of each other. As experimental values of the ZPR are obtained from extrapolation procedures rather than from direct measurements, we have to keep in mind when analyzing the accuracy of theoretical results that there is an experimental uncertainty which can be quite substantial, especially when few experimental datasets are available for a given material. See the Supplementary Note 1 of Ref.~\cite{miglio_predominance_2020} for a detailed discussion about the uncertainties associated with the experimental values of the ZPR.

The results without SOC (blue circles) are equivalent to the non-adiabatic AHC data shown in Fig.~2 of Ref.~\cite{miglio_predominance_2020} for the ten materials considered. 
When SOC is taken into account (red diamonds), all materials now lie within the tolerance criterion, including CdTe, which is largely overestimated without SOC. Thus, SOC does not alter the quantitative agreement between first principles and experiment, although Ge and GaAs reach the lower limit of the tolerance criterion. 

One can also wonder if the greater predictability of the nonadiabatic AHC approach compared to the adiabatic supercell method claimed in Fig.~2 of Ref.~\cite{miglio_predominance_2020} (see empty red triangles, labeled ASC-DFT) would remain upon inclusion of SOC. Although we have not attempted any adiabatic supercell calculation with SOC, our results suggest that the inclusion of SOC would not reduce the significant underestimation of the ZPR by adiabatic supercells for the lighter, more ionic materials. For intermediate to strong SOC, our data show a reduction of the total band gap ZPR ranging between 8\%--34\% (see the rightmost column of Table~S3 of the Supplemental Material~\cite{supplemental_zprsoc}). Should we infer that a similar effect would be observed in adiabatic supercell calculations, one could expect the result obtained from adiabatic supercells based on DFT calculations for CdTe to lie inside the tolerance criterion. At the same time, the underestimation would worsen for CdSe, and the ZnSe adiabatic supercell result would likely exit the shaded area. Our results, therefore, support the general conclusion of Miglio \textit{et al.}~\cite{miglio_predominance_2020} regarding the adiabatic supercell method being outperformed by the non-adiabatic AHC approach. 

\subsubsection{Origin of the SOC-induced ZPR decrease}\label{sec:histograms}

\begin{figure*}
    \centering
    \includegraphics[width=0.49\textwidth]{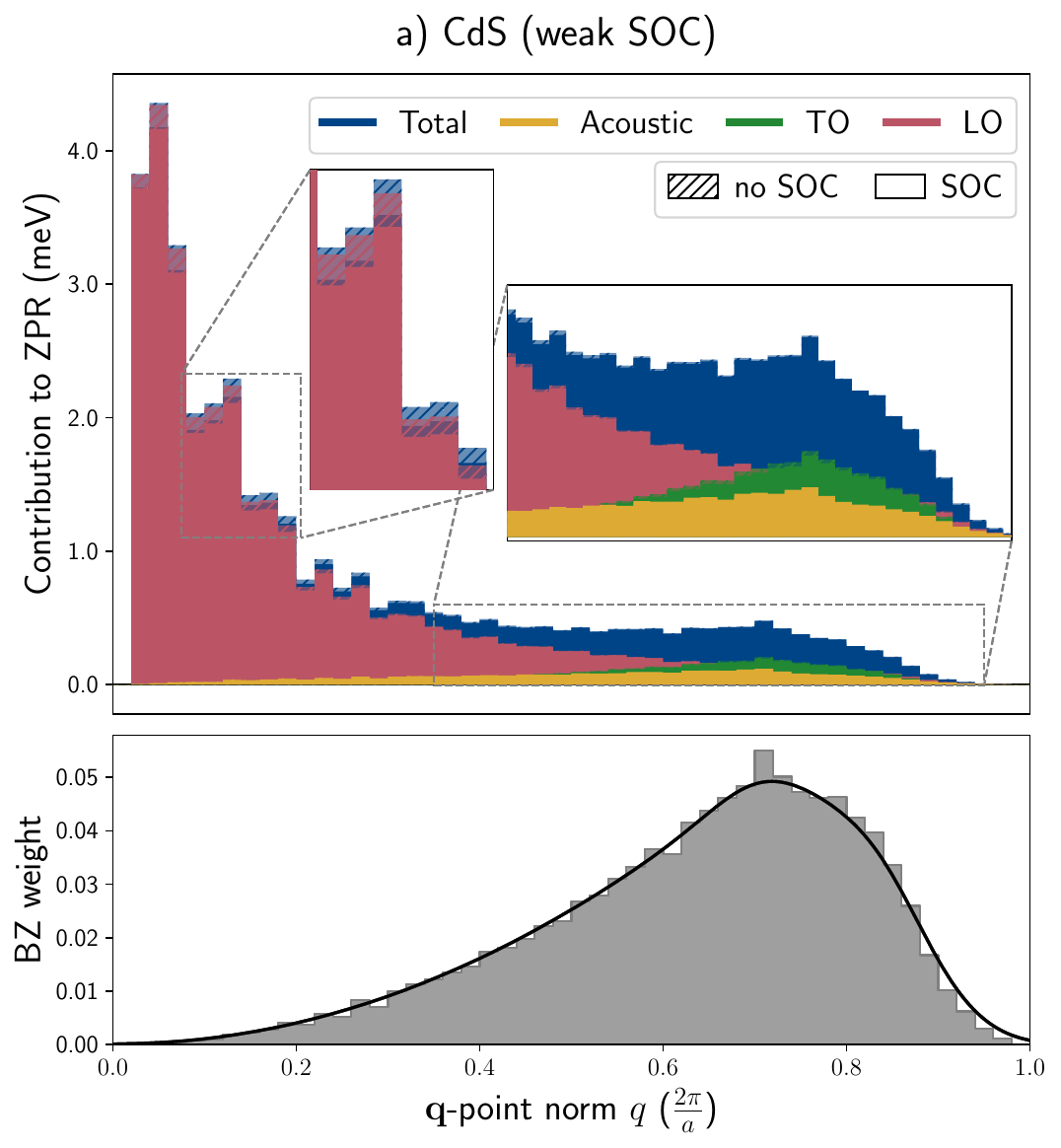}\label{fig:histogram_cds}
    \includegraphics[width=0.49\textwidth]{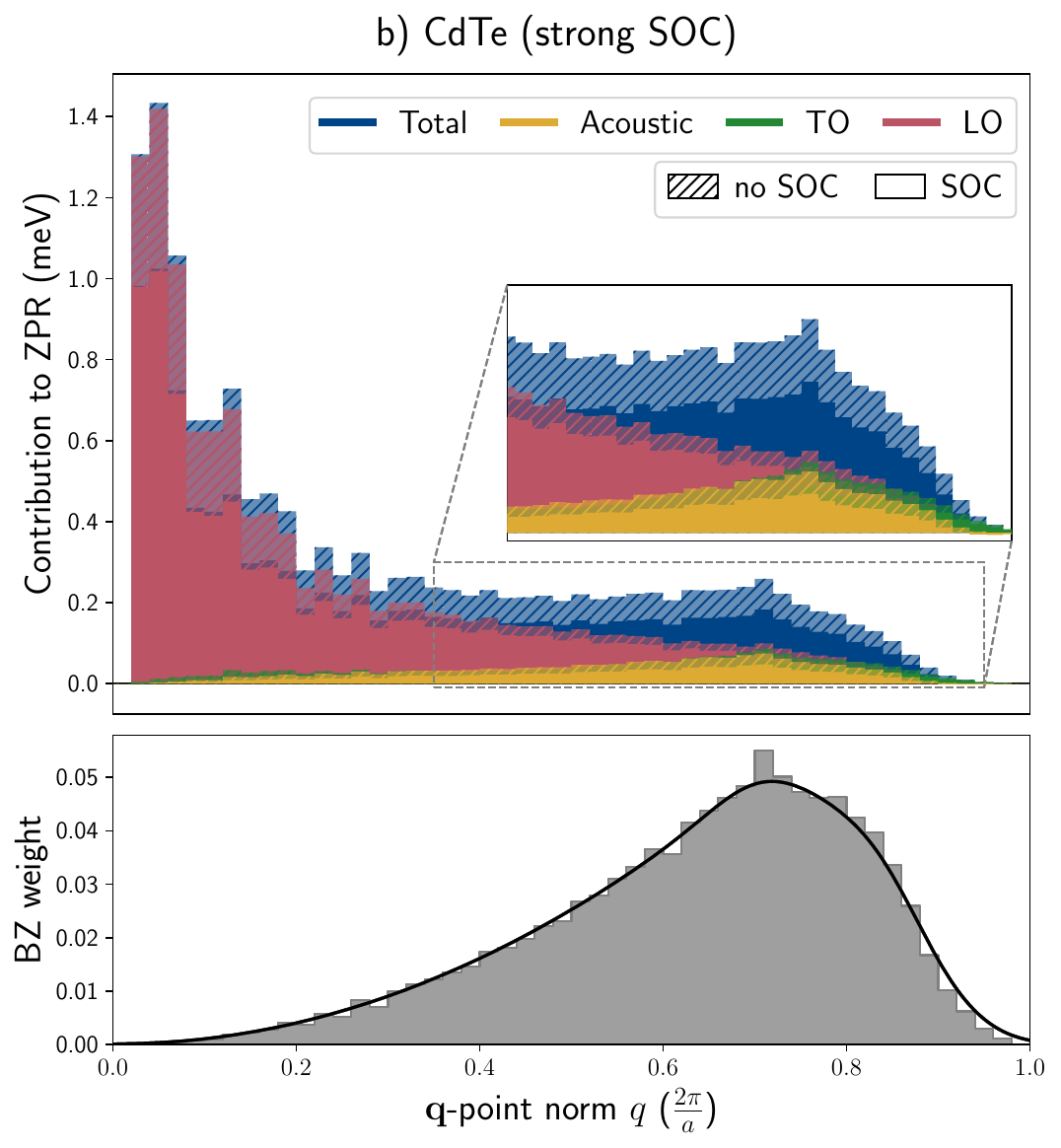}\label{fig:histogram_cdte}
    \caption{\textbf{Total and mode contribution to ZPR$_{\rm{v}}$ for different \qpoint norms, for polar materials,} for a \mbox{$48\times48\times48$} $\Gamma$-centered \qpoint grid, for a) CdS (weak SOC), b) CdTe (strong SOC). Solid histograms include SOC while the shaded hatched ones do not. Lower panels show the Brillouin zone weight distribution for the different \qpoints.  Both materials display a clear signature of the \fro interaction, as the LO mode (red) at small $q$ accounts for the greatest part of the total ZPR. Notice the different $y$-axis scaling for both materials. The total contribution at large $q$ (blue) is more equally splitted between the LO, acoustic (yellow) and TO (green) modes. For CdTe (right), the SOC-induced decrease of the ZPR originates from the whole Brillouin zone, and can be heuristically understood in terms of the variation of the effective masses (small $q$) and of a global decrease of the eigenenergies of the occupied bands (large $q$). Only the small $q$ behavior is observed in CdS (left), as emphasized by the insets.}
    \label{fig:histograms}
\end{figure*}
\begin{figure}
    \centering
    \includegraphics[width=\linewidth]{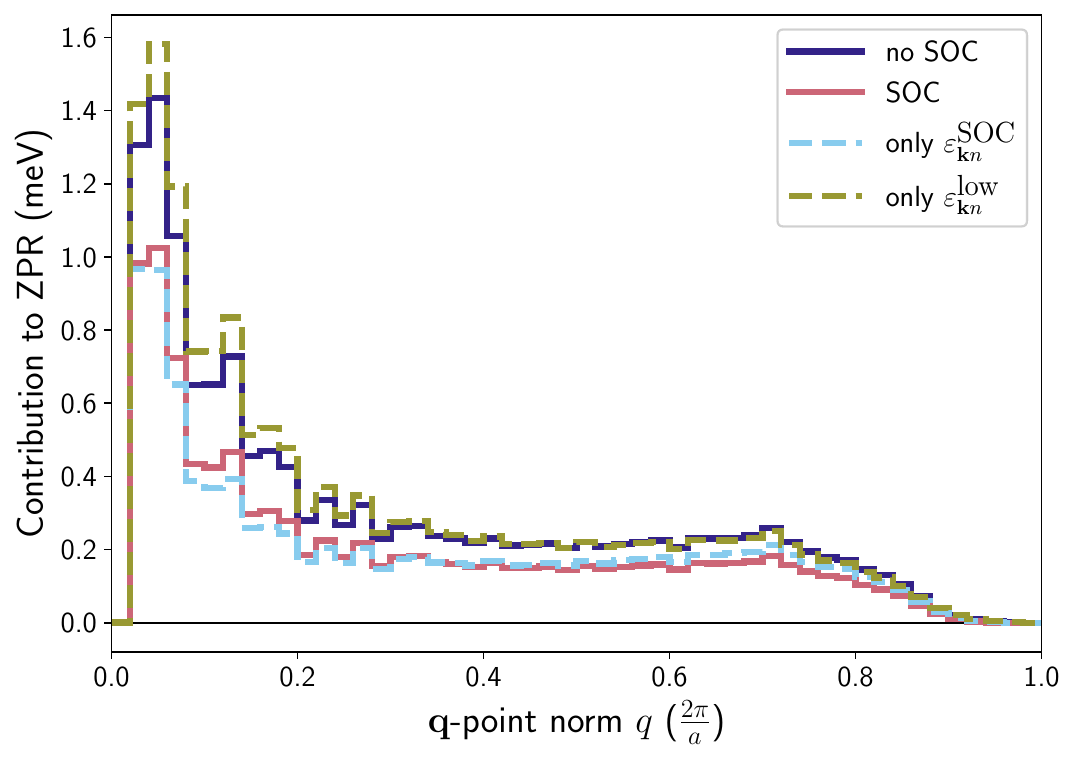}
    \caption{\textbf{Contribution to the total ZPR$_{\rm{v}}$ of CdTe with varying SOC strength for different \qpoint norms,}  for a \mbox{$48\times48\times48$} $\Gamma$-centered \qpoint grid. The different physical quantities entering $\Sigma\kn^{\rm{EPI}}$ (Eq.~(\ref{eq:sigmafan}) and~(\ref{eq:sigmadw})) are computed without SOC (dark indigo), with SOC (light red), and from mixed combinations of full SOC and 1\% SOC quantities (cyan and olive dashes, see text for details). The four histograms can be qualitatively grouped into two classes according to the strength of SOC used to evaluate the electronic eigenvalues, in agreement with the heuristic explanation developed in Fig.~\ref{fig:histograms}.}
    \label{fig:cdte_lowsoc_histograms}
\end{figure}

In Sec.~\ref{sec:res-vbmcbm}, we discussed the effect of SOC on the ZPR in terms of the variation of the electronic eigenvalues. We now refine this analysis by constructing histograms of the different \qpoints contribution to ZPR$_{\rm{v}}$ with respect to the norm of the phonon wavevector, for a \mbox{$48\times48\times48$} \qpoint grid. Figure~\ref{fig:histograms} displays such decomposition for two polar materials, CdS (left) where SOC has little effect on the calculated ZPR, and CdTe (right) where, on the contrary, the calculated ZPR is more strongly reduced by SOC. The bottom panels show the distribution of the Brillouin zone weight for the different wavevector bins. The solid and shaded hatched histograms refer to the ZPR contributions computed respectively with SOC and without SOC.

A first observation that emerges from this figure is the extremely similar shape of the mode histograms, apart from their respective energy scale. For both materials, the vast majority of the ZPR originates from the LO phonon mode (red) in a very small portion of the Brillouin zone, near the zone center (small $q$). This behavior is a clear signature of the \fro interaction. The contribution of the large $q$ modes, which cover most of the Brillouin zone, is significantly smaller and split more equally between the acoustic (yellow), transverse optical (TO, green) and LO modes. 

We now compare the solid and shaded histograms to identify how the different $q$ regimes are affected. The total contribution to the ZPR (blue) for CdTe shows that strong SOC reduces the contribution of all phonon modes throughout the Brillouin zone. This suggests that, in the large $q$ regime, the ZPR decrease can be associated with the global reduction of the electronic eigenvalues in the occupied subset of bands. In contrast, in the small $q$ regime, it can be linked to the decrease of the effective masses. On the contrary, for CdS (weak SOC), the large $q$ regime is unaffected by SOC; thus, only the variation of the band curvatures in the vicinity of the VBM seems to be responsible for reducing the ZPR.

At this point, we insist that our interpretation of the SOC-induced ZPR reduction in terms of the variation of electronic eigenvalues remains a heuristic analysis, since the full ZPR expressions for a given \qpoint (Eq.~(\ref{eq:sigmafan}) and~(\ref{eq:sigmadw}))  include other physical quantities which are, in principle, affected by SOC: the phonon frequencies, $\omega\qv$, and the squared EPI matrix elements, $|g\subkn{k}{n n'}^{\rm{Fan}}(\mathbf{q}\nu)|^2$ and $|g\subkn{k}{n n'}^{\rm{DW}}(\mathbf{q}\nu)|^2$.
To test our hypothesis, we computed all the physical quantities entering $\Sigma\kn^{\rm{Fan}}$ and $\Sigma\kn^{\rm{DW}}$ for CdTe while artificially reducing SOC to 1\% of its full strength. This allows us to decompose the electronic states correctly in terms of the double group irreducible representations while still reproducing the electronic and phononic dispersions without SOC adequately (see Sec.~S1 of the Supplemental Material~\cite{supplemental_zprsoc} for more details). With these in hand, we can precisely control which ingredients of the self-energy are affected by SOC. While such arbitrary combinations of full SOC and low SOC quantities have no physical meaning per se, they will prove insightful for understanding our previous results.

Figure~\ref{fig:cdte_lowsoc_histograms} show the histogram decomposition of the total ZPR$_{\rm{v}}$ (the equivalent of the blue histograms of Fig.~\ref{fig:histograms}) for different combinations of contributions. Table~\ref{tab:cdte_lowsoc} contains the numerical results for ZPR$_{\rm{v}}$, ZPR$_{\rm{c}}$ and ZPR$_{\rm{g}}$ for the same combinations. The data labeled \enquote{only $\varepsilon\kn^{\rm{low}}$} (olive dashes) refers to $\Sigma\kn^{\rm{EPI}}$ being computed with the full SOC except for the electronic eigenenergies, which are taken at 1\% SOC. In constrast, \enquote{only $\varepsilon\kn^{\rm{SOC}}$} (cyan dashes) is computed with 1\% SOC except for the electronic eigenenergies, which are taken at full SOC.

From Fig.~\ref{fig:cdte_lowsoc_histograms} and Table~\ref{tab:cdte_lowsoc}, the histograms can be grouped in two categories. On the one hand, the data computed using low SOC eigenvalues, \enquote{only $\varepsilon\kn^{\rm{low}}$}, yield both histograms and total ZPR values which qualitatively reproduce the calculation without SOC (dark indigo). On the other hand, the data computed by including SOC only through the electronic eigenvalues, \enquote{only $\varepsilon\kn^{\rm{SOC}}$}, capture almost all the ZPR decrease of the SOC data throughout the Brillouin zone (see the light red reference line). These results validate our heuristic explanation and confirm that the modification of the electronic eigenvalues dominates the SOC-induced decrease of ZPR$_{\rm{v}}$ in the different $q$ regimes. While the phonon frequencies and the EPI matrix elements undoubtedly influence the quantitative results, we argue here that the SOC-induced variation of the $\varepsilon\kn^0$ are sufficient to estimate the effect of SOC on the VBM ZPR and, by extension, on the total band gap ZPR for this class of cubic materials (see also Sec.~S1 of the Supplemental Material~\cite{supplemental_zprsoc}). 

We note that our conclusions differ significantly from those of Ref.~\cite{Saidi_temperature_2016}, who reported that SOC enhances the electron-phonon coupling strength in perovskite methylammonium lead iodine (MAPbI$_3$), thus \textit{increasing} the temperature-dependent band gap renormalization compared to a scalar relativistic calculation. The electronic structure of  MAPbI$_3$ is, however, very different from the zincblende, diamond and rocksalt structures we investigated; the VBM is non-degenerate and reasonably well isolated in energy, while the CBM, of Pb character, can couple to more electronic states within a small energy window (see Fig.~4 of their Supplemental Material), resulting in the band gap opening with increasing temperature, in contrast with our set of materials.
We observe some variations of the ZPR$_{\rm{v}}$ as well as in the small $q$ regime of its histogram decomposition when including SOC in the EPI matrix elements and excluding it in the eigenvalues (see \enquote{only $\varepsilon\kn^{\rm{low}}$} data in Table~\ref{tab:cdte_lowsoc} and Fig.~\ref{fig:cdte_lowsoc_histograms}).  However, the effect is too small to allow us to draw conclusions.
In fact, for our set of materials, any effect of SOC on the EPI matrix elements 
is entirely washed out by the variation of the static electronic eigenvalues. 
We also verified that the band gap correction at $T=300$~K decreased by a similar ratio as the ZPR$_{\rm{g}}$ when including SOC (see Table~S4 
of the Supplemental Material), thus confirming that our conclusions hold beyond $T=0$~K. 

\begin{table}[]
    \centering
    \caption{\textbf{VBM, CBM and band gap ZPR of CdTe for different combinations of SOC strength.} See Sec.~\ref{sec:histograms} and the caption of Fig.~\ref{fig:cdte_lowsoc_histograms} for details.}\label{tab:cdte_lowsoc}
\sisetup{
table-format = 3.2 ,
table-number-alignment = center ,
}
    \setlength\extrarowheight{2pt}
    \begin{tabularx}{\columnwidth}{p{20mm} 
    *{3}{Y}
    }
    \hline\hline
    \multirow{2}{*}{Combination} & \multicolumn{3}{c}{ZPR (meV)}\\
    \cmidrule(lr){2-4}
    & VBM & CBM & ZPR$_{\rm{g}}$ \\
        \hline   
    noSOC & 16.4 & -0.4 & -16.8\\
    SOC & 11.4 & ~0.4 & -11.0\\
    1\% SOC & 16.2 & -0.5 & -16.7\\
    only $\varepsilon\kn^{\textrm{low}}$ & 17.4 & -0.8 & -18.3\\
    only $\varepsilon\kn^{\textrm{SOC}}$ & 11.1 & -0.8 & -11.9\\
    \hline\hline
    \end{tabularx}
\end{table}

\subsection{\label{sec:res-gfr}Generalized Fr\"ohlich model with SOC}

\begin{figure}
    \centering
    \includegraphics[width=\linewidth]{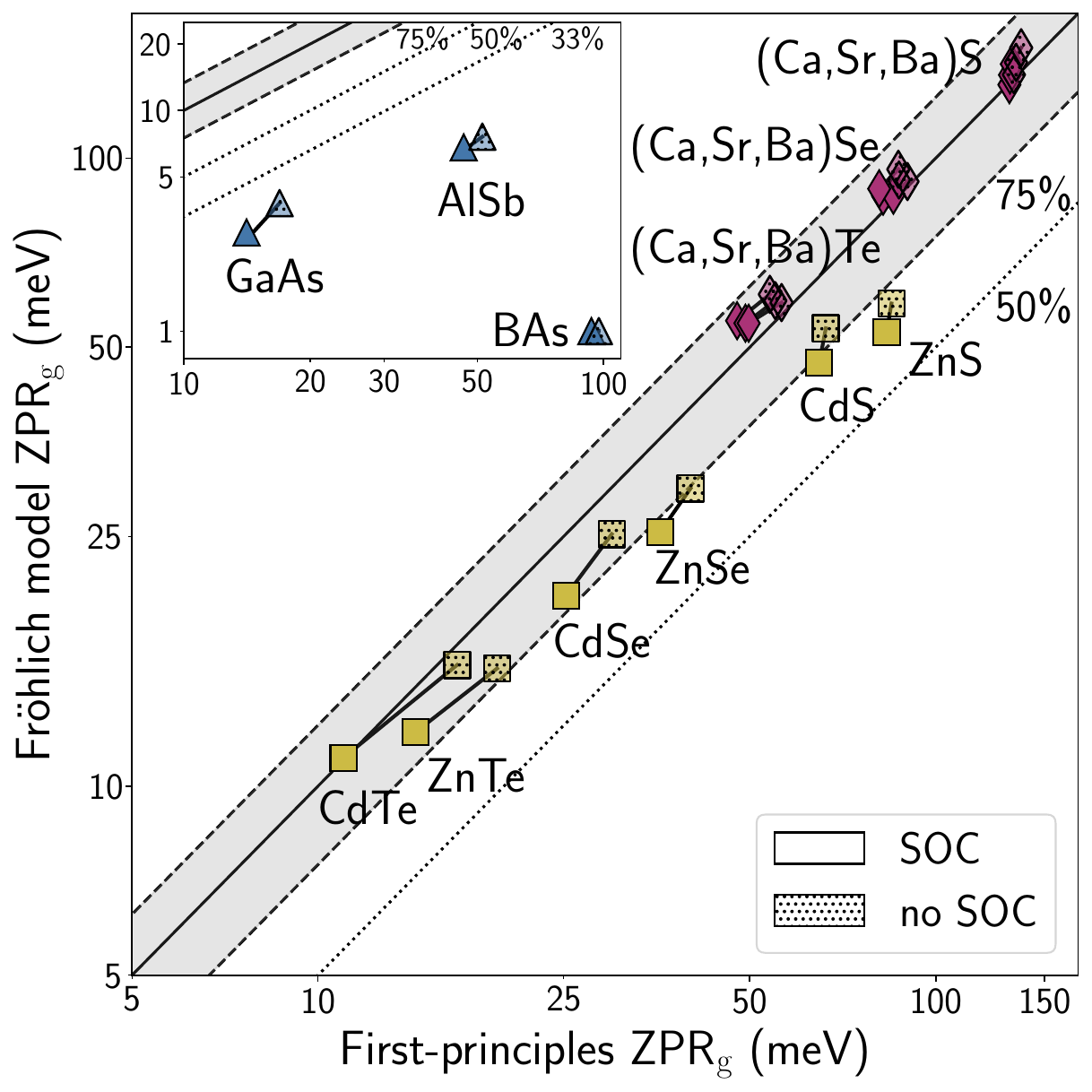}
    \caption{\textbf{Comparison between the generalized \fro model ZPR$_{\rm{g}}$ and the first principles AHC ZPR$_{\rm{g}}$, with and without SOC}. The different markers indicate materials with similar ionicity and band orbital characters: Zn and Cd chalcogenides (yellow squares), alkaline earth chalcogenides (purple diamonds) and group \MakeUppercase{\romannumeral 3}-\MakeUppercase{\romannumeral 5} materials. Solid (dotted) markers include (exclude) SOC. 
    Note that the scales are logarithmic. The dashed lines and shaded gray area delimits the region where they agree within 25\% of each other, while the dotted lines indicate where the ZPR$_{\rm{g}}^{\rm{gFr}}$ captures 50\% (also 33\% for the inset) of the AHC value. Tables~S3, S6 and~S7 of the Supplemental Material~\cite{supplemental_zprsoc} respectively report the numerical values for the ZPR$_{\rm{g}}$ obtained from AHC and the gFr model, as well as 
    the physical parameters entering Eq.~(\ref{eq:zpr-gfr}).
    }
    \label{fig:gfrvalues}
\end{figure}
\begin{figure}
    \centering
    \includegraphics[width=\linewidth]{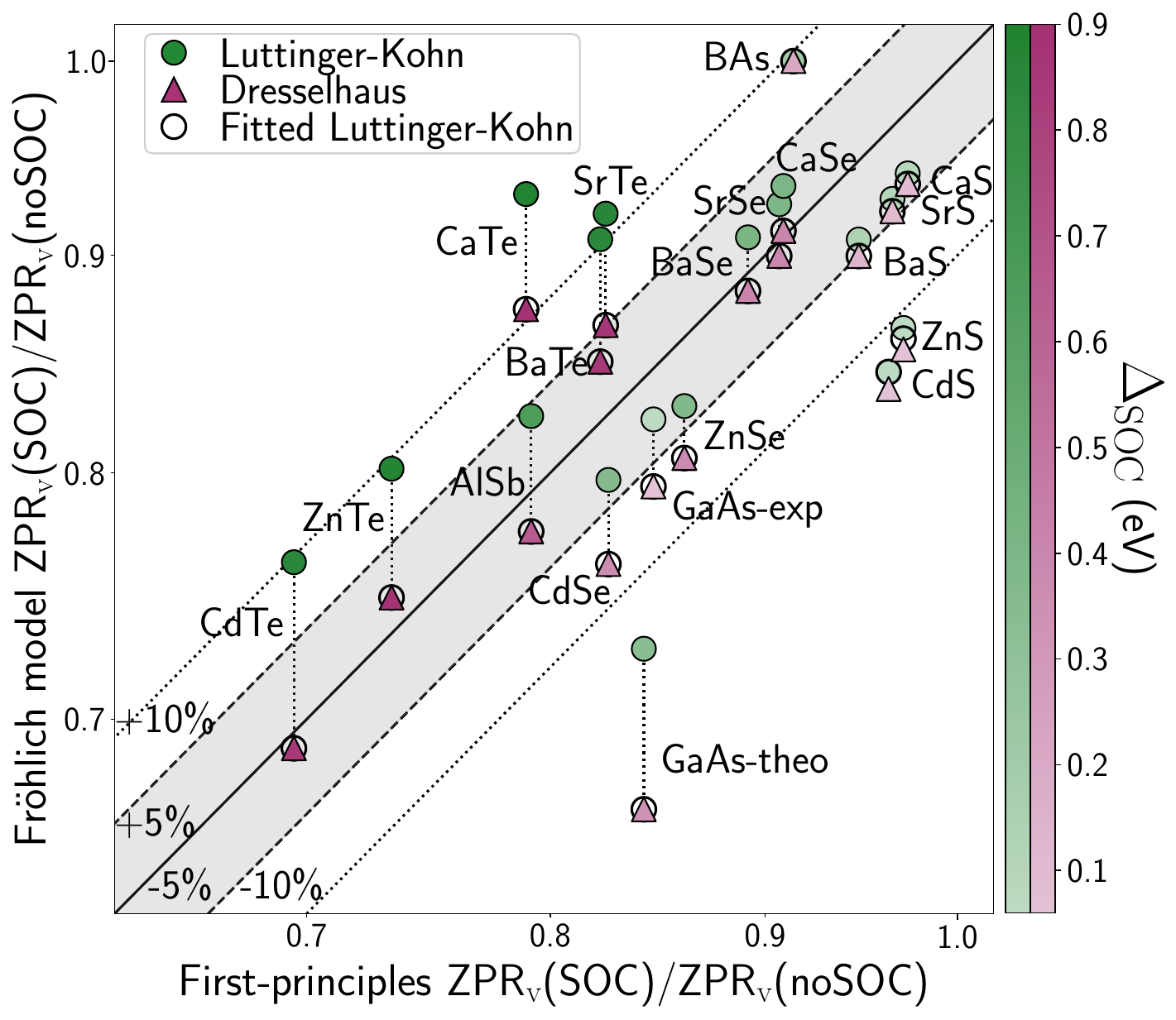}
    \caption{\textbf{Comparison of the SOC-induced reduction of the ZPR$_{\rm{v}}$ obtained with the generalized \fro model and with first-principles AHC methodology.} The electronic dispersion used to compute the angular-averaged effective masses entering the ZPR$_{\rm{v}}^{\rm{gFr}}$ (see Eq.~(\ref{eq:zpr-gfr})) was obtained either from the Luttinger-Kohn (green circles) Hamiltonian or from the Dresselhaus (purple triangles) Hamiltonian. The color intensity of the markers is proportional to the split-off energy. 
    Note that the scales are logarithmic.
    The dashed (dotted) lines indicate where the ratio computed with the gFr model deviates from the first-principles ratio by 5\% (10\%). For most materials, the gFr model provides a fairly good estimate of the first-principles ZPR$_{\rm{v}}$ decrease, although using effective masses from the Dresselhaus Hamiltonian delivers superior result. The discrepancy between results from both Hamiltonians increases with SOC, as the Luttinger-Kohn Hamiltonian tends to overestimate the light hole effective masses (see text). The Luttinger-Kohn Hamiltonian can be reparameterized with values extracted from the Dresselhaus model which reproduce more accurately the curvature of light holes around $\Gamma$. This delivers the SOC-induced reduction represented by empty circle, that agree better with each Dresselhaus-based results. Numerical values 
    are provided in Table~S6 
    of the Supplemental Material~\cite{supplemental_zprsoc}.
    }
    \label{fig:gfr_vs_fp_vb}
\end{figure}

We now present the results from our gFr model with SOC. As mentioned in Sec.~\ref{sec:gfr-with-dresselhaus} and~\ref{sec:computation}, and discussed thoroughly in Appendix~\hyperref[sec:dressmodel]{B}, the VBM of zincblende materials with SOC requires a special treatment, as it displays Dresselhaus splitting.
As a consequence, the VBM in a generic direction $\hat{\mathbf{k}}$ is slightly shifted from $\Gamma$, both in momentum and in energy.
We found an energy offset \mbox{$\Delta E_n(\hat{\mathbf{q}})$} smaller than $0.5$~meV for all materials, and an average momentum offset \mbox{$k^0_n(\hat{\mathbf{q}})$} of \mbox{$\sim5\times10^{-3}$\AA$^{-1}$}, at most \mbox{$10^{-2}$\AA$^{-1}$} for CdS. 
For comparison, the change in momentum expected from a photon doing a vertical optical transition in a semiconductor with a band gap energy of $E_g=1$~eV would be $\Delta k \sim E_g/hc\sim 10^{-4}$\AA$^{-1}$, with $h$ the Planck constant and $c$ the speed of light. 
As we argue in Sec.~\ref{sec:gfr-with-dresselhaus}, the momentum offset has no consequence on the ZPR$^{\rm{gFr}}$. Furthermore, for our set of materials, the largest value of $\Delta E_n(\hat{\mathbf{q}})$ is at least two orders of magnitude smaller than the LO frequency. Hence, we can safely neglect the energy offset and use Eq.~(\ref{eq:zpr-gfr}).

Figure~\ref{fig:gfrvalues} compares ZPR$_{\rm{g}}$ computed with the gFr model to the first-principles AHC result, both with SOC (solid markers) and without SOC (shaded dotted markers). Note that the scales are logarithmic. The shaded gray area delimits the region where the gFr value deviates from the first-principles ZPR by at most 25\%. In this figure, the VBM contribution to the ZPR$_{\rm{g}}$ with SOC was computed with the Dresselhaus Hamiltonian (see Appendix~\hyperref[sec:dressmodel]{B} for details). The Luttinger-Kohn Hamiltonian yields qualitatively equivalent results. The materials are grouped into three sets: the alkaline earth chalcogenides (purple diamonds), the Zn/Cd chalcogenides (yellow squares), both fairly ionic, and the less ionic \MakeUppercase{\romannumeral 3}-\MakeUppercase{\romannumeral 5} materials (blue triangles). Si and Ge are not considered in this Section, as their vanishing Born effective charges yield a null ZPR$_{\rm{g}}$ in the \fro picture.

The alkaline earth chalcogenides, which display a larger ZPR$_{\rm{g}}$ ranging from 50 to 150~meV, are very well described by the gFr model, which captures more than 75\% of the AHC ZPR$_{\rm{g}}$ both without and with SOC. The Zn and Cd chalcogenides ZPR$_{\rm{g}}$ is also reasonably well captured by the model, which accounts for more than two-thirds of the AHC value. The absolute value of their ZPR$_{\rm{g}}$ is smaller compared to the isoelectronic alkaline earth chalcogenides, which can be attributed to a smaller band gap at the DFT level that strengthens the contribution of interband couplings between valence and conduction states, thus reducing ZPR$_{\rm{g}}$. In contrast, the gFr model captures less than one third of the ZPR$_{\rm{g}}$ for \MakeUppercase{\romannumeral 3}-\MakeUppercase{\romannumeral 5} materials.

Upon inclusion of SOC (going from the dotted to solid markers), the gFr model qualitatively retains the same proportion of the AHC ZPR$_{\rm{g}}$ value for all three material families. However, we observe a very small decrease for the sulfides (upper rightmost groups of yellow and purple markers). 
These results confirm that the claim of Ref.~\cite{miglio_predominance_2020} is robust to the inclusion of SOC: the ZPR of ionic materials, here both chalcogenide families, is dominated by the physical picture of a large polaron, where the movement of the slow electron is correlated to the dynamically adjusting phonon cloud, as emphasized by their Fig.~5(b)~\cite{miglio_predominance_2020}. 

As in Ref.~\cite{miglio_predominance_2020}, we stress that a perfect agreement between the gFr and AHC methodologies was not expected, hence the labeling of the reference lines on Fig.~\ref{fig:gfrvalues} emphasize the fraction of the total first-principles ZPR$_{\rm{g}}$ captured by the gFr model rather than the level of agreement between the two values. By construction, the gFr model does not include the contribution from acoustic and TO modes, nor the Debye-Waller contribution, as its purpose is to solely capture the contribution of the nonadiabatic \fro interaction to the ZPR. Moreover, all interband couplings outside the fourfold degenerate subset at the VBM are neglected. Discrepancies could also arise from nonparabolic behavior of the electronic bands, which will naturally occur at some point since the Brillouin zone is finite and periodic, and from LO phonon dispersion.

On a different note, 
one can wonder if our gFr model with SOC would provide a reliable estimate of the SOC-induced decrease of ZPR$_{\rm{v}}$, without resorting to a full AHC calculation with SOC. In the spirit of the discussion presented in Sec.~\ref{sec:histograms}, the gFr model would capture the decrease of the hole effective masses near the $\Gamma$ point. To answer this question, Fig.~\ref{fig:gfr_vs_fp_vb} compares the ratio ZPR(SOC)/ZPR(noSOC) for the VBM obtained from the gFr model, using both the Luttinger-Kohn (green circles) and Dresselhaus (purple triangles) Hamiltonians, to the AHC value displayed in Fig.~\ref{fig:fp_results_lin}. The color intensity of the markers is proportional to the split-off energy, $\Delta_{\rm{SOC}}$. 
Note that the scales are logarithmic. 
The shaded gray area shows the region where both ratios agree within 5\% of each other. 

Two observations can be drawn from Fig.~\ref{fig:gfr_vs_fp_vb}. On the one hand, the two ratios agree within 5\% of each other for about half of the 19 materials (dashed lines):  11 for the Dresselhaus model and 9 for the Luttinger-Kohn model. The discrepancy with the first-principles ratio is below 10\% for most materials (dotted lines), thus providing a reasonable estimate of the SOC-induced decrease of the VBM observed in the full AHC calculation, from a single phonon calculation at the $\Gamma$ point. The main exceptions are CaTe with the Luttinger-Kohn model, GaAs at the theoretical lattice parameter, ZnS and CdS. 

On the other hand, the difference between both models increases with SOC, the Dresselhaus model being more accurate for heavier materials. This discrepancy can be attributed to the construction of the models: as it couples the fourfold heavy hole and light hole bands to the twofold split-off bands, the Luttinger-Kohn model intends to describe the effect of SOC in a broader region of the Brillouin zone, thus capturing the warping of the light hole bands around the split-off energy observed in many cubic materials (see, for example, Fig.~2 of Ref.~\cite{laflamme_janssen_precise_2016}). At stronger SOC, we observed that the cost of qualitatively capturing the correct band warping is a less accurate representation of the curvature of the light hole bands, yielding an overestimated $m^\ast$ compared to the first-principles dispersion. In contrast, the Dresselhaus Hamiltonian treats only the fourfold VBM and provides an accurate description of the bands in the vicinity of the $\Gamma$ point, at the cost of not describing the band warping further away in the Brillouin zone (see Fig.~\ref{fig:sm-modifiedlk} of Appendix~\hyperref[sec:sm-modifiedlk]{C}). Moreover, while our Dresselhaus model parameters are fitted to the first-principles dispersion with SOC, the Luttinger-Kohn Hamiltonian relies on the theoretical Luttinger parameters obtained without SOC, $\Delta_{\rm{SOC}}$ being the only parameter related to SOC. The original work by Luttinger and Kohn already cautioned about a decreased accuracy of their model for strong SOC~\cite{luttinger_motion_1955}. 

To validate this interpretation, we also compute the Luttinger-Kohn model using fitted Luttinger parameters extracted from the Dresselhaus model (empty circles, see Appendix~\hyperref[sec:sm-modifiedlk]{C} for details). The purpose of these parameters is to reproduce more accurately the curvature of the light hole bands around $\Gamma$, at the cost of deviating more severely from the first-principles dispersion around the split-off energy and underestimating the $m^\ast$ of the split-off bands compared to the original Luttinger parameters. In that case, both models agree very well. The fundamental differences between the two models suggest that the modified Luttinger-Kohn Hamiltonian constructed from fitted Luttinger parameters would be a more suitable choice to include SOC in the \fro polaron effective mass theory developed in Guster \textit{et al.}~\cite{guster_frohlich_2021}, as it simultaneously captures the correct curvature for the light hole bands in presence of SOC and lacks the numerical complications arising from the Dresselhaus splitting. One should nevertheless make sure that the agreement region between the fitted Luttinger-Kohn model and the first-principles dispersion covers at least a few $\omega_{\rm{LO}}$ for the predicted polaron effective mass to be reliable.

Lastly, we come back to the underestimation of the ratio observed  in Fig.~\ref{fig:gfr_vs_fp_vb} for ZnS and CdS, despite both models agreeing quite well with each other. To understand this result, recall the radial integral that occurred  during the derivation of the gFr model with SOC, Eqs.~(\ref{eq:gfr_beforeradial}) and~(\ref{eq:radialintegral}). At this point, we suppose that the parabolic behavior of the electronic bands can be extended to infinity to replace the integral with its asymptotic solution. By doing so, we assume that the Brillouin zone region where the effective mass approximation holds is sufficiently large such that one has reached a significant fraction of the asymptotic value once the bands start to deviate from parabolicity. Such fraction can be estimated by looking at the analytical solution of Eq.~(\ref{eq:radialintegral}) at a finite upper bound $q_c$:
 \begin{multline}\label{eq:finiteradialintegral}
     \int\limits_0^{q_c} \,dq  \frac{1}{\frac{q^2}{2m^\ast_{n}(\hat{\mathbf{q}})}+\omega_{\bm{0}j}(\mathbf{\hat{q}})}  \\ = \sqrt{\frac{2m^\ast_{n}(\hat{\mathbf{q}})}{\omega_{\bm{0}j}(\mathbf{\hat{q}})}}\;\;\textrm{arctan}\left(\frac{q_c}{\sqrt{2m^\ast_{n}(\hat{\mathbf{q}})\omega_{\bm{0}j}(\mathbf{\hat{q}})}}\right).
 \end{multline} 
Assuming that $q_c$ corresponds to the wavevector where the electronic bands start to deviate from parabolicity, the argument of the arctan function can be recast as
\begin{equation}
\sqrt{\frac{q_c^2/2m^{\ast}_{n}(\mathbf{\hat{q}})}{\omega_{\bm{0}j}(\mathbf{\hat{q}})}}=\sqrt{\frac{E_c}{\omega_{\bm{0}j}(\mathbf{\hat{q}})}},
\end{equation}
namely the square-root of the ratio of the eigenenergy of the electronic bands where they stop being parabolic and the LO frequency. Should the departure from parabolicity occur at an energy smaller than $\omega_{\bm{0}j}(\mathbf{\hat{q}})=\omega_{\rm{LO}}$ with respect to the VBM energy, like for ZnS and CdS, the radial integral would have reached at most arctan$(1)=0.5$, less than half of the asymptotic value, and one can reasonably question the validity of such an approximation. In physical terms, we warn against a possible breakdown of the parabolic approximation within the energy window that is physically relevant to the \fro interaction. As the effective mass approximation is a cornerstone assumption of the original \fro model, one should therefore be particularly careful when introducing SOC in the treatment of polarons for materials with high phonon frequencies.

\section{\label{sec:concl}Conclusion}

In the present study, we investigate the consequences of spin-orbit coupling on the electron-phonon interaction contribution to the zero-point renormalization of cubic materials. Our first-principles calculations show that spin-orbit coupling reduces the zero-point renormalization of the valence band edge by 15\%--30\% for the heavier materials, while the conduction band edge is scarcely affected. The leading mechanism behind this behavior, brought to light by an Allen-Heine-Cardona calculation where the strength of spin-orbit coupling is artificially reduced to 1\%, is revealed to be the variation of the electronic eigenvalues entering the electron-phonon self-energy and the decrease of the hole effective masses near the valence band maximum.

We also extend the generalized \fro model presented in Miglio \textit{et al.}~\cite{miglio_predominance_2020} to include the spin-orbit coupling, revealing some numerical subtleties in the treatment of the valence band maximum of zincblende materials due to Dresselhaus splitting. We show that the predominance of nonadiabatic effects on the zero-point renormalization of ionic materials is robust to the inclusion of spin-orbit coupling and that the generalized \fro model can be used to estimate the magnitude of the SOC-induced ZPR decrease with reasonable accuracy. We finally warn about the accuracy of the Luttinger-Kohn model with spin-orbit coupling for heavier materials and propose a method relying on the Dresselhaus model to extract fitted Luttinger parameters more suitable for the purpose of our generalized \fro model with spin-orbit coupling, as well as highlight a possible breakdown of the parabolic approximation on which the original \fro model with spin-orbit coupling is built for materials with high phonon frequencies.

\begin{acknowledgments}
The authors acknowledge fruitful discussions with M.~J.~ Verstraete and M.~Giantomassi, and thank S.~Ponc{\'e} for constructive comments about the manuscript. This research was financially supported by the Natural Sciences and
Engineering Research Council of Canada (NSERC), under the Discovery Grants
program grant No. RGPIN-2016-06666,
by the Fonds de la Recherche Scientifique (FRS-FNRS Belgium) through
the PdR Grant No.~T.0103.19 -~ALPS and 
by the European Union’s Horizon 2020 research and innovation program under Grant Agreement No. 951786 - NOMAD CoE.
This work is part of the SHAPEme project (EOS ID 400077525) that has received funding from the FWO and F.R.S.-FNRS under the Excellence of Science (EOS) program.
This research was enabled in part by support provided by Calcul Qu\'ebec (\url{www.calculquebec.ca}) and the Digital Research Alliance of Canada (\url{www.alliancecan.ca}). The operation of the supercomputers used for this research is funded by the
Canada Foundation for Innovation (CFI), the Minist\`ere de la Science, de l'\'Economie et de l'Innovation du Qu\'ebec (MESI), and the Fonds de recherche du Qu\'ebec – Nature et technologies (FRQ-NT). V.B.-C. acknowledges support by the NSERC Alexander
Graham Bell Canada Graduate Scholarship doctoral program, the FRQ-NT B2 Doctoral Scholarship and the Hydro-Qu\'ebec Excellence Scholarship. V.B.-C and M.C. are members of the Regroupement qu\'eb\'ecois sur les mat\'eriaux de pointe (RQMP).
\end{acknowledgments}

\section*{A\lowercase{ppendix} A: L\lowercase{uttinger}-K\lowercase{ohn} H\lowercase{amiltonian}} \label{sec:lkmodel}
\setcounter{equation}{0}
\renewcommand{\thetable}{A\arabic{table}} 
\renewcommand{\theequation}{A\arabic{equation}} 
The effective mass theory derived in 1955 by Luttinger and Kohn~\cite{luttinger_motion_1955} describes the behavior of the electronic bands in the vicinity of band extrema, using the second-order $\mathbf{k}\cdot\mathbf{p}$ theory. For the threefold degenerate VBM of cubic materials, the Hamiltonian without SOC writes
 \begin{widetext}
 \begin{equation}\label{eq:lkh-nosoc}
 \renewcommand*{\arraystretch}{1.5}
    H_{n,n'}(\mathbf{k}) = \begin{bmatrix}
                Ak_x^2 + B(k_y^2+k_z^2) & Ck_xk_y & Ck_xk_z\\
                Ck_xk_y & Ak_y^2 + B(k_x^2+k_z^2) & Ck_yk_z\\
                Ck_xk_z &Ck_yk_z & Ak_z^2 + B(k_x^2+k_y^2)
        \end{bmatrix},
\end{equation}
\end{widetext}
where $\mathbf{k}$ is the electronic wavevector, $n,n'$ are band indices, and $H_{n,n'}(\mathbf{k})$ is expressed in the \mbox{$\{\ket{X}, \ket{Y}, \ket{Z}\}$} basis, which forms an irreducible representation of the symmetry group of the wavevector at the VBM. 

The three parameters $A$, $B$ and $C$, known in the literature as the Luttinger parameters, can be deduced from the effective masses, $m_n^\ast$, along the $[100]$, $[110]$ and $[111]$ cartesian directions in reciprocal space,
\begin{align}
        m_n^{\ast -1} [100] &= \begin{cases}2A,\\ 2B\,(\textrm{twofold}),\end{cases}\nonumber \\
        m_n^{\ast -1} [110] &= \begin{cases}(A+B\pm C),\\ 2B,\end{cases}\\
        m_n^{\ast -1} [111] &= \begin{cases}\frac{2}{3}(A+2B+2C),\\ \frac{2}{3}(A+2B-C)\;(\textrm{twofold}).\end{cases}\nonumber
\end{align}

In the presence of SOC, the Hamiltonian contains an additional term, $H^{\rm{SOC}}$ (see Eq.~(\ref{eq:generalsoc})), 
which is treated as a perturbation in the $\mathbf{k}\cdot\mathbf{p}$ expansion. The Hamiltonian is expressed in the basis of the zeroth-order wavefunctions, which are the 4-fold $\{\ket{3/2, m_j}\}$ (heavy holes and light holes) and the 2-fold $\{\ket{1/2, m_j}\}$ (split-off),
\begin{widetext}
\begin{equation}\label{eq:lkh-soc}
\renewcommand*{\arraystretch}{1.5}
         H_{j,m_j}(\mathbf{k}) =  \begin{bmatrix}
                P/2 & L & M & 0 & iL/\sqrt{2} & -i\sqrt{2}M\\
                L^\ast & P/6+2Q/3 & 0 & M & -i(P-2Q)/3\sqrt{2} & i\sqrt{3}L/\sqrt{2}\\
                M^\ast & 0 & P/6+2Q/3 & -L & -i\sqrt{3}L^\ast/\sqrt{2} & -i(P-2Q)/3\sqrt{2}\\
                0 & M^\ast & -L^\ast & P/2 & -i\sqrt{2}M^\ast & -iL^\ast/\sqrt{2}\\
                -iL^\ast/\sqrt{2} & i(P-2Q)/3\sqrt{2}  & i\sqrt{3}L/\sqrt{2} & i\sqrt{2}M &(P+Q)/3 -\Delta_{\text{SOC}} & 0\\
                i\sqrt{2}M^\ast & -i\sqrt{3}L^\ast/\sqrt{2} & i(P-2Q)/3\sqrt{2} & iL/\sqrt{2} & 0 & (P+Q)/3 -\Delta_{\text{SOC}}
            \end{bmatrix},
\end{equation}
\end{widetext}
where the $\mathbf{k}$-dependent parameters $P$, $Q$, $L$ and $M$ are constructed from the Luttinger parameters,
\begin{equation}
    \begin{split}
        P(\mathbf{k}) &= (A+B)(k_x^2+k_y^2) + 2Bk_z^2,\\[2ex]
        Q(\mathbf{k}) &= B(k_x^2+k_y^2)+Ak_z^2,\\[2ex]
        L(\mathbf{k}) &= -\frac{iC}{\sqrt{3}}(k_xk_z -ik_yk_z),\\[1ex]
        M(\mathbf{k}) &= \frac{1}{\sqrt{12}}\left[(A-B)(k_x^2-k_y^2) -2iCk_xk_y\right].
    \end{split}
\end{equation}
The split-off energy, $\Delta_{\rm{SOC}}$, makes the Luttinger-Kohn Hamiltonian with SOC nonhomogenous in \mbox{($k_x, k_y, k_z)$}, resulting in the typical band warping observed around the split-off energy in light hole bands of cubic materials (see, for example, the $\Gamma$-L direction of Fig.~2 of Ref.~\cite{laflamme_janssen_precise_2016}). Around this energy level, the light hole bands typically depart from the quadratic behavior characterized by $m^{\ast}$~(SOC) 
to recover a curvature that resembles $m^{\ast}$~(noSOC). $\Delta_{\rm{SOC}}$ can be extracted either from experiments or from a DFT ground state calculations with SOC. Note that Eq.~(\ref{eq:lkh-soc}) inherently relies on the approximation that SOC is sufficiently small, such that the zeroth-order wavefunctions form a good basis set for the $H$ matrix. Luttinger and Kohn explicitly warned that this approximation would be less accurate in the presence of very strong SOC. In practice, we observe that the light hole bands predicted by the Luttinger-Kohn model with SOC for tellurides meander around the first-principles dispersion and predict larger light hole effective masses compared to the first-principles results (see the blue curve in Fig.~\ref{fig:sm-modifiedlk}, which will be further discussed in Appendix~\hyperref[sec:sm-modifiedlk]{C}).

In the present work, we extract the three Luttinger parameters from the effective mass tensor without SOC using density-functional perturbation theory and evaluate $\Delta_{\rm{SOC}}$ from our ground state calculation with SOC. The angular-averaged effective masses for the heavy hole and light hole bands are computed with order-4 finite differences, using the electronic dispersion calculated by diagonalizing Eq.~(\ref{eq:lkh-soc}).

We note that, by construction, the Luttinger-Kohn Hamiltonian with SOC assumes inversion symmetry (without SOC, time-reversal symmetry can be used to show that the Hamiltonian matrix elements retain the same form). Hence, it should not, in principle, be applied to zincblende materials when SOC is considered. Nevertheless, we found that, despite missing some microscopic features in the very close vicinity of $\Gamma$, which will be described in the following Appendix, the Luttinger-Kohn model with SOC provides a qualitatively reliable description of the electronic bands, which does not invalidate its use for zincblende materials.

\section*{A\lowercase{ppendix} B: D\lowercase{resselhaus} H\lowercase{amiltonian}}\label{sec:dressmodel}
\setcounter{equation}{0}
\renewcommand{\thetable}{B\arabic{table}} 
\renewcommand{\theequation}{B\arabic{equation}} 
A correct treatment of the lack of inversion symmetry in presence of SOC was made by Dresselhaus~\cite{dresselhaus_spin-orbit_1955}. As time-reversal symmetry is preserved, the electronic dispersion without SOC still verifies $\varepsilon_{-\mathbf{k}n} = \varepsilon_{\mathbf{k}n}$, as per Kramer's theorem, but the Bloch functions no longer have to verify $u_{-\mathbf{k}n}(\bm{r}) = u_{\mathbf{k}n}(-\bm{r})$ up to a phase factor. The inclusion of SOC acts as an effective magnetic field which splits the spin-degenerate states at finite crystal momentum, with condition $\varepsilon_{-\mathbf{k}n\uparrow} = \varepsilon_{\mathbf{k}n\downarrow}$. This effect, known as Dresselhaus splitting, creates a well-defined spin texture in reciprocal space, locking the spin orientation to the crystal momentum~\cite{manchon_new_2015}. Dresselhaus  splitting has been recently observed in GaAs and InSb by circular dichroic photoemission~\cite{cho_observation_2021}.
The underlying physical mechanism is similar to the one driving Rashba splitting~\cite{rashba_properties_1984, bihlmayer_focus_2015}, in which an asymmetry in the crystal potential along a preferred axial direction acts as an effective electric field to break inversion symmetry. The resulting spin texture, however, is different from the helical polarization associated with Rashba systems~\cite{manchon_new_2015, bahramy_bulk_2017}. See Ref.~\cite{manchon_new_2015} for more details about the difference between the Rashba and Dresselhaus effects.

For a generic direction in $\mathbf{k}$~space, the eigenvalues for the heavy hole and light hole bands in the Dresselhaus model are given by
\begin{equation}\label{eq:sm-dressenergy}
    E_{\mathbf{k}} = \frac{k^2}{2} + \left(\frac{L+2M}{3}\right)k^2 + y,
\end{equation}
where the bare electron mass $m=1$ in atomic units. The variables $y= y(\mathbf{k}, L, M, N, W)$ are the roots of a fourth-order polynomial,
\begin{widetext}
\begin{equation}\label{eq:ypolynomial}
\begin{split}
    &y^4 -2y^2\left[ \frac{\alpha^2}{9} k^4 + \beta\left(k_x^2k_y^2+k_y^2k_z^2+k_z^2k_x^2\right) + W^2k^2\right] + 4y\,W^2N\left(k_x^2k_y^2+k_y^2k_z^2+k_z^2k_x^2\right)
    + \left[ \frac{\alpha^2}{9} k^4 + \beta\left(k_x^2k_y^2+k_y^2k_z^2+k_z^2k_x^2\right)\right]^2\\ &+ W^4\left[k^4-3\left(k_x^2k_y^2+k_y^2k_z^2+k_z^2k_x^2\right)\right] 
    + 2\frac{\alpha^2}{9} W^2\left(k_x^6+k_y^6+k_z^6\right) - \left[3\frac{\alpha^2}{9} + \frac{2N^2}{3}\right]W^2k^2\left(k_x^2k_y^2+k_y^2k_z^2+k_z^2k_x^2\right)\\
    &+ 21\frac{\alpha^2}{9} W^2k_x^2k_y^2k_z^2 = 0,
    \end{split}
\end{equation}
\end{widetext}
in which we have defined the shorthands
\begin{equation}\label{eq:sm-dresselhaus-alpha2}
\begin{split}
    \alpha^2 &= \left(L-M\right)^2,\\
    \beta &= \frac{N^2-\alpha^2}{3}.
\end{split}
\end{equation}
Equation~(\ref{eq:ypolynomial}) results from the secular determinant of the $\mathbf{k}\cdot\mathbf{p}$ expansion in the fourfold degenerate subspace at the VBM. Note that our $W$ parameter is labeled $C$ in the original work from Dresselhaus~\cite{dresselhaus_spin-orbit_1955}. We renamed it to avoid confusion with the third Luttinger parameter.

\begin{figure}
    \centering
    \includegraphics[width=\columnwidth]{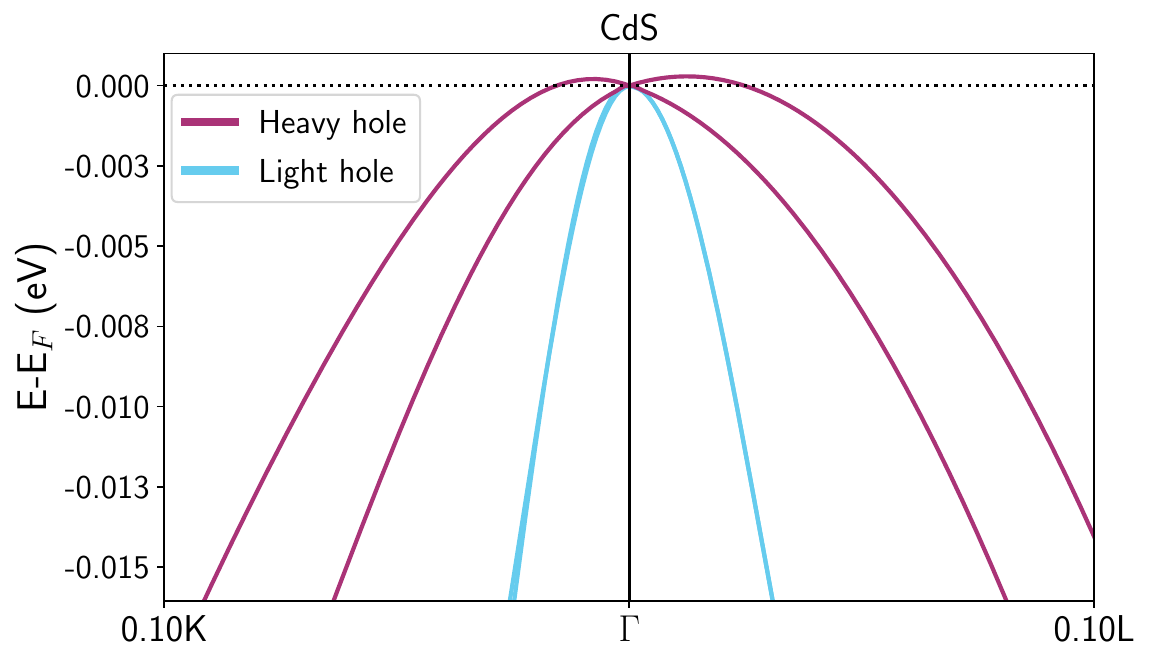}
    \caption{\textbf{Dresselhaus splitting in CdS}. In noncentrosymmetric materials, SOC shifts the band extrema away from the $\Gamma$ point, both in momentum and energy. As a consequence, numerical difficulties arise in the evaluation of the angular-averaged effective masses required by the gFr model with SOC. Zincblende CdS displays the largest effect detected in our dataset, with a momentum offset of $\sim0.01$~\AA$^{-1}$ and an energy offset of $\sim0.5$~meV. As these features occur very close to $\Gamma$ and since $\omega_{\rm{LO}}=34.4$~meV, we retain the physical picture of $\Gamma$-centered parabolic bands inherent to the \fro model (see text).}
    \label{fig:cds-dresselhaus}
\end{figure}

The Dresselhaus model is thus parametrized by four real numbers; three of them, $L$, $M$ and $N$, play a similar role to the Luttinger parameters~\cite{dresselhaus_cyclotron_1955}, while the last one, $W$, captures information about SOC and the breaking of inversion symmetry in the crystal potential~\cite{dresselhaus_spin-orbit_1955}. Rewriting Eq.~(\ref{eq:sm-dressenergy}) as 
\begin{equation}\label{eq:sm-dressenergy-lambda}
  E_{\mathbf{k}} = \lambda\,k^2 + y,  
\end{equation}
hence defining
\begin{equation}\label{eq:sm-dresslambda}
    \lambda = \frac{1}{2} + \left(\frac{L+2M}{3}\right),
\end{equation}
one can conveniently parametrize the Dresselhaus model by $\alpha^2$, $\lambda$, $W$ and $|N|$.
These four parameters can be extracted from the electronic dispersion in the close vicinity of the $\Gamma$ point along two high-symmetry paths: in cartesian direction $\hat{\mathbf{k}}=[1,0,0]$, Eq.~(\ref{eq:sm-dressenergy}) reduces to
\begin{equation}\label{eq:sm-dress100}
    E_{k[1,0,0]} = \lambda\,k^2 \pm \sqrt{\frac{\alpha^2}{9} k^4 + W^2k^2},
\end{equation}
in which both solutions are doubly degenerate, while for~$\hat{\mathbf{k}}=[1,1,1]$, one obtains
\begin{equation}\label{eq:sm-dress111}
    E_{k[1,1,1]} = \lambda\,k^2 + \begin{cases}
    \frac{|N|}{3}k^2,\,\textrm{(twofold),}\\
    -\frac{|N|}{3}k^2 \pm\sqrt{2}W\,k.
    \end{cases}
\end{equation}
Equations~(\ref{eq:sm-dress100}) and~(\ref{eq:sm-dress111}) reveal two peculiarities of the generic electronic dispersion with SOC near the VBM of zincblende crystals which complicate the numerical evaluation of electronic effective masses. These features are sketched in Fig.~2 and~5 of Ref.~\cite{dresselhaus_spin-orbit_1955} and can be observed in Fig.~\ref{fig:cds-dresselhaus} for CdS. On the one hand, the Dresselhaus field shifts the heavy hole band extrema away from the $\Gamma$ point; this can be read from the second case of Eq.~(\ref{eq:sm-dress111}), which contains a linear term in $k$. This momentum offset is, however, much smaller in magnitude than what is typically observed in Rashba materials~\cite{manchon_new_2015}. In the present work, we found it to be of the order of $5\times10^{-3}$\AA$^{-1}$ on average and at most $10^{-2}$\AA$^{-1}$, about half a percent of the length of the reciprocal lattice vectors or less in all cases. The energy difference between the true VBM and the $\Gamma$ point was at most 0.5~meV. Such a small effect will be barely visible on the band structure but will nevertheless render the finite central differences around $\Gamma$ numerically unreliable, regardless of the fact that the momentum offset does not contribute to the gFr model (see Eq.(\ref{eq:zpr-gfr-dresselhaus-exact})).

On the other hand, if we take the higher energy solution of Eq.~(\ref{eq:sm-dress100}) to be the heavy hole bands and the lower energy solutions to be the light hole bands, their slopes will be discontinuous across $\Gamma$. As a consequence, the effective mass computed from finite differences for these bands will diverge as the \kpoint sampling 
gets denser. A similar feature occurs in for the light hole bands in the $\hat{\mathbf{k}}=[1,1,0]$ cartesian direction (see the upper panel of Fig.~2 of Ref.~\cite{dresselhaus_spin-orbit_1955}), as well as for generic $\hat{\mathbf{k}}$.

Nevertheless, in line with Dresselhaus~\cite{dresselhaus_cyclotron_1955} and as addressed in Sec.~\ref{sec:gfr-with-dresselhaus}, we argue that these features have little to no impact on our current application since 
the energy offset between $\Gamma$ and the true extrema is about two orders of magnitude smaller than the LO frequency. 

With these precautions in mind, we obtain the model parameters from the calculated dispersion along the $[100]$ and $[111]$ directions through the following relations:
\begin{enumerate}
    \item The difference between the dispersion of the two nondegenerate bands given by Eq.~(\ref{eq:sm-dress111}) is linear in $k$, with a slope proportional to $W$.
    \item The average value of these two bands is quadratic in $k$. The difference between this average and the dispersion of the degenerate bands in the same direction is also quadratic in $k$, with a curvature proportional to $|N|$.
    \item Averaging the difference between the non-degenerate bands of Eq.(\ref{eq:sm-dress111}) with the degenerate bands in the same direction results in a quadratic function in $k$, with curvature $\lambda$.
    \item The difference between the dispersion of the two pairs of degenerate states in Eq.~(\ref{eq:sm-dress100}) allows extracting the $\alpha^2$ factor.
\end{enumerate}
To circumvent any numerical instabilities arising from the band peculiarities associated with Dresselhaus splitting, we computed the effective masses by fitting quadratic functions to the Dresselhaus model dispersion rather than relying on finite differences. The different parameters used for the Dresselhaus model are tabulated in Table~S8 of the Supplemental Material~\cite{supplemental_zprsoc}. 
Lastly, we emphasize that our Dresselhaus parameters should be perceived as fitting parameters rather than physical parameters. While they retain some physical essence from their original formulation (see Eq.~(47) of Ref.~\cite{dresselhaus_cyclotron_1955}), we did not evaluate them through the \mbox{$\mathbf{k}\cdot\mathbf{p}$} framework but rather fitted them to the first-principles dispersion with SOC.

\section*{A\lowercase{ppendix} C: F\lowercase{itted} L\lowercase{uttinger parameters from} D\lowercase{resselhaus} H\lowercase{amiltonian}}\label{sec:sm-modifiedlk}
\setcounter{equation}{0}
\renewcommand{\thetable}{C\arabic{table}} 
\renewcommand{\theequation}{C\arabic{equation}} 
\begin{figure}
    \centering
    \includegraphics[width=\columnwidth]{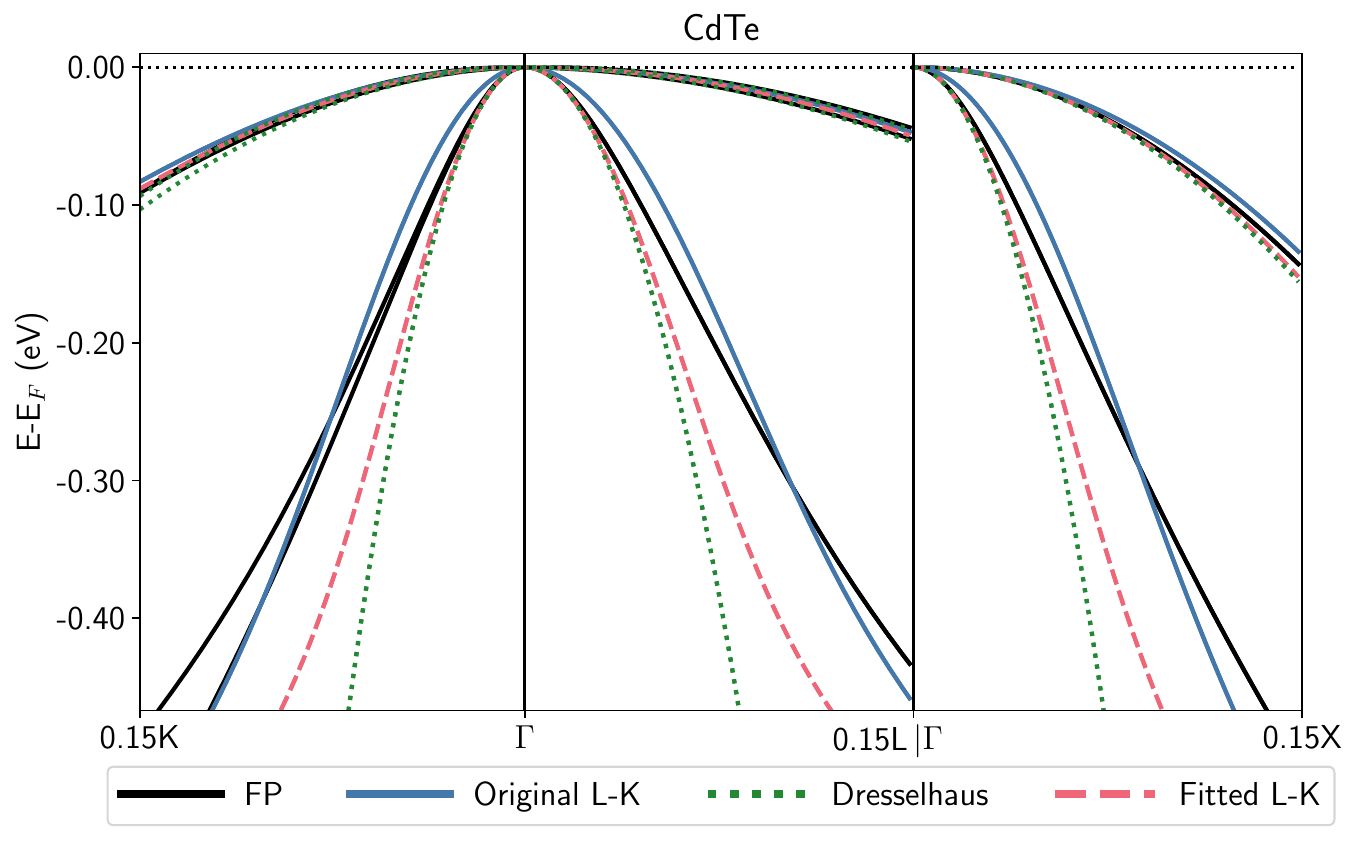}
    \caption{\textbf{Electronic dispersion from the Luttinger-Kohn and Dresselhaus models for CdTe}, computed with the original Luttinger parameters ($(A, B, C)$, solid blue lines), the fitted Dresselhaus model (green dotted lines) and the fitted Luttinger parameters extracted from the Dresselhaus model (($\widetilde{A}, \widetilde{B}, \widetilde{C}$), dashed red lines) as described in Appendix.~\hyperref[sec:sm-modifiedlk]{C}. The first-principles dispersion is shown in black. While the original model qualitatively captures the correct band warping, it overestimates the light holes' effective masses. The modified parameters reproduce the first-principles curvature of the light holes with greater accuracy and produce the same ZPR$^{\rm{gFr}}$ as the Dresselhaus model, without the numerical difficulties.}
    \label{fig:sm-modifiedlk}
\end{figure}
As discussed in Appendix~\hyperref[sec:lkmodel]{A}, the Luttinger-Kohn Hamiltonian becomes less accurate in predicting the light holes effective masses as SOC increases. As a consequence, it yields a smaller decrease of the ZPR$_{\rm{v}}$ (i.e., a larger absolute value of ZPR$_{\rm{v}}$(SOC)) compared to the first-principles result, as emphasized in Sec.~\ref{sec:res-gfr} of the main text, hence the overestimation of the ratios presented in Fig.~\ref{fig:gfr_vs_fp_vb}. 
Nevertheless, it lacks all the numerical difficulties arising from Dresselhaus splitting. In the spirit of the original \fro model, we aim for reliable effective masses with respect to the first-principles dispersion, while remaining in the physical picture of $\Gamma$-centered parabolic bands. The Luttinger-Kohn Hamiltonian thus provides a more efficient framework to evaluate the angular-averaged effective masses required by our gFr model. In this context, we propose a very simple solution: to extract fitted Luttinger parameters from the Dresselhaus model that yield more predictive light hole effective masses compared to the first-principles results in the presence of SOC.

First, we note that the momentum offset in the heavy hole bands of zincblende materials prevents us from directly fitting the Luttinger-Kohn model on the first-principles bands. The Dresselhaus model, in contrast, parametrizes this offset correctly through the $W$ parameter. From our fitted values of $\lambda$ and $\alpha^2$, we can extract fitted values for $L$ and $M$, through Eqs.~(\ref{eq:sm-dresselhaus-alpha2}) and~(\ref{eq:sm-dressenergy-lambda}). To do so, one nevertheless requires prior knowledge of the relative strength of $L$ and $M$, as $\alpha=|L-M|$. 

One can then go back to the original works of Luttinger-Kohn and Dresselhaus (see Eq.~(\MakeUppercase{\romannumeral 5}.10) of Ref.~\cite{luttinger_motion_1955} and Eq.~(47) of Ref.~\cite{dresselhaus_cyclotron_1955}) and note that the definitions of the $(A, B)$ and $(L, M)$ parameters differ only by the bare electron mass term, $\hbar^2/2m_e$, which is equal to $1/2$ in atomic units. The definitions of the $C$ and $N$ parameters are equivalent. As $L$ maps to $A$ and $M$ maps to $B$ up to a constant shift, we can first deduce that ${\rm sgn}(L-M)={\rm sgn}(A-B)$. Then, we can use this mapping to extract fitted Luttinger parameters that are optimized to reproduce the first-principles dispersion with SOC along the $[100]$ and $[111]$ cartesian directions in reciprocal space using the Dresselhaus model. Note that the only purpose of these parameters is to provide an accurate description of the first-principles bands in the close vicinity of the $\Gamma$ point. Hence, the predicted dispersion will rapidly become less accurate than the original Luttinger parameters for larger $|\mathbf{k}|$ values, as well as for the split-off bands.

Figure~\ref{fig:sm-modifiedlk} compares the first-principles dispersion (black) to the Luttinger-Kohn model dispersion obtained with the original set of Luttinger parameters (blue) and the fitted parameters (dashed red) for CdTe. The dispersion obtained from the Dresselhaus model (green dotted lines) is also shown for comparison. 
While the original Luttinger-Kohn model closely follows the band warping for a broader portion of the Brillouin zone, the fitted parameters reproduce the light holes' curvature around $\Gamma$ more accurately, similar to the one predicted by the Dresselhaus model. In turn, they predict a SOC-induced decrease of the ZPR$^{\rm{gFr}}$ in better agreement with the first-principles result, as shown by the open circle markers in Fig~\ref{fig:gfr_vs_fp_vb}. The numerical values of the fitted Luttinger parameters are tabulated in Table~S8 of the Supplemental Material~\cite{supplemental_zprsoc}.

\section*{A\lowercase{ppendix} D: Z\lowercase{ero-point expansion of the lattice}}\label{sec:sm-zple}
\setcounter{equation}{0}
\renewcommand{\thetable}{D\arabic{table}} 
\renewcommand{\theequation}{D\arabic{equation}} 
The zero-point motion of the ions affects the band gap energy in two distinct ways: on the one hand, the electrons interact with the ionic motion through EPI, which is computed at the static lattice geometry, $a^0$  (typically, the lattice parameter obtained from a standard DFT relaxation). On the other hand, the ionic vibrations contribute to the total Helmholtz free energy of the crystal, yielding a small variation of the lattice parameter, called zero-point lattice expansion. In addition to the EPI contribution discussed in Sec.~\ref{sec:theo-ahc}, the zero-point lattice expansion induces a $T=0$~K modification of the band gap energy.

We work within the quasiharmonic approximation, in which the main contribution to the temperature dependence of the phonon frequencies is expressed as their variation with respect to the lattice parameter, which causes the crystal to expand~\cite{grimvall_thermophysical_1986}. The Helmholtz free energy then reads:
\begin{equation}
\begin{split}
    F^{\rm{tot}}(V,T) \simeq F^{\rm{e}}(V)+F^{\rm{vib}}(V,T),\\
    = E_{\rm{stat}}^{\rm{e}}(V) - k_B T \ln Z^{\rm{ph}}(V, T),\\
    \end{split}
\end{equation}
in which we approximate that the electronic and vibrational degrees of freedom can be separated. Since we are dealing with semiconductors and insulators, the entropic contribution of the electrons to the free energy is neglected. $E_{\rm{stat}}^{\rm{e}}(V)$ is the Born-Oppenheimer energy obtained from DFT and $k_B$ is the Boltzmann constant. 

At $T=0$, the contribution of the phonon partition function 
is simply the zero-point energy, such that
\begin{equation}\label{eq:helmholtz}
    F^{\rm{tot}}(V,T=0) = E_{\rm{stat}}^{\rm{e}}(V) + \sum\limits\qv \frac{\wqv(V)}{2}
\end{equation}
For cubic materials, the lattice parameter including zero-point motion, \mbox{$a(T=0)$}, is obtained from the \mbox{$V(T=0)$} volume by minimizing the $T=0$~K Helmholtz free energy. The zero-point lattice expansion is simply the difference between the static and dynamical lattice parameters:
\begin{equation}
    \Delta a(T=0) = a(T=0) - a^0
\end{equation}

Once $a(T=0)$ is known, we approximate the zero-point lattice expansion contribution to the band gap ZPR, labeled ZPR$_{\rm{g}}^{\rm{ZPLE}}$, by computing the difference between the DFT band gap energy, $E_{\rm{g}}^{\rm{DFT}}$, evaluated at the static and \mbox{$T=0$~K} geometries:
\begin{equation}
    \textrm{ZPR}_{\rm{g}}^{\rm{ZPLE}} \simeq E_{\rm{g}}^{\rm{DFT}}(a(T=0)) - E_{\rm{g}}^{\rm{DFT}}(a^0).
\end{equation}

For all materials, we computed the zero-point lattice expansion using an \mbox{$8\times8\times8$} \mbox{$\Gamma$-centered} \qpoint grid. As the construction of the Helmholtz free energy only requires the vibrational spectrum of the crystal, it converges faster with respect to the \qpoint sampling compared to EPI. The values of ZPR$_{\rm{g}}^{\rm{ZPLE}}$ used to obtain Fig.~\ref{fig:experimentlin} can be found in Table~S5 of the Supplemental Material~\cite{supplemental_zprsoc}, along with experimental values of ZPR$_{\rm{g}}$. Note that we did not attempt to evaluate the effect of SOC on ZPR$_{\rm{g}}^{\rm{ZPLE}}$; the same value was used to correct ZPR$_{\rm{g}}^{\rm{EPI}}$ both with and without SOC. As the zero-point lattice expansion depends mostly on the phonon frequencies (see Eq.~(\ref{eq:helmholtz})), which are not significantly affected by SOC for the cubic materials investigated (see Table~S7 of the Supplemental Material~\cite{supplemental_zprsoc}), we expect this approximation to be fairly accurate for the purpose of this work. For a more thorough discussion of the zero-point lattice expansion and its effect on the band gap ZPR, see Ref.~\cite{brousseau-couture_zero-point_2022}.

\section*{SUPPLEMENTAL MATERIAL}

\setcounter{section}{0}
\setcounter{table}{0}
\setcounter{equation}{0}
\setcounter{figure}{0}
\renewcommand{\thesection}{S\arabic{section}} 
\renewcommand{\thetable}{S\arabic{table}} 
\renewcommand{\theequation}{S\arabic{equation}} 
\renewcommand{\thefigure}{S\arabic{figure}}
\section{Low SOC ZPR of \texorpdfstring{C\MakeLowercase{d}T\MakeLowercase{e}}{CdTe}}\label{sec:SMlowsoc}
\begin{figure*}
    \centering
    \includegraphics[width=0.325\textwidth]{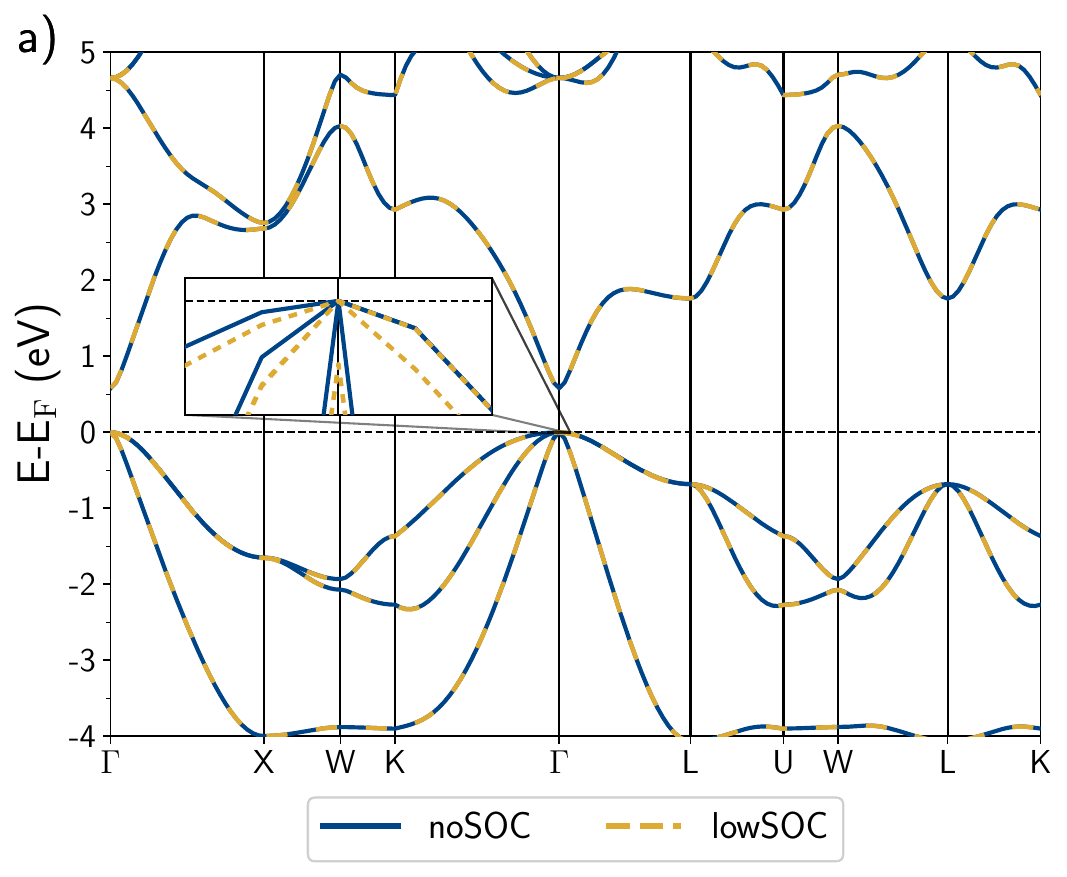}
    \includegraphics[width=0.325\textwidth]{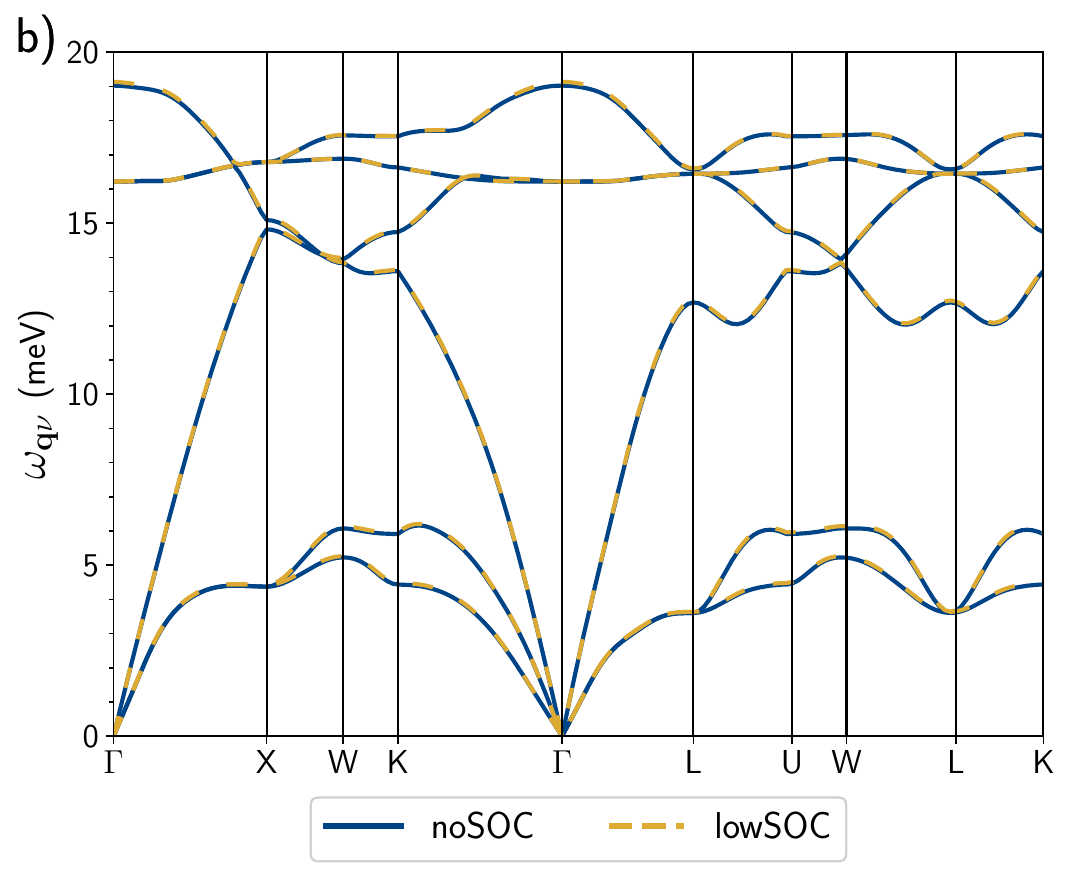}
    \includegraphics[width=0.325\textwidth]{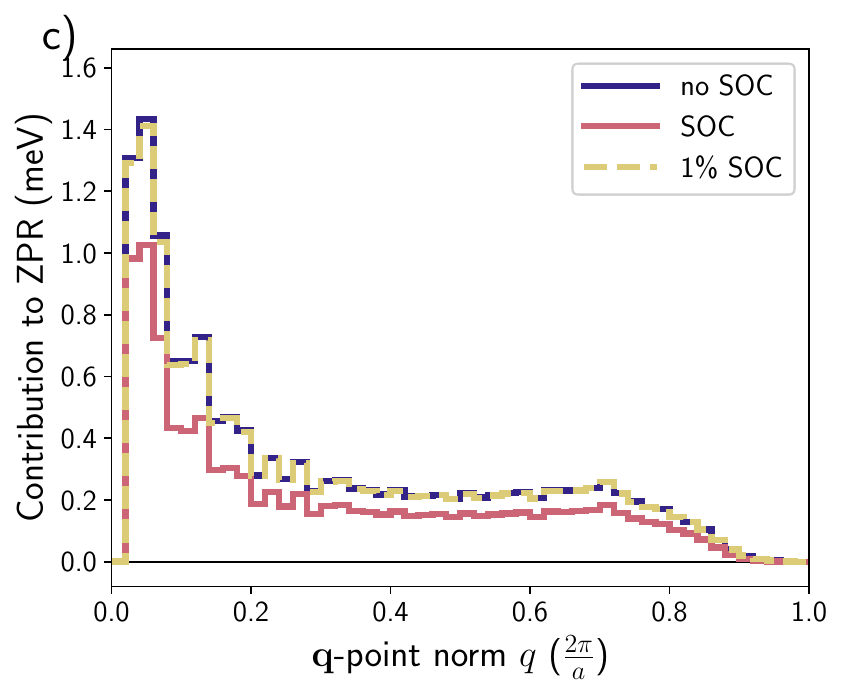}
    \caption{\textbf{Band structure, phonon dispersion and ZPR decomposition of zincblende CdTe,} without SOC (solid blue or indigo lines) and with SOC artificially reduced to 1\% (yellow dashed lines). a)~Both band structures are virtually identical, except in the close vicinity of the $\Gamma$ point (see inset), where a small split-off energy of $\Delta_{\rm{SOC}}^{\rm{low}}\simeq 8$~meV$\approx0.01\Delta_{\rm{SOC}}$ can be observed. b)~The low SOC data adequately reproduces the phonon dispersion without SOC. The maximal difference of 0.1 meV occurs for the LO mode at $\bm{q}=\Gamma$. c)~The contribution of the different phonon modes with wavevector norm $|q|$ to the ZPR$_{\rm{v}}$ computed with low SOC (yellow dashes) reproduces the histogram without SOC (dark indigo) shown in Fig.~6 of the main text. The ZPR decomposition with SOC (light red) is also shown for comparison.}
    \label{fig:cdte_lowsoc}
\end{figure*}

In this Section, we present complementary information regarding the 1\% SOC calculation used to validate our heuristic interpretation of the origin of the SOC-induced ZPR decrease (Sect.~\MakeUppercase{\romannumeral 4}.A.3 of the main text). Fig.~\ref{fig:cdte_lowsoc}a) and~b) respectively show the electronic and phononic dispersions of CdTe without SOC (blue) and with 1\% SOC (yellow dashes). The inset of Fig.~\ref{fig:cdte_lowsoc}a) shows a small residual split-off energy of \mbox{$\sim$8~meV}, namely 1\% of the full $\Delta_{\rm{SOC}}$ for CdTe. Both band structures are virtually identical. The phonon dispersions also show excellent agreement. The largest energy difference between equivalent phonon modes in our denser \mbox{$48\times48\times$48} \qpoint grid is 0.1~meV,  occurring for the LO mode at \mbox{$\mathbf{q}=\Gamma$}. The average energy difference per mode is at most 0.01~meV. Fig.~\ref{fig:cdte_lowsoc}c) confirms that the histogram decomposition of the ZPR$_{\rm{v}}$ computed with 1\% SOC (yellow dashes) is almost identical to the reference histogram without SOC (dark indigo) shown in Fig.~6 of the main text. A negligible discrepancy remains in the small $q$ bins, most likely due to the inevitable non-zero split-off energy in the low SOC framework. The histogram with SOC (light red) is shown for comparison purposes.

\begin{figure*}
    \centering
    \includegraphics[width=0.325\textwidth]{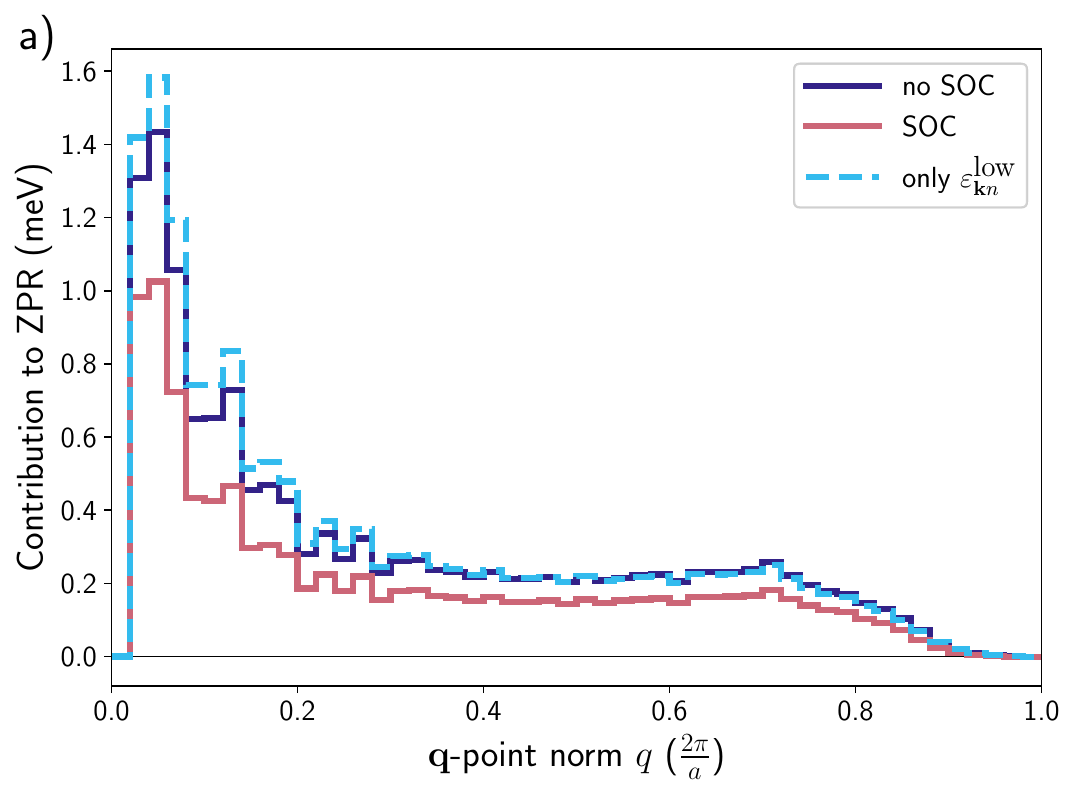}
    \includegraphics[width=0.325\textwidth]{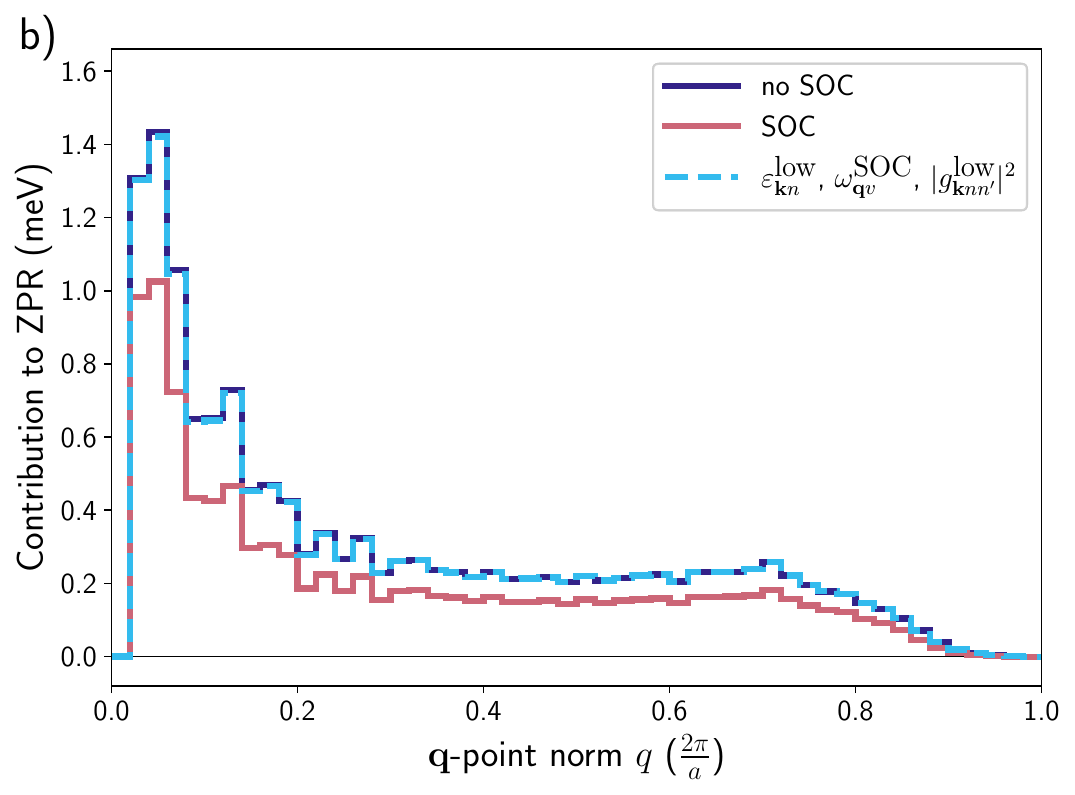}
    \includegraphics[width=0.325\textwidth]{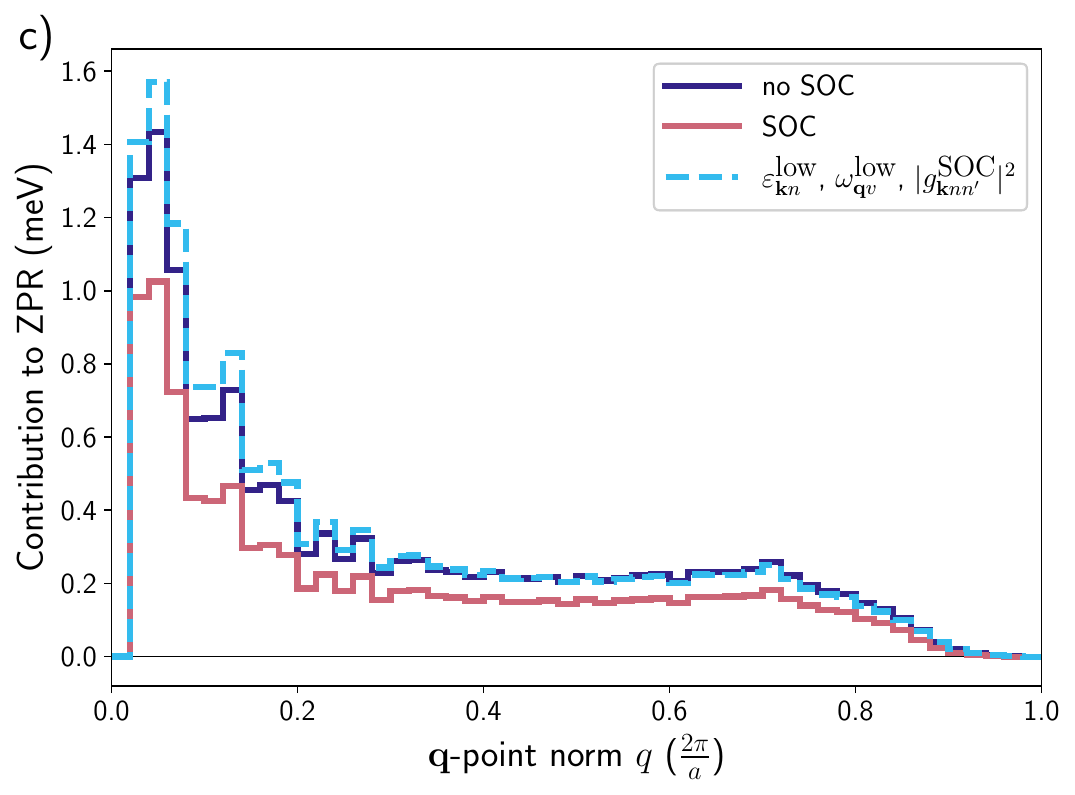}\\
    \includegraphics[width=0.325\textwidth]{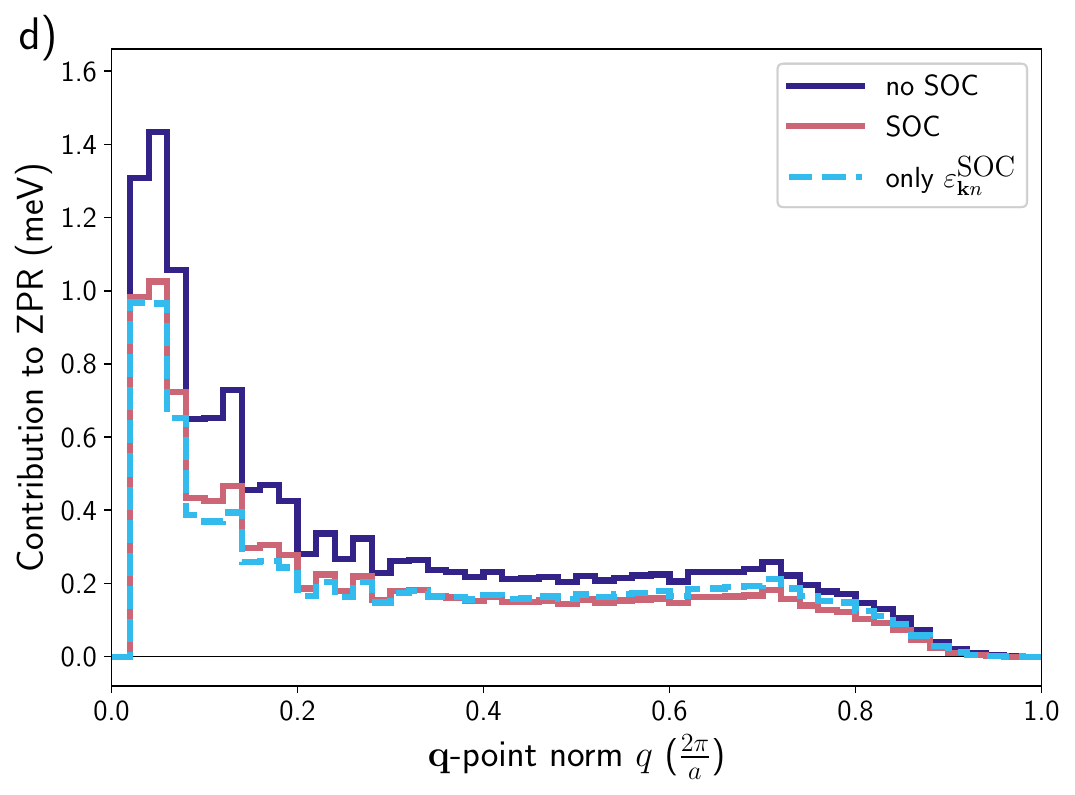}
    \includegraphics[width=0.325\textwidth]{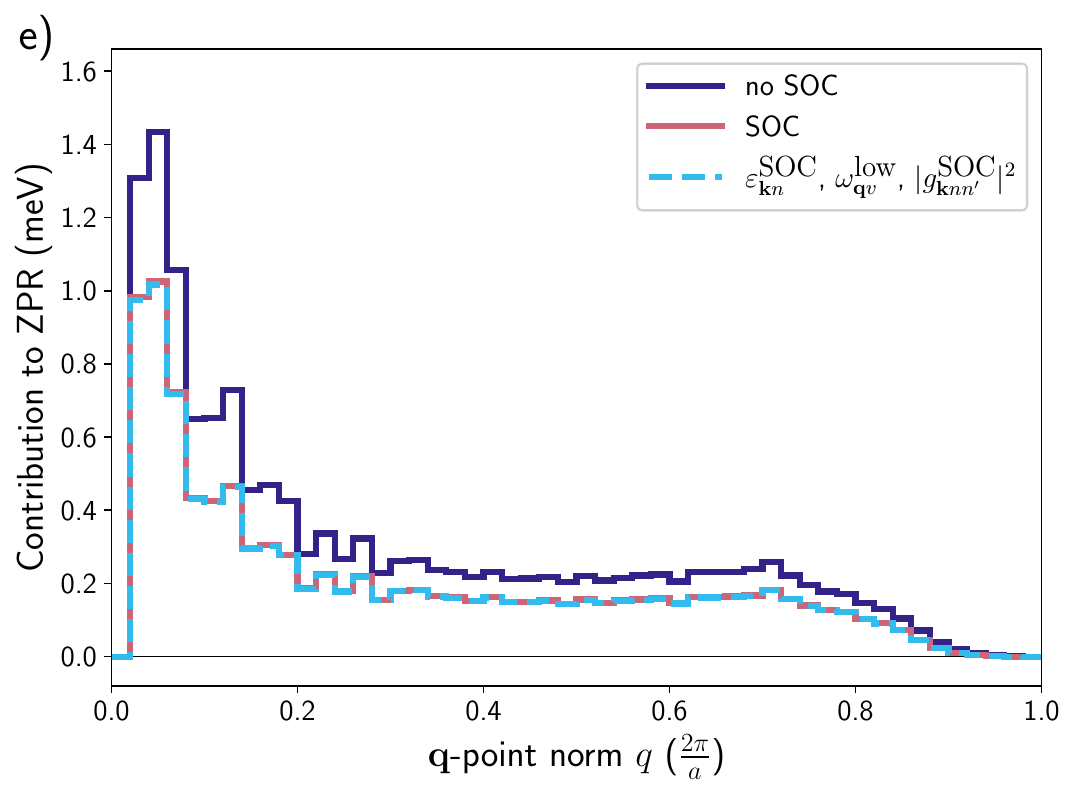}
    \includegraphics[width=0.325\textwidth]{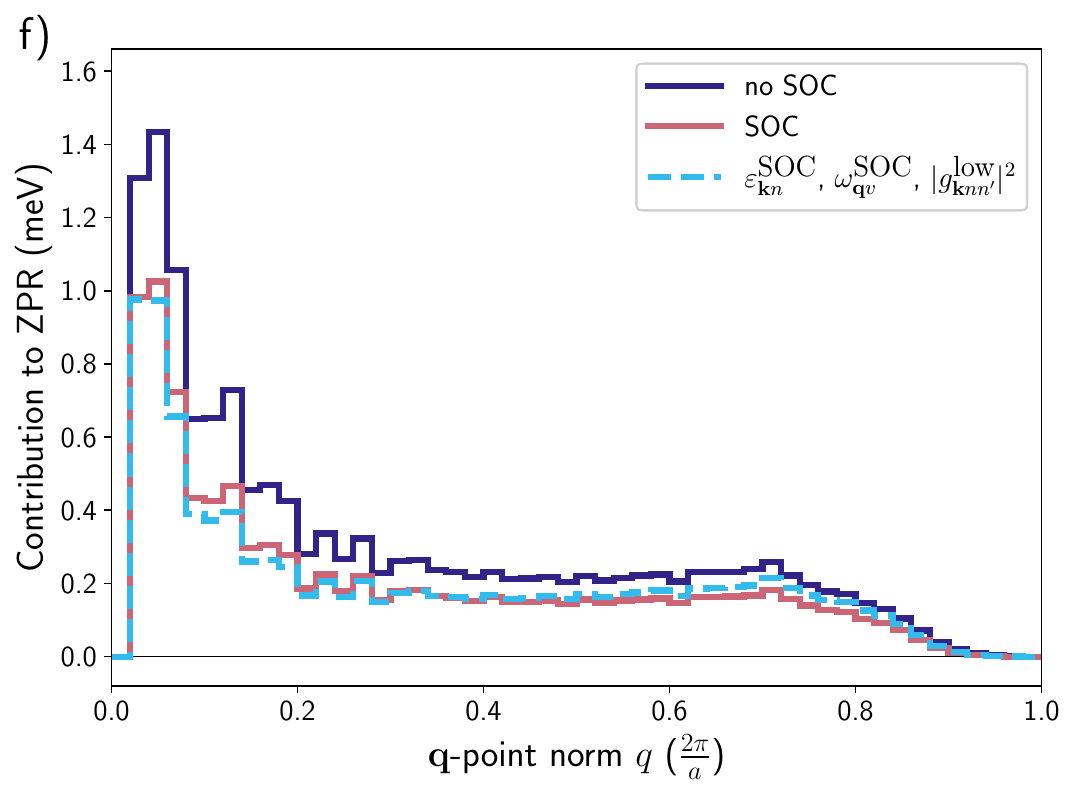}\\
    \caption{\textbf{Contribution to the total ZPR$_{\rm{v}}$ of CdTe for different \qpoint norms,}  for a \mbox{$48\times48\times48$} $\Gamma$-centered \qpoint grid. In extension to Fig.~6 of the main text, all possible combinations of $\varepsilon\kn^0$, $\omega\qv$ and $|g\subkn{k}{n n'}|^2$ computed with either full SOC (\enquote{SOC} superscripts) or artificially reduced 1\% SOC (\enquote{low} superscripts) are displayed with cyan dashes in the different subfigures (see subfigure legends for details). The histograms with SOC (light red) and without SOC (dark indigo) serve as references. The strength of SOC used to evaluate the electronic eigenvalues is sufficient to group the different histograms into two classes: one with 1\% SOC (upper panels a), b) and c)), and one with full SOC (lower panels, d), e) and f)). The small variations with respect to the reference lines  can most likely be attributed to small discrepancies in EPI matrix elements.}
    \label{fig:sm-histograms_alldata}
\end{figure*}

Table~\ref{tab:cdte_lowsoc_sm} extends Table~\MakeUppercase{\romannumeral 1} of the main text to all six possible combinations of the self-energy ingredients: $\varepsilon\kn$, $\omega\qv$ and $|g\subkn{k}{n n'}|^2$, are computed with either full SOC (\enquote{SOC} superscripts) or artificially reduced 1\% SOC (\enquote{low} superscripts). The corresponding histograms are shown in Fig.~\ref{fig:sm-histograms_alldata}. Note that the data labelled \enquote{only $\varepsilon\kn^{\rm{low}}$} and \enquote{only $\varepsilon\kn^{\rm{SOC}}$} is respectively equivalent to \enquote{$\varepsilon\kn^{\rm{low}}$, $\omega\qv^{\rm{SOC}}$, $|g\subkn{k}{n n'}^{\rm{SOC}}|^2$} and \enquote{$\varepsilon\kn^{\rm{SOC}}, \omega\qv^{\rm{low}}$, $|g\subkn{k}{n n'}^{\rm{low}}|^2$}.

As discussed in the main text, the different datasets can be grouped according to whether SOC is included or not in the bare electronic eigenvalues (see the caption of Fig.~\ref{fig:sm-histograms_alldata} and notice the difference between the upper and lower panels). When either including or excluding SOC only in $\varepsilon\kn^0$ (resp. Fig.~\ref{fig:sm-histograms_alldata}a) and~d)), some small discrepancies yet remain when compared to the reference lines (SOC (dark indigo) and noSOC (light red)), which can be attributed to 
small variations of the EPI matrix elements. Including or excluding SOC only in the phonon frequencies has virtually no effect on the final results for this material, as their corresponding histograms (resp. Fig.~\ref{fig:sm-histograms_alldata}b) and~e)) reproduce almost perfectly the corresponding reference lines.

\begin{table}[t]
    \centering
    \caption{\textbf{VBM, CBM and band gap ZPR for different combinations of SOC strength.} In extension to Table~\MakeUppercase{\romannumeral 1} of the main text, all possible combinations of $\varepsilon\kn$, $\omega\qv$ and $|g\subkn{k}{n n'}|^2$ computed with either full SOC (\enquote{SOC} superscripts) or artificially reduced 1\% SOC (\enquote{low} superscripts) are displayed.}\label{tab:cdte_lowsoc_sm}
\sisetup{
table-format = 3.2 ,
table-number-alignment = center ,
}
    \setlength\extrarowheight{2pt}
    \begin{tabularx}{\columnwidth}{p{30mm}
    *{3}{Y}
    }
    \hline\hline
    \multirow{2}{*}{Combination} & \multicolumn{3}{c}{ZPR (meV)}\\
    \cmidrule(lr){2-4}
    & VBM & CBM & ZPR$_\textrm{g}$ \\
        \hline
    noSOC & 16.4 & -0.4 & -16.8\\
    SOC & 11.4 & ~0.4 & -11.0\\
    1\% SOC & 16.2 & -0.5 & -16.7\\
    only $\varepsilon\kn^{\textrm{low}}$ & 17.4 & -0.8 & -18.3\\
    only $\varepsilon\kn^{\textrm{SOC}}$ & 11.1 & -0.8 & -11.9\\
    $\varepsilon\kn^{\textrm{low}}, \omega\qv^{\textrm{SOC}}$, $|g\subkn{k}{n n'}^{\textrm{low}}|^2$ & 16.3 & -0.6 & -16.9\\
    $\varepsilon\kn^{\textrm{SOC}}, \omega\qv^{\textrm{low}}$, $|g\subkn{k}{n n'}^{\textrm{SOC}}|^2$ & 11.3 & 0.4 & -10.9\\
    $\varepsilon\kn^{\textrm{low}}, \omega\qv^{\textrm{low}}$, $|g\subkn{k}{n n'}^{\textrm{SOC}}|^2$ & 17.3 & -0.8 & -18.1\\
    $\varepsilon\kn^{\textrm{SOC}}, \omega\qv^{\textrm{SOC}}$, $|g\subkn{k}{n n'}^{\textrm{low}}|^2$ & 11.1 & -0.9 & -12.0\\
    \hline\hline
    \end{tabularx}
\end{table}
\section{Impact of SOC on the real and imaginary parts of the electron-phonon self-energy}\label{sec:realimag}
In the current work, the effect of SOC we observe on the ZPR of the VBM, hence on the real part of the electron-phonon self-energy, is smaller than the relative impact on the hole mobility reported in the literature for several semiconductors~\cite{ma_first-principles_2018, ponce_towards_2018, ponce_first-principles_2021}. As discussed in Sec.~\mbox{\MakeUppercase{\romannumeral 4}.A.1} of the main text, the imaginary part of the self-energy is proportional to the inverse relaxation time for a given electronic state $\ket{\mathbf{k}n}$. For its parts, the mobility depends on the relaxation time of all the electronic states contributing to the scattering channels~\cite{giustino_electron-phonon_2017}. Hence, when including SOC in the calculations, the imaginary part of the self-energy of a given electronic state can be expected to decrease by a similar ratio as the mobility when SOC is neglected. When comparing those quantities, one has nevertheless to keep in mind that, in contrast with the self-energy, the mobility is a global quantity which is integrated on the Brillouin zone.

\begin{figure*}
    \centering
    \includegraphics[width=0.49\linewidth]{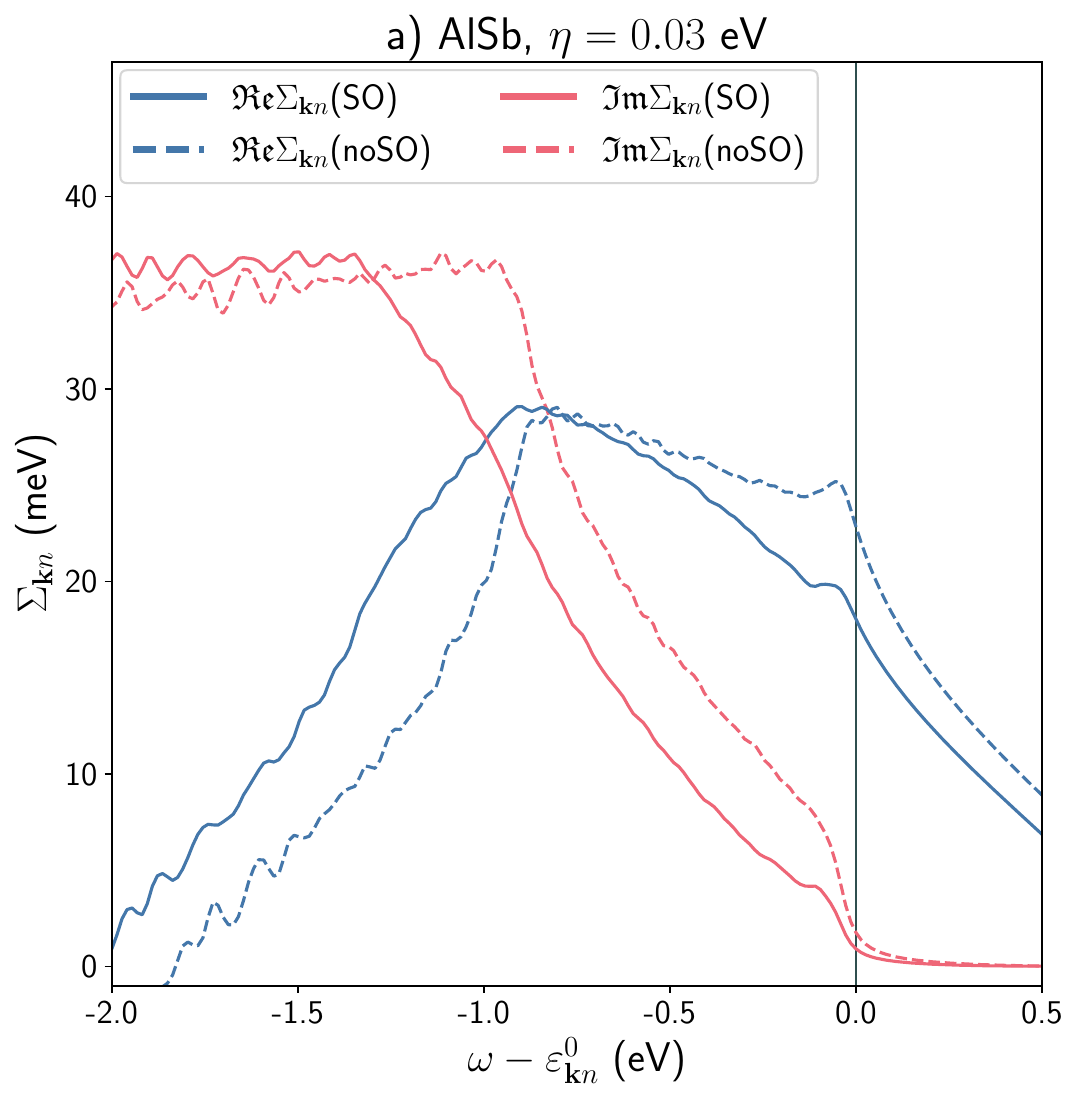}
     \includegraphics[width=0.49\linewidth]{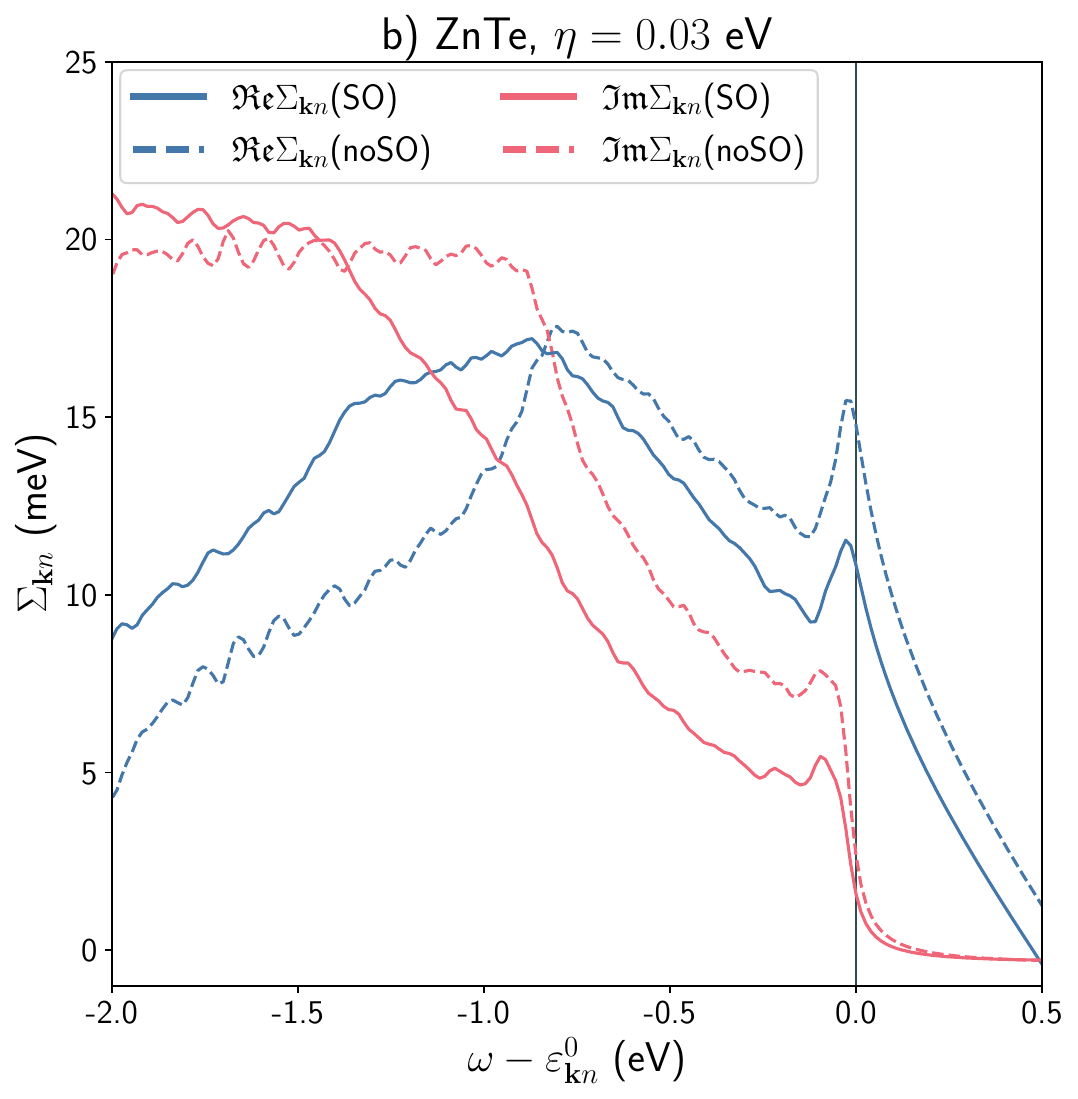}\\
    \includegraphics[width=0.49\linewidth]{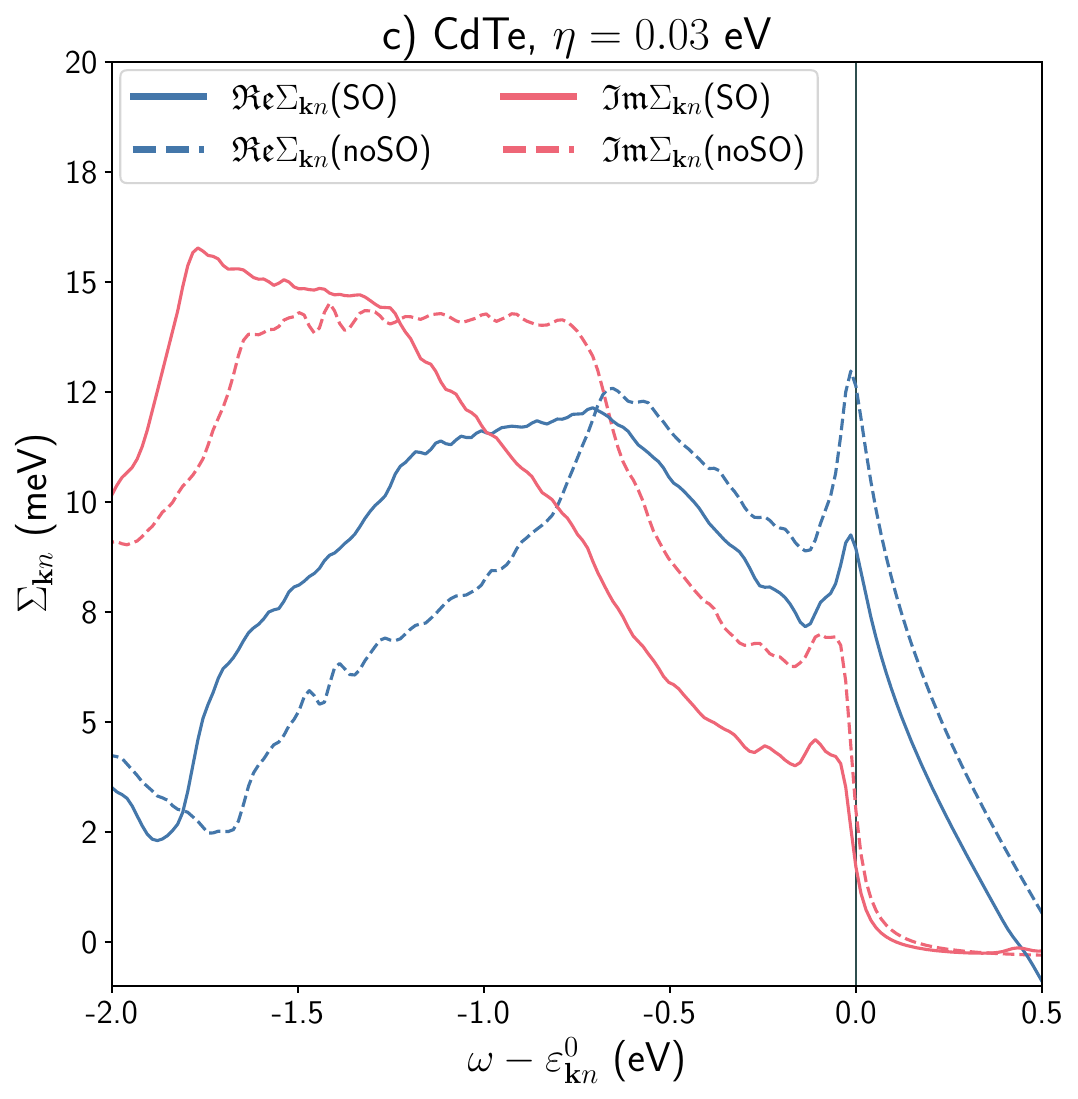}
      \includegraphics[width=0.49\linewidth]{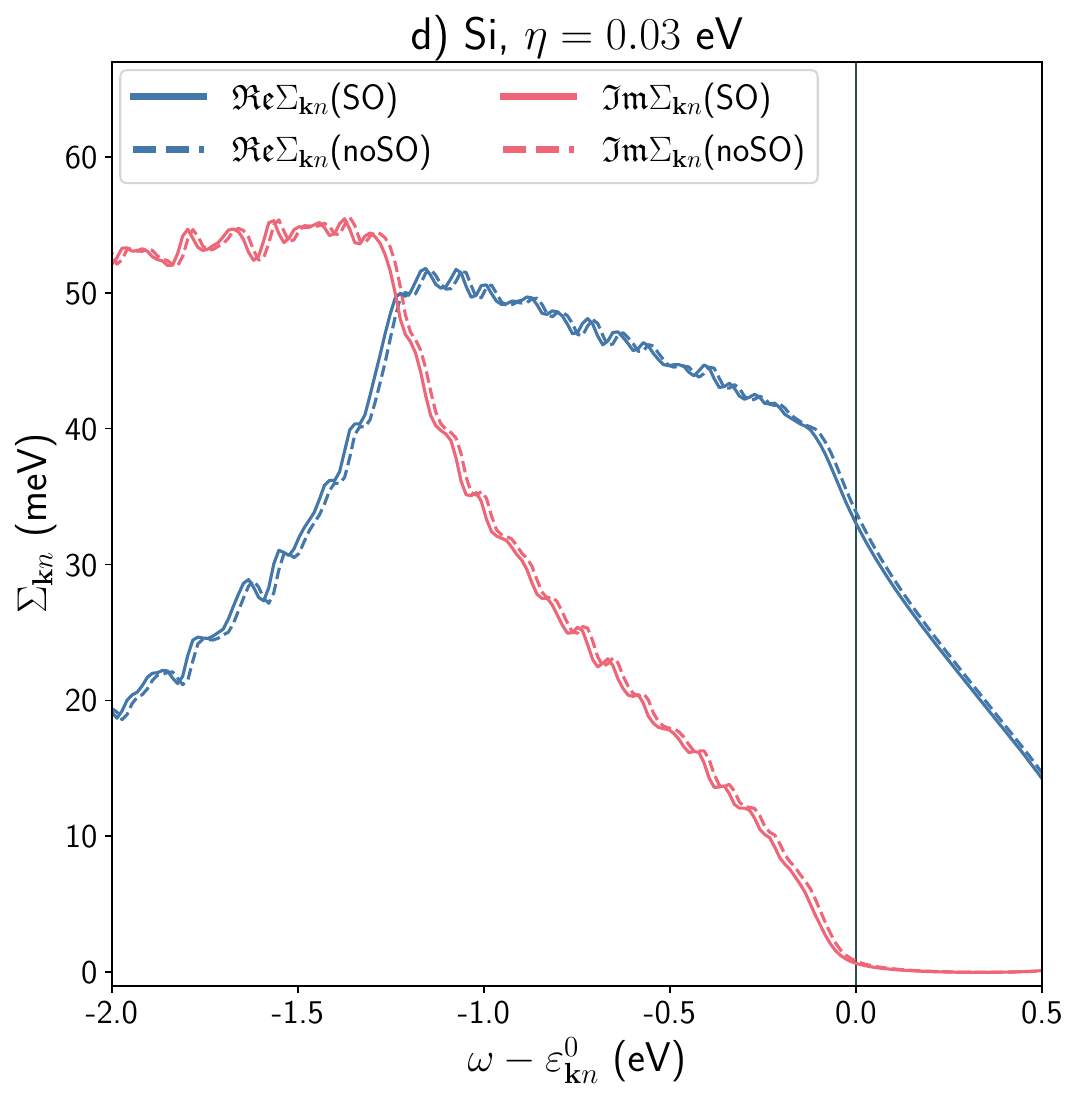}
    \caption{\textbf{Frequency-dependent electron-phonon self-energy for AlSb, ZnTe, CdTe and Si} for a \mbox{$48\times48\times48$} \qpoint grid. Solid lines include SOC while dashed lines do not. 
    For a significant frequency range below the bare VBM energy, the inclusion of SOC has a larger relative impact on the imaginary part (red), which is linked to the hole mobility, than on the real part (blue), which governs the ZPR. As the magnitude of $\mathfrak{Im}\Sigma\kn$ is much smaller than $\mathfrak{Re}\Sigma\kn$, the relative impact on the imaginary part is significant despite the absolute variations being small (see text).}
    \label{fig:selfenergy}
\end{figure*}

Figure~\ref{fig:selfenergy} shows the real (blue) and imaginary (red) parts of the frequency-dependent electron-phonon self-energy (i.e. retaining the frequency dependence in Eq.~(5) of the main text) for AlSb~(a), ZnTe~(b), CdTe~(c) and Si~(d), computed with a \mbox{$48\times48\times48$} \qpoint grid. Solid lines include SOC while dashed lines do not. Note that for this Figure, the imaginary parameter $\eta$ was raised to $0.03$~eV, in order for the self-energy to be a smooth function of the frequency. For all four materials, the introduction of SOC has a larger absolute effect on the real part of the self-energy close to the bare eigenvalue. Nevertheless, as the magnitude of the imaginary part of the self-energy is much smaller than its real part, we observe that the SOC-induced \textit{relative} decrease of the self-energy is more significant for the imaginary part than for the real part for a substantial frequency range below the bare VBM energy, which could lead to a stronger relative effect on transport properties. For this region, we find that the relative decrease of the imaginary part of the self-energy in the vicinity of the bare VBM eigenvalue is of the order of 45\% for AlSb, 35\% for ZnTe, 40\% for CdTe and 20\% for Si, which is larger than the ratios reported in Fig.~2a) of the main text for the ZPR$_{\text{v}}$ (respectively 21\%, 27\%, 30\% and 3\%). Our results thus agree with the trends reported in the literature for the hole mobility. Further investigation will be required to fully understand how the SOC-induced decrease of the band gap ZPR and increase of the hole mobility should correlate.

\vfill
\section{Numerical tables}\label{sec:sm-datatables}
Table~\ref{tab:Calcdata} regroups the relevant calculation parameters for all materials investigated in the present work, as well as the band gap location and band gap energy with and without SOC. Table~\ref{tab:zpr_fp_extrap} provides the numerical values of the ZPR for the VBM, CBM and band gap with and without SOC computed from first principles, and the ratio R=ZPR(SOC)/ZPR(noSOC) which is shown in Fig.~2 of the main text. Table~\ref{tab:dedt_ratio} compares the band edge and band gap renormalization ratio R computed at $T=300$~K and $T=0$~K, and confirms that the main conclusions of this article, obtained at $T=0$~K, remain valid at finite temperature. Table~\ref{tab:zprte} reports the contribution of the zero-point lattice expansion to the band gap ZPR, which was added to the EPI contribution to produce Fig.~4 of the main text, as well as the experimental values used for comparison. Table~\ref{tab:gfr_compare} contains numerical results for the VBM, CBM and band gap ZPR from the generalized \fro model (gFr, Eq.~(34) of the main text), evaluated using the physical parameters reported in Table~\ref{tab:gfr_model_params}. Table~\ref{tab:sm-modelparams} reports the model parameters used to construct the Luttinger-Kohn and Dresselhaus Hamiltonians, enabling the computation of the angular-averaged effective masses for the VBM of zincblende materials. The fitted Luttinger parameters obtained using the procedure described in Appendix~C of the main text are also reported. Table~\ref{tab:gfr_angular_efmas} finally reports angular averaged effective masses for the different hole bands, as well as their band average. Note that these quantities differ from the angular and band averaged square root effective masses entering the gFr model (reported in Table~\ref{tab:gfr_model_params}).

\begin{table*}[t]
    \centering
    \caption{\textbf{Calculations parameters for twenty materials.} From left to right, material ID on the Materials Project~\cite{MatProj}, cubic lattice parameter ($a$), cut-off energy for the plane-wave basis set (E$_{\rm{cut}}$), band gap edges (VBM-CBM), band gap energy (E$_{\rm{g}}$) with and without SOC, and \qpoint sampling for the BZ ({$N\times N\times N$} $\Gamma$-centered Monkhorst-Pack grids). All materials were computed with {($6\times6\times6)\times4$ shifts} Monkhorst-Pack \kpoint grids. The broadening parameter $\eta$ for the self-energy was set to 0.01~eV for all materials.}
    \label{tab:Calcdata}
     \setlength\extrarowheight{2pt}
    \begin{tabularx}{\textwidth}{p{15mm} p{15mm}     S[table-column-width=1.9cm]
    S[table-column-width=1.6cm, table-number-alignment=center,
    table-format=2.4
    ]
    S[table-column-width=1.5cm] 
    S[table-column-width=2.0cm]
    S[table-column-width=2.0cm, table-format=1.3]
    S[table-column-width=2.0cm, table-format=1.3]
    S[table-column-width=3.0cm]}
    \hline\hline
      {Material}  & {MP-ID} &{Space group} & {$a$ (bohr)} & {E$_{\textrm{cut}}$ (Ha)}  & {Gap} & {E$_{\rm{g}}^{\textrm{noSOC}}$ (eV)} & {E$_{\rm{g}}^{\textrm{SOC}}$ (eV)} & {\qpoint sampling}\\[2pt]
       \hline
       Si & 149 &{Fd$\bar{3}$m [227]} & 10.3348 &25    & {$\Gamma-\Delta$} & 0.607 & 0.591  & {$48\times 48\times 48$}\\
       ZnS & 10695 & {F$\bar{4}3$m [216]}& 10.2859 &40 & {$\Gamma-\Gamma$} & 2.020& 2.000& {$48\times 48\times 48$}\\
    CdS &  2469 &{F$\bar{4}3$m [216]} & 11.2021 & 45 & {$\Gamma-\Gamma$} & 1.040& 1.024&{$48\times 48\times 48$}\\       
     CaS & 1672& {Fm$\bar{3}$m [225]} & 10.7900 & 45& {$\Gamma-X$} & 2.382& 2.350&{$48\times 48\times 48$}\\
     SrS & 1087& {Fm$\bar{3}$m [225]} & 11.4434 & 45& {$\Gamma-X$} & 2.520& 2.484&{$48\times 48\times 48$}\\
     BaS & 1500& {Fm$\bar{3}$m [225]} & 12.1734 & 45& {$\Gamma-X$} & 2.185& 2.140&{$48\times 48\times 48$}\\
    BAs & 10044 & {F$\bar{4}3$m [216]} & 9.0880 &40 & {$\Gamma-\Delta$}  & 1.216& 1.149&{$48\times 48\times 48$}\\  
    Ge & 32 & {Fd$\bar{3}$m [227]} & 10.5220 &40 & {$\Gamma-L$} & 0.333& 0.234&{$48\times 48\times 48$}\\
    GaAs & 2534 &{F$\bar{4}3$m [216]} & 10.8627&40 & {$\Gamma-\Gamma$} & 0.149& 0.044&{$64\times 64\times 64$}\\
    GaAs-exp & 2534 &{F$\bar{4}3$m [216]} & 10.6770 &40 & {$\Gamma-\Gamma$} & 0.531& 0.424&{$64\times 64\times 64$}\\
    ZnSe & 1190 & {F$\bar{4}3$m [216]}& 10.8330 & 40 & {$\Gamma-\Gamma$}& 1.155& 1.031& {$48\times 48\times 48$}\\       
    CdSe & 2691 & {F$\bar{4}3$m [216]}& 11.7114 & 50 & {$\Gamma-\Gamma$} & 0.493& 0.374&{$48\times 48\times 48$}\\        
    SrSe & 2758& {Fm$\bar{3}$m [225]}& 11.8947 &45& {$\Gamma-X$} & 2.252& 2.118&{$48\times 48\times 48$}\\
    CaSe & 1415& {Fm$\bar{3}$m [225]} & 11.2563 &45& {$\Gamma-X$}& 2.082& 1.946& {$48\times 48\times 48$}\\
    BaSe & 1253& {Fm$\bar{3}$m [225]}& 12.6054 &45& {$\Gamma-X$} & 1.979& 1.842&{$48\times 48\times 48$}\\
    AlSb & 2624 &{F$\bar{4}3$m [216]} & 11.7621 &40 &  {$\Gamma-\Delta$} & 1.268& 1.059&{$48\times 48\times 48$}\\    
    ZnTe & 2176 &{F$\bar{4}3$m [216]} & 11.6819 &40 & {$\Gamma-\Gamma$} & 1.071& 0.797&{$48\times 48\times 48$}\\        
    CdTe & 406 & {F$\bar{4}3$m [216]}& 12.5133 &50 & {$\Gamma-\Gamma$} & 0.580& 0.315&{$48\times 48\times 48$}\\        
    BaTe & 1000& {Fm$\bar{3}$m [225]}&13.3727 &45& {$\Gamma-X$} & 1.620& 1.347&{$48\times 48\times 48$}\\
    SrTe & 1958& {Fm$\bar{3}$m [225]}& 12.6950&45& {$\Gamma-X$} & 1.785& 1.506&{$48\times 48\times 48$}\\
    CaTe & 1519& {Fm$\bar{3}$m [225]}& 12.0793&45& {$\Gamma-X$} & 1.563& 1.275&{$48\times 48\times 48$}\\
    \hline\hline
    \end{tabularx}
\end{table*}

\begin{table*}[]
    \centering
    \caption{\textbf{First-principles ZPR} for the VBM, CBM and band gap, split-off energy  $\Delta_{\rm{SOC}}$ and LO phonon frequency $\omega_{\rm{LO}}$ for the twenty materials in our set. R=ZPR(SOC)/ZPR(noSOC) is unitless. The ratio for the CBM of CdTe is not meaningful for our study as its ZPR changes sign upon inclusion of SOC. For GaAs, results obtained with both theoretically relaxed lattice parameter (\enquote{GaAs-theo}) and experimental lattice parameter~\cite{kittel_condensed_2004} (\enquote{GaAs-exp}) are provided. Note also that the values without SOC are slightly different but very close to those reported in Ref.~\cite{miglio_predominance_2020}, as we use a different method to extrapolate the converged value to an infinite number of $\mathbf{q}$-points.
    }\label{tab:zpr_fp_extrap}
\sisetup{
table-format = 3.2 ,
table-number-alignment = center ,
table-column-width = 1.4cm ,
}
    \setlength\extrarowheight{2pt}
    \begin{tabularx}{\textwidth}{
    p{15mm} 
    C{1.2cm}
    C{1.2cm}
    *{2}{S}
    S[table-format = 1.3]
    *{2}{S}
    S[table-format = 1.3]
    *{2}{S}
    S[table-format = 1.3]
    }
    \hline\hline
    \multirow{2}{*}{Material} & \multirow{2}{*}{\shortstack{$\Delta_{\text{SOC}}$\\ (meV)}}& \multirow{2}{*}{\shortstack{{$\omega_{\text{LO}}$}\\ (meV)}}&
\multicolumn{3}{c}{VBM  (meV)} & \multicolumn{3}{c}{CBM (meV)} &
\multicolumn{3}{c}{Total ZPR (meV)}\\
\cmidrule(lr){4-6}\cmidrule(lr){7-9}\cmidrule(lr){10-12}
 & &  &noSOC & SOC & R & noSOC & SOC & R &  noSOC & SOC & R\\
\midrule
Si & 47& 62.4 &34.1 &33.2&0.974&-25.4&-25.3&0.996&-59.5&-58.5&0.983\\
ZnS & 60 &40.6& 48.1 &46.7  & 0.971 & -36.7 &-36.6 & 0.997& -84.8& -83.3& 0.982\\
CdS & 49 &34.4 &43.3 &41.7 & 0.963& -23.1 & -23.1& 1.000& -66.4& -64.8& 0.976\\
CaS & 93 & 41.6 & 78.3 & 76.2& 0.973&-55.5 & -55.4& 0.998 & -133.8 & -131.7& 0.984\\
SrS & 106 & 34.3 & 83.0 & 80.1 & 0.965& -51.9& -51.8 & 0.998& -134.9&-132.0 & 0.978\\
BaS & 137 & 30.1& 87.5& 82.9 & 0.947& -50.0& -49.9 & 0.998&-137.5& -132.7 & 0.966\\
BAs & 209 &84.4 &45.4 &41.5 & 0.914& -52.0 & -52.0 & 1.000& -97.4& -93.5 & 0.960 \\
Ge & 302 &38.7 &16.4 &14.0 & 0.854& -15.9 & -15.9 & 1.000& -32.5 & -30.1 & 0.926\\
GaAs & 328 &33.5 &19.0 &16.0 & 0.842& 2.1 & 1.9 & 0.905& -16.9& -14.1 &  0.834\\
GaAs-exp & 335 &35.4 &19.0 &16.0 & 0.842& -4.1 & -4.0 & 0.976& -23.1& -20.0 &  0.866\\
ZnSe & 381 &29.3 &29.5  & 25.4 & 0.861& -10.6 &-10.4 & 0.981& -40.1& -35.8 & 0.893\\
CdSe & 364& 23.6&24.7 &20.4 &  0.826 &-5.2 & -4.8&  0.923& -29.9& -25.2 & 0.843\\
SrSe & 403 & 24.0& 51.6 & 46.8& 0.907& -35.6 & -35.4& 0.994&-87.2 & -82.2& 0.943\\
CaSe & 407 & 31.3& 49.5 & 45.0& 0.909&-40.5 & -40.4& 0.998& -90.0 & -85.4&0.949\\
BaSe & 416 & 20.3& 53.5 & 47.7& 0.892& -33.5 & -33.3& 0.994& -87.0 & -81.0 & 0.931\\
AlSb & 661 &39.8 &24.0&19.0 & 0.792& -27.4 & -27.4 & 1.000& -50.8& -46.4 & 0.903 \\
ZnTe & 881& 24.1&18.0& 13.2 & 0.733& -1.5& -1.2& 0.800& -19.5& -14.4 & 0.738\\
CdTe & 849 &19.1&16.4  &11.4 & 0.695& -0.4& 0.4& {n/a}& -16.8 & -11.0 & 0.655\\
BaTe & 850 & 16.3& 31.5 & 25.9& 0.822& -22.4 & -21.8& 0.973& -53.9 & -47.6 & 0.885\\
SrTe & 864 & 19.5& 30.8 & 25.4& 0.825& -24.2 & -23.8&  0.983& -55.0 & -49.2 & 0.895\\
CaTe & 891 & 26.1& 28.5 & 22.5& 0.789& -27.8 & -27.3& 0.982& -56.3 & -49.8& 0.885\\
    \hline\hline
    \end{tabularx}
\end{table*}

\begin{table}[]
    \centering
    \caption{\textbf{Ratio of eigenenergy renormalization with and without SOC, $\bm{\Delta \varepsilon(\textrm{SOC})/\Delta \varepsilon(\textrm{noSOC})}$ evaluated at $\bm{T=300}$~K and $\bm{T=0}$~K,} for the VBM, CBM and band gap. All ratios are unitless. Note that $\Delta \varepsilon$($T=0$~\textrm{K}) corresponds by definition to the ZPR reported in Table~\ref{tab:zpr_fp_extrap}; hence, the ratios at $T=0$~K are identical. We include them here for ease of comparison. The CBM ratios for CdTe are absent from this Table for the same reason as in  Table~\ref{tab:zpr_fp_extrap}. }\label{tab:dedt_ratio}
\sisetup{
table-format = 3.1 ,
table-number-alignment = center ,
table-column-width = 1.05cm ,
}
    \setlength\extrarowheight{2pt}
    \begin{tabularx}{\columnwidth}{p{14mm}
    *{2}{S[table-format=1.3]}
    *{2}{S[table-format=1.3]}
    *{2}{S[table-format=1.3]}
    }
    \hline\hline
    \multirow{3}{*}{Material} & \multicolumn{6}{c}{$\Delta \varepsilon(\textrm{SOC})/\Delta \varepsilon(\textrm{noSOC})$ }\\
&\multicolumn{2}{c}{{VBM}} & \multicolumn{2}{c}{{CBM}}& \multicolumn{2}{c}{{Total ZPR}} \\
\cmidrule(lr){2-3}\cmidrule{4-5}\cmidrule(lr){6-7}
& {0~K} & {300~K} & {0~K} & {300~K}& {0~K} & {300~K}\\
\hline
Si & 0.974 & 0.979 &0.996 & 1.000 & 0.983 & 0.988\\
ZnS &  0.971 & 0.948 & 0.997 & 0.996 & 0.982 & 0.973\\
CdS & 0.963 & 0.930 & 1.000 & 1.021 & 0.976 & 0.970\\
CaS & 0.973 & 0.944 & 0.998 & 0.998 & 0.984 & 0.963\\
SrS & 0.965 & 0.917 & .998 & 1.000 & 0.978 & 0.942\\
BaS & 0.947 & 0.906 & 0.998 & 0.997 & 0.966 & 0.930\\
BAs &  0.914 & 0.907 & 1.000& 1.000 & 0.960 & 0.958\\
Ge & 0.847 & 0.837 & 1.000& 1.002 & 0.917 & 0.937\\
GaAs &  0.842 & 0.801 & 0.905 & 1.112 & 0.834 &  0.641\\
ZnSe &  0.861 & 0.875 & 0.981 & 1.000 & 0.893 & 0.921\\
CdSe & 0.826 & 0.828 & 0.923 & 0.918 & 0.843 & 0.845\\
SrSe & 0.907 & 0.874 & 0.994 & 0.995 & 0.943 & 0.916\\
CaSe & 0.909 & 0.827 & 0.998 & 1.002 & 0.949 & 0.889\\
BaSe &  0.892 & 0.878 & 0.994 & 0.995 & 0.931 & 0.916\\
AlSb & 0.792 & 0.799 & 1.000 & 1.007 & 0.903 & 0.909\\
ZnTe & 0.733 & 0.704 & 0.800 & 0.954 & 0.738 & 0.747\\
CdTe & 0.695 & 0.651 & {n/a} & {n/a} & 0.655 & 0.555\\
BaTe & 0.822 & 0.824 & 0.973 & 0.981 & 0.885 & 0.881\\
SrTe & 0.825 & 0.822 & 0.983 & 0.982 & 0.895 & 0.885\\
CaTe &  0.789 & 0.736 & 0.982 & 0.982 & 0.885 & 0.838\\
    \hline\hline
    \end{tabularx}
\end{table}

\begin{table*}[]
    \centering
    \caption{\textbf{Comparison between the total band gap ZPR and experimental values.} The contribution of the zero-point lattice expansion to the band gap ZPR, ZPR$_{\rm{g}}^{\rm{ZPLE}}$, is added to the EPI contribution, ZPR$_{\rm{g}}^{\rm{EPI}}$ (see Table~\ref{tab:zpr_fp_extrap}),
    to obtain the total band gap ZPR, ZPR$_{\rm{g}}^{\rm{tot}}$, which is compared to experimental data shown in Fig.~4 of the main text, ZPR$_{\rm{g}}^{\rm{exp}}$. Note that the same value of ZPR$_{\rm{g}}^{\rm{ZPLE}}$, computed without SOC, was applied to our ZPR$_{\rm{g}}^{\rm{EPI}}$ results, both with and without SOC. See Appendix~D of the main text for more details. All ZPR values are in meV. 
    R=ZPR$^{\rm{tot}}$/ZPR$^{\rm{exp}}$ is unitless. ZPR$^{\rm{tot}}$ is rounded to 1~meV before computing R.
    }\label{tab:zprte}
\sisetup{
table-format = 3.2 ,
table-number-alignment = center ,
table-column-width = 1.75cm ,
}
    \setlength\extrarowheight{2pt}
    \begin{tabularx}{0.8\textwidth}{p{15mm}
*{1}{S[table-format=3.1]}
    *{1}{S}
    *{1}{S[table-format=3.0, table-column-width=2.5cm]}
    *{1}{S[table-column-width=1.8cm]}
    *{1}{S[table-format=3.0, table-column-width=2.5cm]}
    *{1}{S[table-column-width=1.8cm]}
    }
    \hline\hline
    Material & {ZPR$_{\rm{g}}^{\rm{ZPLE}}$} & ZPR$_{\rm{g}}^{\rm{exp}}$ & {ZPR$_{\rm{g}}^{\rm{tot}}$ (noSOC )}     & {R (noSOC)} & {ZPR$_{\rm{g}}^{\rm{tot}}$ (SOC)} & {R (SOC)} \\[1.5pt]
    \hline
    Si & 8.7 & {-59~\cite{karaiskaj_photoluminescence_2002}}& -51 & 0.86& -50 &  0.85\\
    ZnS  &-24.2 & {-105~\cite{manjon_effect_2005}} & -109 & 1.04& -108 & 1.02\\
    CdS & -10.5 & {-62~\cite{zhang_isotope_1998}} & -77 & 1.24& -76 & 1.24\\
    Ge  &-9.3 & {-52~\cite{parks_electronic_1994}} & -42 & 0.81& -39 & 0.75\\
    GaAs  &-31.4 & {-60~\cite{passler_dispersion_2002}} & -48 & 0.80& -46 & 0.77\\
    ZnSe  &-17.2 & {-55~\cite{passler_dispersion_2002}} & -57 & 1.04& -53 & 0.96\\
    CdSe &-7.2 & {-38~\cite{passler_dispersion_2002}} & -37 & 0.97& -32 & 0.84\\
    AlSb  &5.5 & {-35~\cite{passler_dispersion_2002}} & -45 & 1.24& -40 & 1.11\\
    ZnTe & -17.9 & {-40~\cite{passler_dispersion_2002}} & -37 & 0.93& -32 & 0.80\\
    CdTe & -9.0 & {-16~\cite{passler_dispersion_2002}} &-26 & 1.63& -20 & 1.25\\
    \hline\hline
    \end{tabularx}
\end{table*}

\begin{table*}[]
    \centering
    \caption{\textbf{Generalized \fro model ZPR.} R=ZPR(SOC)/ZPR(noSOC) is unitless. When SOC is included, the angular-averaged effective masses entering Eq.~(34) of the main text were evaluated from the electronic dispersion obtained either with the Luttinger-Kohn Hamiltonian (SOC-LK), from DFT (SOC-DFT), or from the Luttinger-Kohn Hamiltonian with fitted Luttinger parameters (SOC-$\widetilde{\textrm{LK}}$, see Appendix~C of the main text). For the VBM, the special treatment of the SOC-DFT data based on the Dresselhaus Hamiltonian is described in Appendix~B of the main text. The different ratios (R-LK, R-DFT, R-$\widetilde{\textrm{LK}}$) refer to the calculation method with SOC. All effective masses without SOC were computed with density-functional perturbation theory. For GaAs, results obtained with both theoretically relaxed lattice parameter (\enquote{GaAs}) and experimental lattice parameter (\enquote{GaAs-exp}) are provided.}\label{tab:gfr_compare}
\sisetup{
table-format = 3.1 ,
table-number-alignment = center ,
table-column-width = 1.10cm ,
}
    \setlength\extrarowheight{2pt}
    \begin{tabularx}{\textwidth}{p{14mm}
    S[table-column-width = 0.95cm]
    S[table-column-width = 1.15cm]
    S[table-column-width = 1.30cm]
    S[table-column-width = 1.15cm]
    *{2}{S[table-format = 1.3, table-column-width = 1.00cm]}
    S[table-format = 1.3, table-column-width = 0.95cm]
    S[table-column-width = 1.15cm]
    S[table-column-width = 1.30cm]
    S[table-format = 1.3, table-column-width = 1.15cm]
    S[table-column-width = 1.15cm]
    S[table-column-width = 1.30cm]
    S[table-format = 1.3, table-column-width = 1.15cm]   
    }
    \hline\hline
    \multirow{2}{*}{Material} & 
\multicolumn{7}{c}{VBM  (meV)} & \multicolumn{3}{c}{CBM (meV)} &
\multicolumn{3}{c}{Total ZPR (meV)}\\[2pt]
\cmidrule(lr){2-8}\cmidrule(lr){9-11}\cmidrule(lr){12-14}\\[-10pt]
 & {\footnotesize{noSOC}} & {\footnotesize{ SOC-LK}} & {\footnotesize{SOC-DFT}} & {\footnotesize{SOC-$\widetilde{\textrm{LK}}$}}& {\footnotesize{R-LK}} & {\footnotesize{R-DFT}} & {\footnotesize{R-$\widetilde{\textrm{LK}}$}}& {\footnotesize{noSOC}} &  {\footnotesize{SOC-DFT}} & {\footnotesize{R-DFT}} &   {\footnotesize{noSOC}} & {\footnotesize{SOC-DFT}} & {\footnotesize{R-DFT}}\\[2pt]
\midrule
ZnS &  40.1 & 34.7& 34.3& 34.5&0.865 & 0.855& 0.860& -18.6 & -18.6& 1.000& -58.7 & -52.9 & 0.901\\
CdS &  38.7 & 32.9& 32.4 & 32.7&0.850 & 0.837& 0.845& -14.9 & -14.9& 1.000& -53.6 & -47.3 & 0.882\\
CaS &  74.5 & 70.1& 69.7& 69.7&0.941 & 0.936& 0.936& -61.3 & -61.2& 0.998& -135.8 & -130.9 & 0.964\\
SrS &  80.6 & 74.8& 74.3& 74.3&0.928& 0.922&0.922 & -61.5 & -61.4& 0.998& -142.1 & -135.7 & 0.955\\
BaS &   86.8 & 78.8& 78.1& 78.1&0.908 & 0.900& 0.900& -63.1 & -63.1& 1.000& -149.9 & -141.2 & 0.942\\
BAs &  0.5 & 0.5& 0.5& 0.5&1.000& 1.000&1.000 &-0.5& -0.5& 1.000& -1.0 & -1.0 & 1.000\\
GaAs & 3.3 & 2.4& 2.2& 2.2&0.727 & 0.667&0.667 &-0.5 & -0.4& 0.800& -3.8 & -2.6 & 0.690\\
GaAs-exp &  3.4 & 2.9& 2.7& 2.7& 0.853 & 0.794 & 0.794& -1.0 & -1.0 & 1.000 & -4.4 & -3.7 & 0.841\\
ZnSe &  21.7 & 18.0& 17.5& 17.5&0.829 & 0.806& 0.806& -8.1& -7.9& 0.975& -29.8 & -25.4 & 0.852\\
CdSe &  19.7 & 15.7& 15.0& 15.0&0.797 & 0.761& 0.761& -5.5 & -5.1& 0.927& -25.2 & -20.1 & 0.798\\
SrSe &  50.9 & 47.1& 45.8 & 45.8&0.925 & 0.900& 0.900& -41.4 & -41.2& 0.995& -92.3 & -87.0 & 0.943\\
CaSe &  48.9 & 45.5& 44.6& 44.6&0.935 & 0.912& 0.912& -42.9 & -42.7& 0.995& 91.8 & -87.3 & 0.951\\
BaSe & 53.8 & 48.9& 47.5& 47.5&0.909 & 0.883& 0.883& -42.2 & -41.9& 0.993& -96.0 & -89.4 & 0.931\\
AlSb &  4.0 & 3.3& 3.1& 3.1&0.825& 0.775& 0.775& -3.6 & -3.5& 0.972& -7.6 & -6.6 & 0.868 \\
ZnTe &  11.1 & 8.9& 8.3& 8.3&0.802 & 0.748&0.748 &-4.3 &-3.9& 0.907& -15.4 & -12.2 & 0.792\\
CdTe &  11.9 & 9.1& 8.2& 8.2&0.765& 0.689&0.689 &-3.7 & -2.9& 0.784& -15.6 & -11.1 & 0.712\\
BaTe &   32.6 & 29.6& 27.7& 27.7&0.908 & 0.850& 0.850& -28.1& -27.4& 0.975& -60.7 & -55.1 & 0.908\\
SrTe &  31.5 & 29.0& 27.3& 27.3&0.921 & 0.867& 0.867& -27.8 & -27.3& 0.982& -59.3 & -54.6 & 0.921\\
CaTe &  30.2 & 28.1& 26.4 & 26.4&0.930 & 0.874& 0.874& -28.7 & -28.2& 0.983& -58.9 & -54.6 & 0.927\\
    \hline\hline
    \end{tabularx}
\end{table*}

\begin{table*}[]
    \centering
    \caption{\textbf{Generalized \fro model parameters} used in the evaluation of Eq.~(34) of the main text: LO phonon frequency $\omega_{\rm{LO}}$ (in meV), optical dielectric constant $\epsilon^\infty$, Born effective charges $Z^\ast$ and absolute value of the angular and band averaged square root effective masses $\braket{\braket{(m^\ast)^{1/2}}}$ for the VBM and CBM (in units of square-root of the bare electron mass), with and without SOC. Square root effective masses with SOC were computed with the Dresselhaus model.}\label{tab:gfr_model_params}
\sisetup{
table-number-alignment = center
}
    \setlength\extrarowheight{2pt}
    \begin{tabularx}{\textwidth}{p{22mm}
    *{3}{S[table-column-width = 1.0cm]}
    *{2}{S[table-column-width = 2.0cm]}
    *{3}{S[table-column-width = 1.0cm]}
    *{2}{S[table-column-width = 2.0cm]}
    }
    \hline\hline
    \multirow{2}{*}{Material}& \multicolumn{5}{c}{noSOC} &\multicolumn{5}{c}{SOC}\\
   \cmidrule(lr){2-6}\cmidrule(lr){7-11}\\[-10pt]
  &{$\omega_{\rm{LO}}$} &{$\epsilon^\infty$} & {$Z^\ast$} & {$\braket{\braket{(m^\ast)^{1/2}}}_{\rm{v}}$}&{$\braket{\braket{(m^\ast)^{1/2}}}_{\rm{c}}$}& {$\omega_{\rm{LO}}$} &{$\epsilon^\infty$} & {$Z^\ast$}& {$\braket{\braket{(m^\ast)^{1/2}}}_{\rm{v}}$}&{$\braket{\braket{(m^\ast)^{1/2}}}_{\rm{c}}$}\\
        \midrule   
ZnS & 40.6 & 5.97 & 2.03 & 0.882 & 0.409 &
40.6 & 5.97 & 2.03 & 0.756 & 0.409\\
CdS & 34.4 & 6.21 & 2.23 & 0.893 & 0.343 &
34.4 & 6.22 & 2.23 & 0.749 & 0.343\\
CaS & 41.6 & 4.99 & 2.37 & 0.836 & 0.688 &
41.6 & 4.99 & 2.37 & 0.784 & 0.688\\
SrS & 34.2 & 4.63 & 2.40 & 0.893 & 0.681 &
34.3 &4.63 & 2.40 & 0.824 & 0.681\\
BaS & 30.1 & 4.78 & 2.60 & 0.950 & 0.691 &
30.0 & 4.78 & 2.60 & 0.855 & 0.690\\
BAs & 84.4 & 9.81 & 0.45 & 0.596 & 0.576 &
84.4 & 9.81 & 0.45 & 0.527 & 0.575\\
GaAs & 33.5 & 15.31 & 2.23 & 0.581 & 0.095 &
33.5 & 15.42 & 2.23 & 0.395 & 0.063\\
GaAs-exp & 35.4 & 13.78 & 2.12 & 0.594 & 0.175&
35.5 & 13.85 & 2.12 & 0.454 & 0.169\\
ZnSe & 29.3 & 7.35 & 2.10 & 0.802 & 0.299&
29.3 & 7.38 & 2.10 & 0.651 & 0.296\\
CdSe & 23.6 & 7.83 & 2.31 & 0.812 & 0.226&
23.5 & 7.91 & 2.31 & 0.630 & 0.215\\
SrSe & 24.0 & 5.15 & 2.39 & 0.812 & 0.660&
24.0 & 5.17 & 2.40 & 0.734 & 0.660\\
CaSe & 31.3 & 5.61 & 2.38 & 0.761 &0.668&
31.4 & 5.62 & 2.38 & 0.696 & 0.667\\
BaSe & 20.3 & 5.26 & 2.59 & 0.857 & 0.672&
20.3 & 5.28 & 2.59 & 0.760 & 0.671\\
AlSb & 39.8 & 12.02 & 1.83 & 0.656 & 0.584 &
39.8 & 12.13 &1.84 & 0.514 & 0.585\\
ZnTe & 24.1 & 9.05 & 2.09 & 0.713 & 0.276&
24.1 & 9.25 & 2.09 & 0.554 & 0.261\\
CdTe & 19.1 & 8.89 & 2.29 & 0.738 & 0.228&
19.0 & 9.27 & 2.29 & 0.545 & 0.191\\
BaTe & 16.3 & 5.93 & 2.59 & 0.749 & 0.645&
16.3 & 6.02 & 2.59 & 0.651 & 0.642\\
SrTe & 19.5 & 5.90 & 2.42 & 0.719 & 0.633&
19.4 & 5.96 & 2.42 & 0.633 & 0.632\\
CaTe & 26.1 & 6.57 & 2.42 & 0.673 & 0.640&
26.1 & 6.63 & 2.42 & 0.598 & 0.638\\
    \hline\hline
    \end{tabularx}
\end{table*}

\begin{table*}[]
    \centering
    \caption{\textbf{Luttinger-Kohn and Dresselhaus model parameters.} This table regroups the parameters used in Eq.~(A3)--(A4) and (B1)--(B3) of the main text to evaluate the electronic dispersion from which one can extract the angular-averaged effective masses entering our gFr model with SOC (Eq.~(34) of the main text). Note that the $W$ parameter, labeled $C$ in the original work from Dresselhaus~\cite{dresselhaus_cyclotron_1955}, was renamed to avoid confusion with the $3^{\rm{rd}}$ Luttinger parameter. To extract these parameters from our fitting parameters, $\lambda$ and $\alpha^2$ (see Eq.~(B3) and~(B5)), keep note that Eq.~(B1) is always negative for hole-like bands, and that we suppose that $|L|>|M|$, as observed for the Luttinger parameters. Note that $W=0$ indicates a centrosymmetric materials. See Appendix~C for more details. We also provide the fitted Luttinger parameters ($\widetilde{A}$, $\widetilde{B}$,  $\widetilde{C}$) derived from the Dresselhaus model. All parameters are in atomic units. The split-off energy used in the Luttinger-Kohn Hamiltonian can be found in Table~(\ref{tab:zpr_fp_extrap}). Note that the Luttinger parameters ($A$, $B$, $C$) listed here for the zincblende materials were originally published by us in Ref.~\cite{guster_frohlich_2021}. We include them here for completeness.
    }\label{tab:sm-modelparams}
    \setlength\extrarowheight{2pt}
    \begin{tabularx}{\textwidth}{p{18mm}
    *{10}{R{1.45cm}}
    }
    \hline\hline
    \multirow{2}{*}{Material} & \multicolumn{4}{c}{Dresselhaus} & \multicolumn{3}{c}{Luttinger-Kohn} & \multicolumn{3}{c}{Fitted Luttinger-Kohn}\\
    \cmidrule(lr){2-5} \cmidrule(lr){6-8} \cmidrule(lr){9-11}\\[-12pt]
   {} & {$L$~~~~} & {$M$~~~~} & {$N$~~~~} & {$W$ ($10^{-3}$)} & {$A$~~~~} & {$B$~~~~} & {$C$~~~~} & {$\widetilde{A}$~~~~}& {$\widetilde{B}$~~~~}& {$\widetilde{C}$~~~~}\\
        \midrule
ZnS & -3.293& -1.206& -3.234& 1.352& -2.751 & -0.694& -3.170& -2.793& -0.706& -3.234\\
CdS & -4.575& -1.124& -4.443& 2.495& -3.999 & -0.605& -4.321& -4.075& -0.624& -4.443\\
CaS & -2.715& -0.912& -1.290& 0.000& -2.186 & -0.409& -1.276 & -2.215& -0.412& -1.290\\
SrS & -2.774& -0.799& -1.139& 0.000& -2.220 & -0.300& -1.122& -2.274& -0.299& -1.139\\
BaS & -2.859& -0.694& -0.753& 0.000& -2.249 & -0.204& -0.736& -2.359& -0.194& -0.753\\
BAs & -2.853& -2.598& -3.925& 0.183& -2.337 & -2.104& -3.912 & -2.353& -2.098& -3.925\\
GaAs & -189.915& -1.957& -190.432& 0.323& -54.896 & -1.362& -55.859 & -189.415& -1.457& -190.432\\
GaAs-exp & -21.263& -2.049& -21.968& 0.360&-16.371 & -1.450& -17.412 & -20.763& -1.549& -21.968\\
ZnSe & -6.589& -1.327& -6.630& 1.074& -5.340 & -0.791& -5.834 & -6.089& -0.827& -6.630\\
CdSe & -13.236& -1.231& -13.202& 2.120& -9.504 & -0.684 &-9.881  & -12.736& -0.731& -13.202\\
SrSe & -3.440& -0.894& -1.703& 0.000& -2.752 & -0.380& -1.592 & -2.940& -0.394& -1.703\\
CaSe & -3.428& -1.032& -1.924& 0.000& -2.769 & -0.508& -1.817 & -2.928& -0.532& -1.924\\
BaSe & -3.416& -0.773& -1.131& 0.000& -2.680 & -0.270& -1.047 & -2.916& -0.273& -1.131\\
AlSb & -8.382& -1.992& -9.040& 0.109& -6.473 &-1.372 & -7.520 & -7.882& -1.492& -9.040\\
ZnTe & -9.533& -1.641& -9.832& 0.769& -6.495 & -1.032& -7.174 & -9.033& -1.141& -9.832\\
CdTe & -18.628& -1.486& -18.819& 1.711& -9.517 & -0.867& -10.033 & -18.128& -0.986& -18.819\\
BaTe & -4.176& -0.941& -1.543& 0.000& -3.164 & -0.399& -1.329 & -3.676& -0.441& -1.543\\
SrTe & -4.198 & -1.094& -2.291 & 0.000& -3.226 & -0.534& -1.984 & -3.698 & -0.594& -2.291\\
CaTe & -4.209& -1.293& -2.587& 0.000& -3.261 & -0.707& -2.273 & -3.709& -0.793& -2.587\\
\hline\hline
    \end{tabularx}
\end{table*}

\begin{table*}[]
    \centering
    \caption{\textbf{Absolute value of the angular averaged effective masses $\bm{\braket{(m_n^\ast)^{1/2}}^2}$ and absolute value of the band and angular averaged effective masses $\bm{\braket{\braket{(m^\ast)^{1/2}}}^2}$} for the VBM, with and without SOC (in units of the bare electron mass). Effective masses with SOC were computed with the Dresselhaus model.
 }\label{tab:gfr_angular_efmas}

\sisetup{
table-number-alignment = center
}
    \setlength\extrarowheight{2pt}
    \begin{tabularx}{\textwidth}{p{22mm}
    *{3}{S[table-column-width = 1.2cm]}
    S[table-column-width = 3.0cm]
    *{4}{S[table-column-width = 1.2cm]}
    S[table-column-width = 3.0cm]
    }
    \hline\hline
    \multirow{2}{*}{Material}& \multicolumn{4}{c}{noSOC} &\multicolumn{5}{c}{SOC}\\
   \cmidrule(lr){2-5}\cmidrule(lr){6-10}\\[-10pt]
  &\multicolumn{3}{c}{$\braket{(m_n^\ast)^{1/2}}_{\rm{v}}^2$} & {$\braket{\braket{(m^\ast)^{1/2}}}_{\rm{v}}^2$} & \multicolumn{4}{c}{$\braket{(m_n^\ast)^{1/2}}_{\rm{v}}^2$} & {$\braket{\braket{(m^\ast)^{1/2}}}_{\rm{v}}^2$}\\
        \midrule   
ZnS & 0.156 & 0.842 & 1.777 & 0.778 & 0.205 & 0.224 & 1.097 & 1.097 & 0.571 \\
CdS & 0.114 & 0.956 & 1.861 & 0.798 & 0.156 & 0.165 & 1.204 & 1.204 & 0.561\\
CaS & 0.251 & 0.896 & 1.123 & 0.699 & 0.330 & 0.330 & 0.987 & 0.987 & 0.614\\
SrS & 0.262 & 0.956 & 1.414 & 0.797 & 0.343 & 0.343 & 1.128 & 1.128 & 0.678\\
BaS & 0.285 & 0.953 & 1.791 & 0.902 & 0.360 & 0.360 & 1.232 & 1.232 & 0.731\\
BAs & 0.126 & 0.285 & 0.809 & 0.355 & 0.158 & 0.160 & 0.429 & 0.429 & 0.278\\
GaAs & 0.009 & 0.433& 0.983 & 0.338 & 0.004 & 0.004 & 0.521 & 0.538 & 0.156\\
GaAs-exp & 0.029 & 0.408 & 0.950 & 0.353 & 0.033 & 0.034 & 0.526 & 0.526 & 0.206\\
ZnSe & 0.085 & 0.740 & 1.569 & 0.643 & 0.107 & 0.110 & 0.946 & 0.946 & 0.424\\
CdSe & 0.050 & 0.849 & 1.668 & 0.660 & 0.055 & 0.056 & 1.047 & 1.047 & 0.396\\
SrSe & 0.204 & 0.827 & 1.154 & 0.659 & 0.259 & 0.259 & 0.918 & 0.918 & 0.538\\
CaSe & 0.193 & 0.777 & 0.924 & 0.579 & 0.245 & 0.245 & 0.805 & 0.805 & 0.485\\
BaSe & 0.232 & 0.807 & 1.419 & 0.734 & 0.284 & 0.284 & 0.977 & 0.977 & 0.578\\
AlSb & 0.067 & 0.434 & 1.100 & 0.430 & 0.079 & 0.079 & 0.556 & 0.556 & 0.264\\
ZnTe & 0.070 & 0.570 & 1.254 & 0.508 & 0.073 & 0.074 & 0.699 & 0.699 & 0.306\\
CdTe & 0.050 & 0.672 & 1.374 & 0.545 & 0.039 & 0.040 & 0.794 & 0.794 & 0.296\\
BaTe & 0.191 & 0.640 & 1.020 & 0.561 & 0.219 & 0.219 & 0.703 & 0.703 & 0.423\\
SrTe & 0.170 & 0.671 & 0.854 & 0.516 & 0.199 & 0.199 & 0.672 & 0.672 & 0.401\\
CaTe & 0.159 & 0.629 & 0.687 & 0.454 & 0.187 & 0.187 & 0.584 & 0.584 & 0.358\\
    \hline\hline
    \end{tabularx}
\end{table*}

\FloatBarrier
%
\end{document}